\documentclass[twocolumn]{aastex631}
\usepackage{natbib}
\usepackage{amssymb}
\usepackage{amsmath}
\def\be{\begin{eqnarray}}
\def\ee{\end{eqnarray}}
\usepackage{comment}
\usepackage{graphicx,subfigure}
\usepackage{enumerate}
\usepackage{rotating}
\usepackage{longtable}
\usepackage{booktabs}
\usepackage{float}

\usepackage{hyperref}
\hypersetup{colorlinks,urlcolor={blue},linkcolor={blue},citecolor={blue}}

\shorttitle{A search for relativistic SN\lowercase{e} Ic-BL}

\shortauthors{Corsi et al.}

\begin{document}

\title{A search for relativistic ejecta in a sample of ZTF broad-lined Type Ic supernovae}

\author[0000-0001-8104-3536]{Alessandra~Corsi}
\email{alessandra.corsi@ttu.edu}
\affiliation{Department of Physics and Astronomy, Texas Tech University, Box 1051, Lubbock, TX 79409-1051, USA.}

\author[0000-0002-9017-3567]{Anna~Y.~Q.~Ho}
\affiliation{Department of Astronomy, Cornell University, Ithaca, NY 14853, USA}
\affiliation{Miller Institute for Basic Research in Science, 468 Donner Lab, Berkeley, CA 94720, USA}
\affiliation{Department of Astronomy, University of California, Berkeley, Berkeley, CA, 94720, USA}
\affiliation{Lawrence Berkeley National Laboratory, 1 Cyclotron Road, MS 50B-4206, Berkeley, CA 94720, USA}

\author[0000-0003-1673-970X]{S.~Bradley Cenko}
\affiliation{Astrophysics Science Division, NASA Goddard Space Flight Center, MC 661, Greenbelt, MD 20771, USA}
\affiliation{Joint Space-Science Institute, University of Maryland, College Park, MD 20742, USA}

\author[0000-0001-5390-8563]{Shrinivas~R.~Kulkarni}
\affiliation{Division of Physics, Mathematics and Astronomy,  California Institute of Technology, Pasadena, CA 91125, USA}

\author[0000-0003-3768-7515]{Shreya Anand}
\affiliation{Division of Physics, Mathematics and Astronomy,  California Institute of Technology, Pasadena, CA 91125, USA}

\author[0000-0002-2898-6532]{Sheng Yang}
\affiliation{The Oskar Klein Center, Department of Astronomy, Stockholm University, AlbaNova, 10691 Stockholm, Sweden}

\author[0000-0003-1546-6615]{Jesper Sollerman}
\affiliation{The Oskar Klein Center, Department of Astronomy, Stockholm University, AlbaNova, 10691 Stockholm, Sweden}

\author[0000-0002-6428-2700]{Gokul~P.~Srinivasaragavan}
\affiliation{Department of Astronomy, University of Maryland, College Park, MD 20742, USA}

\author[0000-0002-9646-8710]{Conor~M.~B.~Omand}
\affiliation{The Oskar Klein Center, Department of Astronomy, Stockholm University, AlbaNova, 10691 Stockholm, Sweden}

\author[0000-0003-0477-7645]{Arvind Balasubramanian}
\affiliation{Department of Physics and Astronomy, Texas Tech University, Box 1051, Lubbock, TX 79409-1051, USA}

\author{Dale A.~Frail}
\affiliation{National Radio Astronomy Observatory, Socorro, NM 87801,
USA}

\author[0000-0002-4223-103X]{Christoffer~Fremling}
\affiliation{Division of Physics, Mathematics and Astronomy,  California Institute of Technology, Pasadena, CA 91125, USA}
\affiliation{Caltech Optical Observatories, California Institute of Technology, Pasadena, CA 91125, USA}

\author[0000-0001-8472-1996]{Daniel~A.~Perley}
\affiliation{Astrophysics Research Institute, Liverpool John Moores University, IC2, Liverpool Science Park, 146 Brownlow Hill, Liverpool L3 5RF, UK}

\author[0000-0001-6747-8509]{Yuhan Yao}
\affiliation{Division of Physics, Mathematics and Astronomy,  California Institute of Technology, Pasadena, CA 91125, USA} 

\author{Aishwarya~S.~Dahiwale}
\affiliation{Department of Physics, Michigan Technological University, Fisher Hall 118, 1400 Townsend Drive Houghton, MI 49931}

\author[0000-0002-3961-1365]{Kishalay De}
\affiliation{MIT---Kavli Institute for Astrophysics and Space Research, 77 Massachusetts Ave., Cambridge, MA 02139, USA}

\author{Alison Dugas}
\affiliation{Department of Physics and Astronomy, Watanabe 416, 2505 Correa Road, Honolulu, HI 96822, USA}

\author{Matthew Hankins}
\affiliation{Department of Physics, Arkansas Tech University, 1701 N Boulder Avenue Russellville, AR 72801, USA}

\author{Jacob Jencson}
\affiliation{Steward Observatory, University of Arizona, 933 North Cherry Avenue, Rm. N204, Tucson, AZ 85721-0065, USA}

\author[0000-0002-5619-4938]{Mansi~M.~Kasliwal}
\affiliation{Division of Physics, Mathematics and Astronomy,  California Institute of Technology, Pasadena, CA 91125, USA}

\author[0000-0003-0484-3331]{Anastasios Tzanidakis}
\affiliation{Department of Astronomy, University of Washington, 3910 15th Avenue NE, Seattle, WA 98195, USA}

\author[0000-0001-8018-5348]{Eric C. Bellm}
\affiliation{DIRAC Institute, Department of Astronomy, University of Washington, 3910 15th Avenue NE, Seattle, WA 98195, USA}

\author[0000-0003-2451-5482]{Russ~R.~Laher}
\affiliation{IPAC, California Institute of Technology, 1200 E. California Blvd, Pasadena, CA 91125, USA}

\author[0000-0002-8532-9395]{Frank~J.~Masci}
\affiliation{IPAC, California Institute of Technology, 1200 E. California Blvd, Pasadena, CA 91125, USA}

\author{Josiah~N.~Purdum}
\affiliation{Division of Physics, Mathematics and Astronomy,  California Institute of Technology, Pasadena, CA 91125, USA}

\author{Nicolas Regnault}
\affiliation{LPNHE, CNRS/IN2P3 \& Sorbonne Universit\`e, 4 place Jussieu, 75005 Paris, France}

\begin{abstract}
\label{abstract}
The dividing line between gamma-ray bursts (GRBs) and ordinary stripped-envelope core-collapse supernovae (SNe) is yet to be fully understood. Observationally mapping the variety of ejecta outcomes (ultra-relativistic, mildly-relativistic or non-relativistic) in SNe of Type Ic with broad lines (Ic-BL) can provide a key test to stellar explosion models. However, this requires large samples of the rare Ic-BL events with follow-up observations in the radio, where fast ejecta can be probed largely free of geometry and viewing angle effects. Here, we present the results of a radio (and X-ray) follow-up campaign of 16 SNe Ic-BL detected by the Zwicky Transient Facility (ZTF). Our radio campaign resulted in 4 counterpart detections and 12 deep upper limits. None of the events in our sample is as relativistic as SN\,1998bw and we constrain the fraction of SN\,1998bw-like explosions to $< 19\%$ (3$\sigma$ Gaussian equivalent), a factor of $\approx 2$ smaller than previously established. We exclude relativistic ejecta with radio luminosity densities in between $\approx 5\times10^{27}$\,erg\,s$^{-1}$\,Hz$^{-1}$ and  $\approx 10^{29}$\,erg\,s$^{-1}$\,Hz$^{-1}$ at $t\gtrsim 20$\,d since explosion for $\approx 60\%$ of the events in our sample. This shows that SNe Ic-BL similar to the GRB-associated SN\,1998bw, SN\,2003lw, SN\,2010dh, or to the relativistic SN\,2009bb and iPTF17cw, are rare. Our results also exclude an association of the SNe Ic-BL in our sample with largely off-axis GRBs with energies $E\gtrsim 10^{50}$\,erg. The parameter space of SN\,2006aj-like events (faint and fast-peaking radio emission) is, on the other hand, left largely unconstrained and systematically exploring it represents a promising line of future research.
\end{abstract}

\keywords{\small supernovae: general -- supernovae: individual -- radiation mechanisms: general  -- radio continuum: general}

\section{Introduction}
\label{intro}
Massive stars contribute to the chemical composition of matter as we know it in the universe, and their deaths are accompanied by energetic core-collapse supernovae (SNe) that seed our universe with black holes (BHs) and neutron stars (NSs) -- the most exotic objects of the stellar graveyard. Large time-domain surveys of the sky \citep[e.g.,][]{York2000,Drake2009,Law2009,Kaiser2010,Shappee2014,Abbott2016,Tonry2018,Bellm2019}, paired with targeted follow-up efforts, have greatly enriched our view on the final stages of massive star evolution. Yet, a lot remains to be understood about the diverse paths that bring massive stars toward
their violent deaths \citep{Langer2012}.  

Core-collapse SNe can occur in stars with a hydrogen envelope
(Type II) or in stars where hydrogen is almost or completely missing \citep[Type Ib/c, also referred to as stripped-envelope SNe;][]{Filippenko1997,Matheson2001,Li2011,Modjaz2014,Perley2020b,Frohmaier2021}. Type Ib/c SNe constitute approximately 25\% of all massive star explosions \citep{Smith2011}, and their pre-SN progenitors are thought to be either massive ($M\gtrsim 20-25$\,M$_{\odot}$) and metal-rich single stars that have been stripped through stellar mass loss; or the mass donors in close binary systems (at any metallicity) that have initial masses $\gtrsim 8$\,M$_{\odot}$ \citep[e.g.,][and references therein]{Langer2012}. 

A small fraction of Type Ib/c SNe show velocities in their optical spectra that are systematically higher than those measured in ordinary SNe Ic at similar epochs. Hence, these explosions are referred to as SNe of Type Ic with broad lines  \citep[hereafter, Ic-BL; e.g.,][]{Filippenko1997,Modjaz2016,GalYam2017}. Compared to Type Ib/c SNe, broad-lined events are found to prefer environments with lower metallicity (in a single star scenario, mass loss mechanisms also remove angular momentum and are enhanced by higher metallicities), and in galaxies with higher star-formation rate density. Thus, it has been suggested that SN Ic-BL progenitors may be stars younger and more massive than those of normal Type Ic (more massive progenitors can lose their He-rich layers to winds at lower metallicity due to the higher luminosities driving the winds), and/or tight massive binary systems that can form efficiently in dense stellar clusters \citep[e.g.,][]{Kelly2014,Japelj2018,Modjaz2020}. 

The spectroscopic and photometric properties used to classify core-collapse SNe are largely determined by the stars' outer envelopes \citep[envelope mass, radius, and chemical composition;][]{Young2004}. On the other hand, quantities such as explosion energies, nickel masses, and ejecta geometries can be inferred via extensive multi-wavelength and multi-band observations. These quantities, in turn, can help constrain the properties of the stellar cores  \citep[such as mass, density structure, spin, and magnetic fields; see e.g. ][and references therein]{Woosley2002,Burrows2007,Jerkstrand2015} that are key to determine the nature of the explosion. For example, based on nickel masses and ejecta masses derived from bolometric light curve analyses, \citet{Taddia2019} found that $\gtrsim 21\%$ of Ic-BL  progenitors are compatible with massive ($\gtrsim 28$\,M$_{\odot}$), possibly single stars, whereas $\gtrsim 64\%$ could be associated with less massive stars in close binary systems.  

Understanding the progenitor scenario of SNe Ic-BL is particularly important as these SNe challenge greatly the standard explosion mechanism of massive stars \citep[e.g.,][and references therein]{Mezzacappa1998,MacFadyen1999,Heger2003,WoosleyHeger2006,Janka2007,Janka2012,Smith2014,Foglizzo2015,Muller2020,Schneider2021}. The energies inferred from optical spectroscopic modeling of Ic-BL events are of order $\approx 10^{52}\,{\rm erg}$, in excess of the $\approx 10^{51}\,{\rm erg}$ inferred in typical SNe Ib/c, while ejecta masses are comparable or somewhat higher \citep{Taddia2019}. In the traditional core-collapse scenario, neutrino irradiation from the proto-NS revives the core-bounce shock, making the star explode. However, the neutrino mechanism cannot explain the more energetic SNe Ic-BL. Unveiling the nature of an engine powerful enough to account for the extreme energetics of SNe Ic-BL is key to understanding the physics behind massive stellar deaths, and remains as of today an open question. 

A compelling scenario invokes  the existence of a jet or a newly-born magnetar as the extra source of energy needed to explain SNe Ic-BL \citep[e.g.,][]{Burrows2007,Papish2011,Gilkis2014,Mazzali2014,Lazzati2012,Gilkis2016,Soker2017,Barnes2018,Shankar2021}. The rapid rotation of a millisecond proto-NS formed in the collapse of a rotating massive star can amplify the NS magnetic field to $\gtrsim 10^{15}$\,G, creating a magnetar. The magnetar spins down quickly via magnetic braking, and in some cases magneto-rotational instabilities can launch a collimated jet that drills through the outer layers of the star producing a gamma-ray burst \citep[GRB; e.g.,][]{Heger2003,Izzard2004,WoosleyHeger2006,Burrows2007,Bugli2020,Bugli2021}.  These jets can transfer sufficient energy to the surrounding stellar material to explode it into a SN.

The above scenario is particularly interesting in light of the fact that SNe Ic-BL are also the only type of core-collapse events that, observationally, have been unambiguously linked to GRBs \citep[e.g.,][and references therein]{Woosley2006,Cano2017}. GRBs are characterized by bright afterglows that emit radiation from radio to X-rays, and are unique laboratories for studying relativistic particle acceleration and magnetic field amplification processes \citep{Piran2004,Meszaros2006}. In between ordinary SNe Ic-BL and cosmological GRBs is a variety of transients that we still have to fully characterize. Among those are low-luminosity GRBs, of which the most famous example is GRB\,980425, associated with the radio-bright Type  Ic-BL SN\,1998bw \citep{Galama1998,Kulkarni1998}.  

Recently, \citet{Shankar2021} used the jetted outflow model produced from a consistently formed proto-magnetar in a 3D core-collapse SN to extract a range of central engine parameters  (energy $E_{eng}$ and  opening  angle $\theta_{eng}$)  that were then used as inputs to hydrodynamic models of  jet-driven explosions. The output of these models, in turn, were used to derive predicted SN light curves and spectra from different viewing angles, and found to be in agreement with SN Ic-BL optical observables \citep[see also][]{Barnes2018}. It was also shown that additional energy from the engine can escape through the tunnel drilled in the star as an ultra-relativistic jet (GRB) with energy $\approx 10^{51}$\,erg. On the other hand, a SN Ic-BL can be triggered even if the jet engine fails to produce a successful GRB jet. The duration of the central engine, $t_{eng}$, together with $E_{eng}$ and $\theta_{eng}$, are critical to determining the fate of the jet \citep{Lazzati2012}. 

A more general scenario where the high velocity ejecta found in SNe Ic-BL originate from a cocoon driven by a relativistic jet (regardless of the nature of the central engine) is also receiving attention. In this scenario, cosmological long GRBs are explosions where the relativistic jet breaks out successfully from the stellar envelope, while low-luminosity GRBs and SNe Ic-BL that are not associated with GRBs represent cases where the jet is choked \citep[see e.g. ][and references therein]{Piran2019,Eisenberg2022,Gottlieb2022,Pais2022}. 

Overall, the dividing line between successful GRB jets and failed ones is yet to be fully explored observationally, and observed jet outcomes in SNe Ic-BL have not yet been systematically compared to model outputs. While we know that SNe discovered by means of a GRB are all of Type Ic-BL, the question that remains open is whether all SNe Ic-BL make a GRB (jet), at least from some viewing angle, or if instead the jet-powered SNe Ic-BL are intrinsically different and rarer than ordinary SNe Ic-BL. Indeed, due to the collimation of GRB jets, it is challenging to understand whether all SNe Ic-BL are linked to  successful GRBs: a non-detection in $\gamma$- or X-rays could simply be due to the explosion being directed away from us. 
 Radio follow-up observations are needed to probe  the explosions' fastest-moving ejecta ($\gtrsim 0.2c$) largely free of geometry and viewing angle constraints.  Determining observationally what is the fraction of Type Ic-BL explosions that output jets which successfully break out of the star (as mildly-relativistic or ultra-relativistic ejecta), and measuring their kinetic energy via radio calorimetry, can provide jet-engine  explosion models a direct test of their predictions. 
 
 Using one of the largest sample of SNe Ic-BL with deep radio follow-up observations \citep[which included 15 SNe Ic-BL discovered by the Palomar Transient Factory, PTF/iPTF;][]{Law2009}, \citet{Corsi2016}  already established that $<41\%$ of SNe Ic-BL harbor relativistic ejecta similar to that of SN\,1998bw. Here, we present the results of a systematic radio follow-up campaign of an additional 16 SNe Ic-BL (at $z\lesssim 0.05$) detected independently of $\gamma$-rays by the Zwicky Transient Facility \citep[ZTF;][]{Bellm2019,Graham2019}. This study greatly expands our previous works on the subject \citep[][]{Corsi2017,Corsi2016,Corsi2014}. Before the advent of PTF and ZTF, the comparison between jet-engine model outcomes and radio observables was severely limited by the rarity of SN Ic-BL discoveries \citep[e.g.,][]{Berger2003,Soderberg2006} and/or by selection effects \citep[e.g.,][]{Woosley2006}---out of the thousands of jets identified, nearly  all were discovered at large distances via their high-energy emission (implying aligned jet geometry and ultra-relativistic speeds). In this work, we aim to provide a study free of these biases. 
 
 Our paper is organized as follows. In Section \ref{sec:discovery} we describe our multi-wavelength observations; in Section \ref{section:sample} we describe in more details the SNe Ic-BL included in our sample; in Section \ref{sec:modeling} we model the optical, X-ray, and radio properties of the SNe presented here and derive constraints on their progenitor and ejecta properties. Finally, in Section \ref{sec:conclusion} we summarize and conclude. Hereafter we assume cosmological parameters  $H_0 = 69.6$\,km\,s$^{-1}$\,Mpc$^{-1}$, $\Omega_{\rm M }= 0.286$, $\Omega_{\rm vac} = 0.714$ \citep{Bennett2014}. All times are given in UT unless otherwise stated.

\begin{table*}
\begin{center}
\caption{The sample of 16 SNe Ic-BL  analyzed in this work. For each SN we provide the IAU name, the ZTF name, the position, redshift, and luminosity distance. \label{tab:sample}}
\begin{tabular}{lccc}
\hline
\hline
SN (ZTF name)  & RA, Dec (J2000) &  $z$& $d_L$  \\
 & (hh:mm:ss~~dd:mm:ss) & &(Mpc)\\
\hline
2018etk (18abklarx) & 15:17:02.53 +03:56:38.7 & 0.044& 196 \\
2018hom (18acbwxcc)& 22:59:22.96 +08:45:04.6 & 0.030 & 132   \\
2018hxo (18acaimrb) &  21:09:05.80 +14:32:27.8 & 0.048 & 214  \\
2018jex (18acpeekw) & 11:54:13.87 +20:44:02.4 & 0.094 & 434  \\
2019hsx (19aawqcgy) & 18:12:56.22 +68:21:45.2 & 0.021 & 92  \\
2019xcc (19adaiomg) & 11:01:12.39 +16:43:29.1 & 0.029 & 128  \\
2020jqm (20aazkjfv) & 13:49:18.57 $-$03:46:10.4 & 0.037 & 164  \\
2020lao (20abbplei) & 17:06:54.61 +30:16:17.3 & 0.031 & 137  \\
2020rph (20abswdbg) & 03:15:17.83 +37:00:50.8 & 0.042 & 187\\
2020tkx (20abzoeiw) &  18:40:09.01 +34:06:59.5 & 0.027 & 119 \\
2021xv (21aadatfg) & 16:07:32.82 +36:46:46.2 & 0.041 & 182 \\
2021aug (21aafnunh) & 01:14:04.81 +19:25:04.7 & 0.041 & 182 \\
2021epp (21aaocrlm) & 08:10:55.27 $-$06:02:49.3 & 0.038 & 168\\
2021htb (21aardvol) & 07:45:31.19 +46:40:01.3  & 0.035& 155 \\
2021hyz (21aartgiv) &  
09:27:36.51 +04:27:11.0 &  0.046& 205\\
2021ywf (21acbnfos) & 05:14:10.99 +01:52:52.4 & 0.028 & 123 \\
\hline
\end{tabular}
\end{center}
\end{table*}

\begin{figure*}
    \centering
    \includegraphics[width=0.9\textwidth]{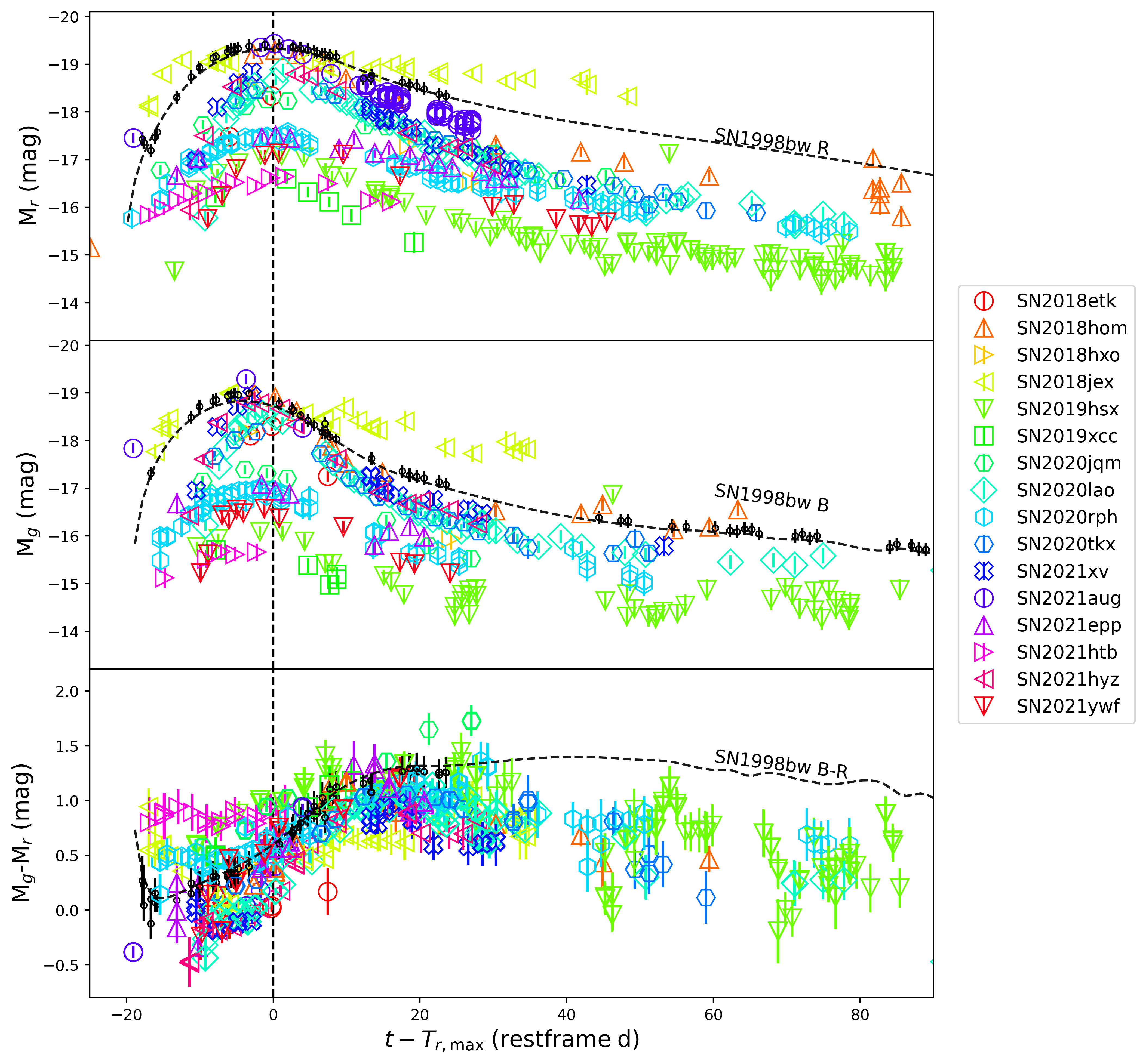}
    \caption{P48 $r$- (top) and $g$-band (middle) light curves for the SNe Ic-BL in our sample, compared with the $R$- and $B$-band light curves of SN\,1998bw, respectively. The bottom panel shows the corresponding color evolution. Observed AB magnitudes are corrected for Milky Way extinction. The archetypal SN\,1998bw is shown in black solid points, and its Gaussian Process interpolation in black dashed lines. See also \citet{Anand2022} and \citet{Gokul2022}.}
    \label{fig:opt-lc-mag}
\end{figure*}

\begin{figure*}
    \centering
    \includegraphics[width=0.9\textwidth]{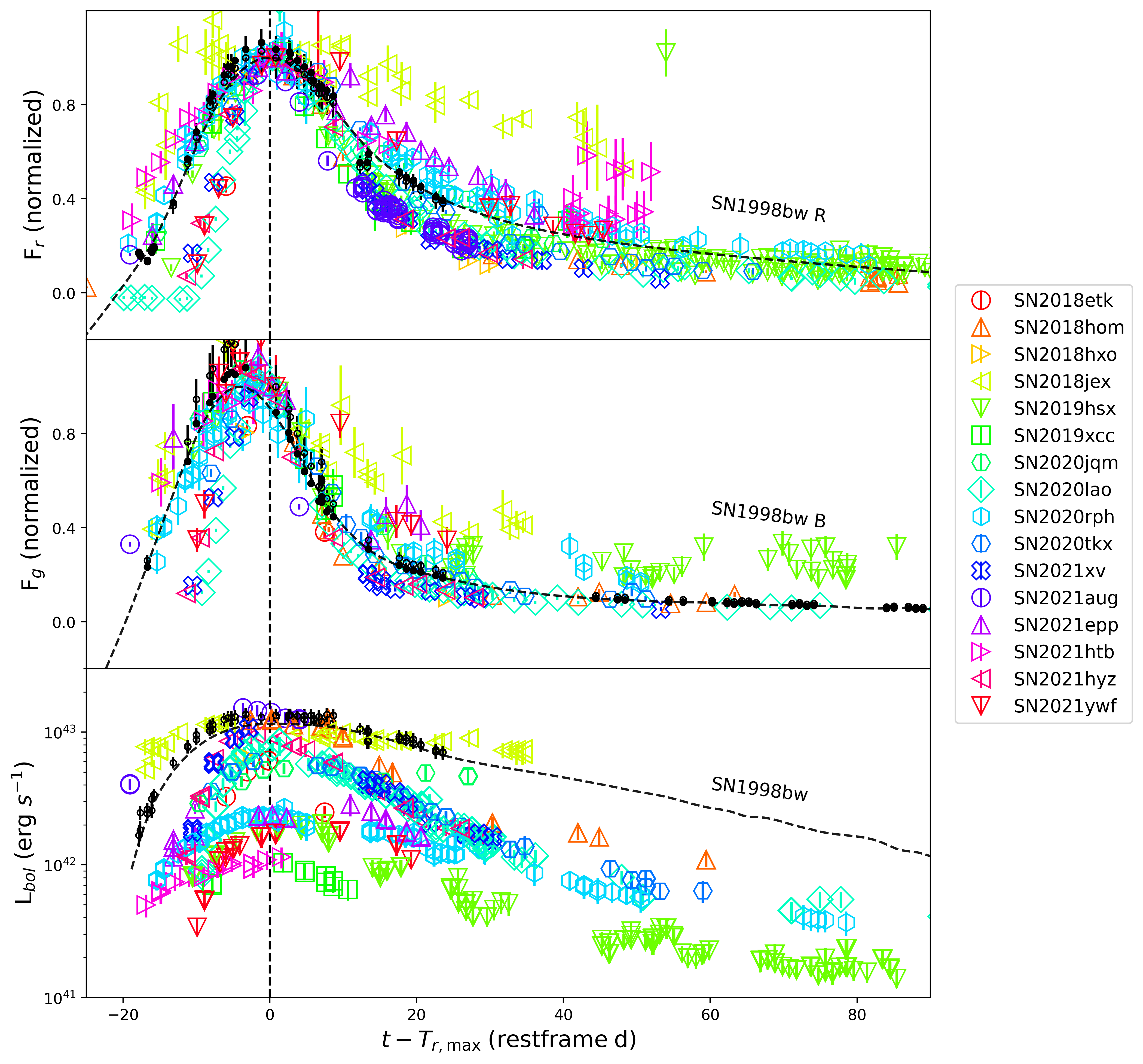}
    \caption{Top and middle panels: same as Figure \ref{fig:opt-lc-mag} but in flux space and with fluxes  normalized to their Gaussian Process maximum. Bottom panel: bolometric light curves. We converted $g-r$ to bolometric magnitudes with the empirical relations by \cite{Lyman2014,Lyman2016}. See also \citet{Anand2022} and \citet{Gokul2022}.
    \label{fig:opt-lc-flux}}
\end{figure*}

\section{Multi-wavelength observations}
\label{sec:discovery}
We have collected a sample of 16 SNe Ic-BL observed with the ZTF and with follow-up observations in the radio. The SNe Ic-BL included in our sample are listed in Table \ref{tab:sample}. We selected these SNe largely based on the opportunistic availability of follow-up observing time on the Karl G. Jansky Very Large Array (VLA).  The sample of SNe presented here doubles the sample of SNe Ic-BL with deep VLA observations presented in \citet{Corsi2016}. 

The SNe considered in this work are generally closer than the PTF/iPTF sample of SNe Ic-BL presented in \citet{Taddia2019}. In fact, their median redshift ($\approx 0.037$) is about twice as small as the median redshift of the PTF/iPTF SN Ic-BL sample \citep[$\approx 0.076$;][]{Taddia2019}. However, the median redshift of the ZTF SNe in our sample is compatible with the median redshift ($\approx 0.042$) of the full ZTF SN Ic-BL population presented in \citet{Gokul2022}. A subset of the SNe Ic-BL presented here is also analyzed in a separate paper and in a different context \citep[r-process nucleosynthesis;][]{Anand2022}. In this work, we report for the first time the results of our radio follow-up campaign of these events. We note that the ``Asteroid Terrestrial-impact Last Alert System''\citep[ATLAS;][]{Tonry2018} has contributed to several of the SN detections considered here (see Section \ref{section:sample}). Three of the Ic-BL in our sample were also reported in the recently released bright SN catalog by the All-Sky Automated Survey for Supernovae \citep[ASAS-SN;][]{Neumann2022}.

In what follows, we describe the observations we carried for this work. In Section \ref{section:sample} we give more details on each of the SNe Ic-BL in our sample. 

\subsection{ZTF photometry}
All photometric observations presented in this work were conducted with the Palomar Schmidt 48-inch (P48) Samuel Oschin telescope as part of the ZTF survey \citep{Bellm2019,Graham2019}, using the ZTF camera \citep{dekany2020}. 
In default observing mode, ZTF uses 30\,s exposures, and survey observations are carried out in $r$ and $g$ band, down to a typical limiting magnitude of $\approx 20.5$\,mag. P48 light curves were derived using the ZTF pipeline \citep{Masci2019}, and forced photometry \cite[see][]{Yao2019}. Hereafter, all reported magnitudes are in the AB system. 
The P48 light curves are shown in Figures~\ref{fig:opt-lc-mag}-\ref{fig:opt-lc-flux}. All the light curves presented in this work will be made public on the Weizmann Interactive Supernova Data Repository (WISeREP\footnote{\url{https://www.wiserep.org/}}).

\begin{table}
 \begin{center}
\caption{\label{tab:xrt_summary}\textit{Swift}/XRT observations of 9 of the 16 SNe Ic-BL in our sample. We provide the MJD of the \textit{Swift} observations, the XRT exposure time, and the 0.3-10\,keV unabsorbed flux measurements (or $3\sigma$ upper-limits for non detections).  \label{tab:x-ray}}
\begin{tabular}{llccc}
\hline
\hline
SN  & T$_{\rm XRT}$  & Exp. & $F_{\rm 0.3-10\,keV}$ \\
 & (MJD)  & (ks) & ($10^{-14}$\,erg\,cm$^{-2}$\,s$^{-1}$)\\
\hline
2018etk & 58377.85 & 4.8 & $< 4.2$ \\
2018hom & 58426.02 & 4.3 & $< 6.4$ \\
2019hsx & 58684.15 & 15 & $6.2_{-1.8}^{+2.3}$ \\
2020jqm & 59002.09 & 7.4 & $< 3.3 $ \\
2020lao & 59007.40 & 14 & $< 2.9 $ \\
2020rph & 59088.89 & 7.5 & $< 3.6 $ \\
2020tkx & 59125.38 & 8.1 & $< 3.3 $ \\
2021hyz & 59373.09 & 4.7 & $< 3.5 $ \\
2021ywf & 59487.60 & 7.2 & $5.3_{-3.3}^{+4.9} $ \\
\hline
\end{tabular}
\end{center}
\end{table}

\subsection{Optical Spectroscopy}
Preliminary spectral type classifications of several of the SNe in our sample were obtained with the Spectral Energy Distribution Machine (SEDM) mounted on the Palomar 60-inch telescope (P60), and quickly reported to the Transient Name Server (TNS). The SEDM is a very low resolution ($R\sim 100$) integral field unit spectrograph optimized for transient classification with high observing efficiency \citep{Blagorodnova2018,Rigault2019}. 

After initial classification, typically further spectroscopic observations are carried out as part of the ZTF transient follow-up programs to confirm and/or improve classification, and to characterize the time-evolving spectral properties of interesting events. Amongst the series of spectra obtained for each of the SNe presented in this work, we select one good quality photospheric phase spectrum  (Figures~\ref{fig:spectra1}-\ref{fig:spectra2}; grey) on which we run SNID \citep{Blondin2007} to obtain the best match to a SN Ic-BL template (black), after clipping the host emission lines and fixing the redshift to that derived either from the SDSS host galaxy or from spectral line fitting (H$\alpha$; see Section \ref{section:sample} for further details).
Hence, in addition to the SEDM, in this work we also made use of the following instruments: the Double Spectrograph \citep[DBSP;][]{Oke1995}, a low-to-medium resolution grating instrument for the Palomar 200-inch telescope Cassegrain focus that uses a dichroic to split light into separate red and blue channels observed simultaneously; the Low Resolution Imaging Spectrometer \citep[LRIS;][]{Oke1982,Oke1995}, a visible-wavelength imaging and spectroscopy instrument operating at the Cassegrain focus of Keck-I; the Alhambra Faint Object Spectrograph and Camera (ALFOSC), a CCD camera and spectrograph installed at the Nordic Optical Telescope \citep[NOT;][]{Djupvik2010}. All spectra presented in this work will be made public on WISeREP.

\subsection{X-ray follow up with \textit{Swift}}
For 9 of the 16 SNe presented in this work we carried out follow-up observations in the X-rays using the X-Ray Telescope  \citep[XRT;][]{Burrows+2005} on the \textit{Neil Gehrels Swift Observatory} \citep{Gehrels+2004}. 

We analyzed these observations using the online XRT tools\footnote{See \url{https://www.swift.ac.uk/user_objects/}.}, as described in \citet{Evans+2009}. 
We correct for Galactic absorption, and adopt a power-law spectrum with photon index $\Gamma = 2$  for count rate to flux conversion for non-detections, and for detections (two out of ninw events) where the number of photons collected is too small to enable a meaningful spectral fit (one out of two detections).  Table~\ref{tab:xrt_summary} presents the results from co-adding all observations of each source.

\subsection{Radio follow up with the VLA}
\label{sec:radioobs}
We observed the fields of the SNe Ic-BL in our sample with the VLA via several of our programs using various array configurations and receiver settings (Table \ref{tab:data}). 

The VLA raw data were calibrated in \texttt{CASA} \citep{McMullin2007} using the automated VLA calibration pipeline. After manual inspection for further flagging, the calibrated data were imaged using the \texttt{tclean} task. Peak fluxes were measured from the cleaned images using \texttt{imstat} and circular regions centered on the optical positions of the SNe, with radius equal to the nominal width (FWHM) of the VLA synthesized beam (see Table \ref{tab:data}). The RMS noise values were estimated with \texttt{imstat} from the residual images. Errors on the measured peak flux densities in Table \ref{tab:data} are calculated adding  a 5\% error in quadrature to measured RMS values. This accounts for systematic uncertainties on the absolute flux calibration. 

We checked all of our detections (peak flux density above $3\sigma$) for the source morphology (extended versus point-like), and ratio between integrated and peak fluxes using the CASA task \texttt{imfit}. All sources for which these checks provided evidence for extended or marginally resolved emission are marked accordingly in Table \ref{tab:data}. For non detections, upper-limits on the radio flux densities are given at the $3\sigma$ level unless otherwise noted.

\renewcommand\arraystretch{1.5}
\setlength\LTcapwidth{2\linewidth}
\begin{longtable*}{ccclccc}
\caption{VLA follow-up observations of the 16 SN\lowercase{e} I\lowercase{c}-BL in our sample. For all of the observations of the SNe in our sample we report: the mid MJD of the VLA observation; the central observing frequency; the measured flux density (all flux density upper-limits are calculated at $3\sigma$ unless otherwise noted); the VLA array configuration; the FWHM of the VLA nominal synthesized beam; the VLA project code under which the observations were conducted. See Sections \ref{sec:radioobs} and \ref{sec:radioanalysis} for discussion.} \label{tab:data}\\
\toprule
\toprule
SN & ${\rm T_{VLA}}$$^{a}$    & $\nu$  & $F_{\nu}$ & Conf. & Nom.Syn.Beam  & Project \\
 & (MJD)  & (GHz) & ($\mu$Jy) &  & (FWHM; arcsec) &  \\
\midrule
2018etk & 58363.08  & 6.3&  $90.1\pm8.7$$^{b}$   & D & 12&  VLA/18A-176$^{d}$\\
 & 58374.09   &14 & $41\pm11$ & D  & 4.6&  VLA/18A-176$^{d}$ \\
  & 58375.03   & 6.3& $89.7\pm8.8$$^{b}$  & D  &$12$&  VLA/18A-176$^{d}$ \\
 &  59362.27  &6.2 & $78.5\pm6.3$$^{b}$  & D  &12 &  VLA/20B-149$^{d}$\\
\midrule
2018hom &  58428.04  & 6.6 &  $133\pm11$  & D & 12&  VLA/18A-176$^{d}$\\
\midrule
2018hxo  & 58484.73  & 6.4 & $\lesssim 234$$^{c}$  & C & 3.5 &  VLA/18A-176$^{d}$\\
\midrule
2018jex &  58479.38 &6.4 & $\lesssim 28$ & C & 3.5&  VLA/18A-176$^{d}$\\
\midrule
2019hsx   & 58671.14 & 6.2 & $\lesssim 19$ & BnA & 1.0&  VLA/19B-230$^{d}$\\
\midrule
2019xcc &  58841.43 &6.3 &$62.7\pm 8.7$$^{b}$& D  & 12 &  VLA/19B-230$^{d}$\\
   & 58876.28  & 6.3  &$60.1\pm8.5$$^{b}$ & D  & 12&  VLA/19B-230$^{d}$  \\
  &59363.00 & 6.3 & $50.4\pm8.1$$^{b}$& D & 12&  VLA/20B-149$^{d}$ \\
\midrule
2020jqm &  58997.03  & 5.6& $175\pm13$& C  &  3.5&SG0117$^{d}$\\
  &59004.03 & 5.6 & $310\pm19$&C  & 3.5 &SG0117$^{d}$ \\
 &59028.48  & 5.5& $223\pm18$&B & 1.0& VLA/20A-568$^{d}$\\
 &59042.95  &5.7 & $202\pm15$&B &1.0& VLA/20A-568$^{d}$ \\
  &59066.09  &5.5  &$136\pm13$ &B  & 1.0& VLA/20A-568$^{d}$ \\
 &59088.03  & 5.5  &$168\pm13$ &B  & 1.0& VLA/20A-568$^{d}$ \\
 &59114.74 & 5.5 & $620\pm33$ &B  & 1.0& VLA/20A-568$^{d}$ \\
  &59240.37 & 5.5 & $720\pm37$ &A &0.33& VLA/20B-149$^{d}$\\
\midrule
2020lao &  59006.21 & 5.2& $\lesssim 33$ &C & 3.5& SG0117$^{d}$\\
  & 59138.83 &5.5& $\lesssim 21$&  B & 1.0& SG0117$^{d}$  \\
\midrule
2020rph & 59089.59 &5.5 & $42.7\pm7.4$&   B & 1.0& SG0117$^{d}$\\
  & 59201.28 &  5.5 & $43.9\pm7.0$&  A &0.33 & SG0117$^{d}$ \\
\midrule
2020tkx &   59117.89 & 10 & $272\pm 16$& B &0.6 & VLA/20A-374$^{e}$\\
 & 59136.11  & 10  & $564\pm29$& B  &0.6 & VLA/20A-374$^{e}$ \\
 & 59206.92 & 5.5&$86.6\pm7.3$& A & 0.33 & VLA/20B-149$^{d}$\\
\midrule
2021xv &  59242.42  & 5.5 & $\lesssim 23$& A & 0.33& VLA/20B-149$^{d}$\\
 & 59303.24 & 5.2& $\lesssim 29$& D &12 & VLA/20B-149$^{d}$ \\
 &59353.11 & 5.2 & $34.3\pm8.1$$^{b}$& D & 12 & VLA/20B-149$^{d}$ \\
\midrule
2021aug &  59254.75  & 5.2& $\lesssim 22$& A & 0.33& VLA/20B-149$^{d}$\\
  & 59303.62 &5.4 & $\lesssim 45$& D & 12& VLA/20B-149$^{d}$\\
 & 59353.48 & 5.4 & $\lesssim 30$& D &12 & VLA/20B-149$^{d}$\\
\midrule
2021epp &  59297.06 &5.3 & $(2.62\pm0.13)\times10^3$$^{b}$ & D&12 & VLA/20B-149$^{d}$\\
 &  59302.99 & 5.1 &$(2.82\pm0.18)\times10^3$$^{b}$ & D & 12& VLA/20B-149$^{d}$ \\
 & 59352.83& 5.3  &$(2.75\pm0.20)\times10^3$$^{b}$ & D &12 & VLA/20B-149$^{d}$ \\
\midrule
2021htb  &59324.94 & 5.2& $50\pm10$$^{b}$& D&12 & VLA/20B-149$^{d}$\\
  & 59352.87 & 5.2 &$59.4\pm9.5$$^{b}$ &D  & 12& VLA/20B-149$^{d}$ \\
\midrule
2021hyz & 59326.08  & 5.2& $38\pm11$&D &12 & VLA/20B-149$^{d}$\\
 &  59352.99& 5.5 & $\lesssim 30$& D & 12& VLA/20B-149$^{d}$ \\
\midrule
2021ywf &  59487.57  & 5.0  & $83\pm10$ & B &1.0 &SH0105$^{d}$\\
 &  59646.12 & 5.4 &$19.8\pm 6.3$ & A & 0.33& SH0105$^{d}$\\
\bottomrule
\multicolumn{7}{l}{$^{a}$ Mid MJD time of VLA observation (total time including calibration).}\\
\multicolumn{7}{l}{$^{b}$ Resolved or marginally resolved with emission likely dominated by the host galaxy.}\\
\multicolumn{7}{l}{$^{c}$ Image is dynamic range limited due to the presence of a bright source in the field.}\\
\multicolumn{7}{l}{$^{d}$ PI: Corsi.}\\
\multicolumn{7}{l}{$^{e}$ PI: Ho.}
\end{longtable*}

\begin{figure*}
    \centering
    \includegraphics[width=15cm]{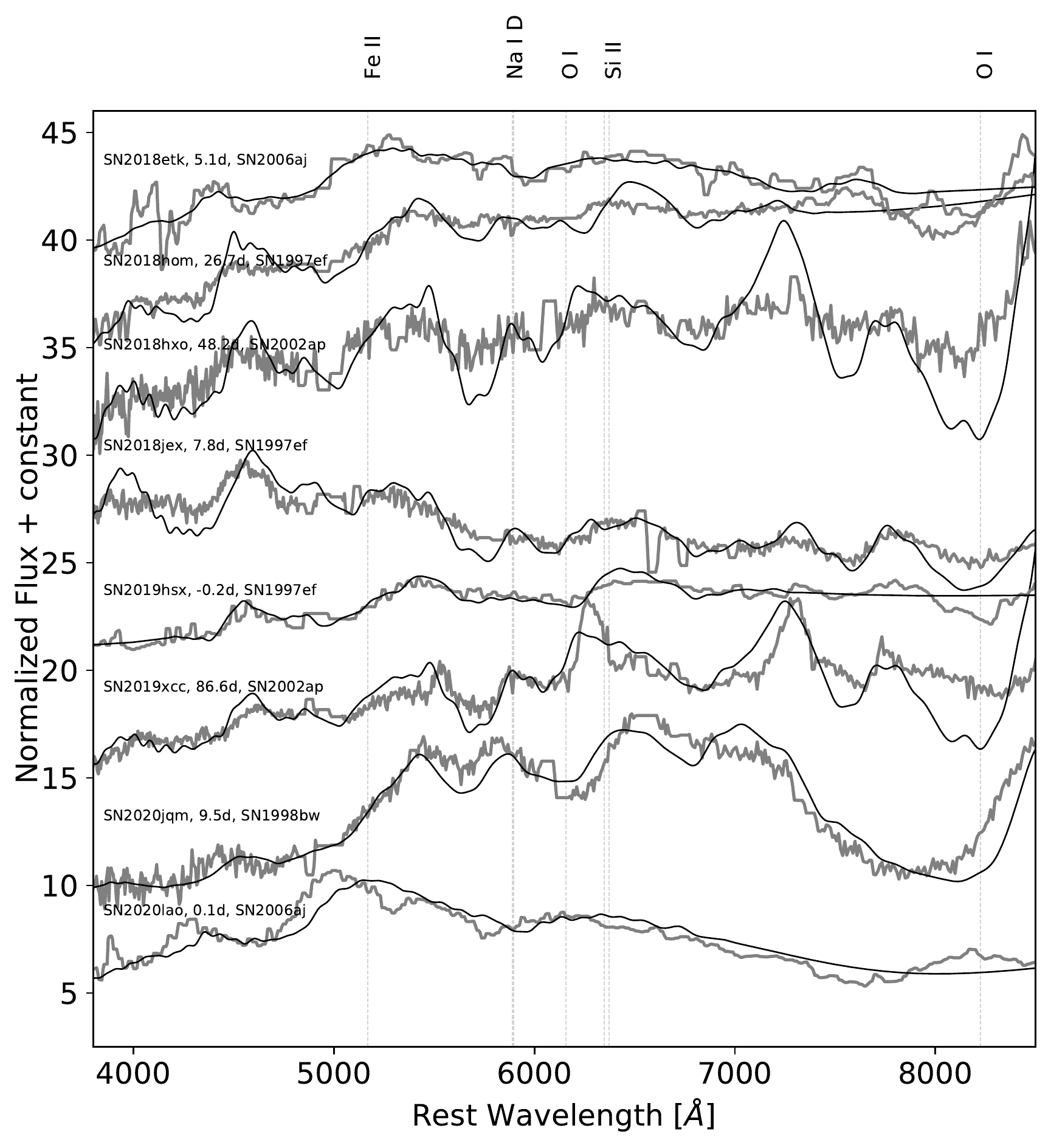}
    \caption{Photospheric phase spectra (grey) plotted along with their SNID best match templates (black) for the first half of the SNe Ic-BL in our sample. Spectra are labeled with their IAU name and spectroscopic phase (since $r$-band maximum; see Table \ref{tab:opt_data}). We note that the spectra used to estimate the photospheric velocities of SN\,2019xcc, SN\,2020lao, and SN\,2020jqm presented in Table \ref{tab:opt_data} are different from the ones shown here for classification purposes. This is because for spectral classification we prefer later-time but higher-resolution spectra, while for velocity measurements we prefer earlier-time spectra even if taken with the lower-resolution SEDM. All spectra presented in this work will be made public on WISeREP. }
    \label{fig:spectra1}
\end{figure*}

\begin{figure*}
    \centering
    \includegraphics[width=15cm]{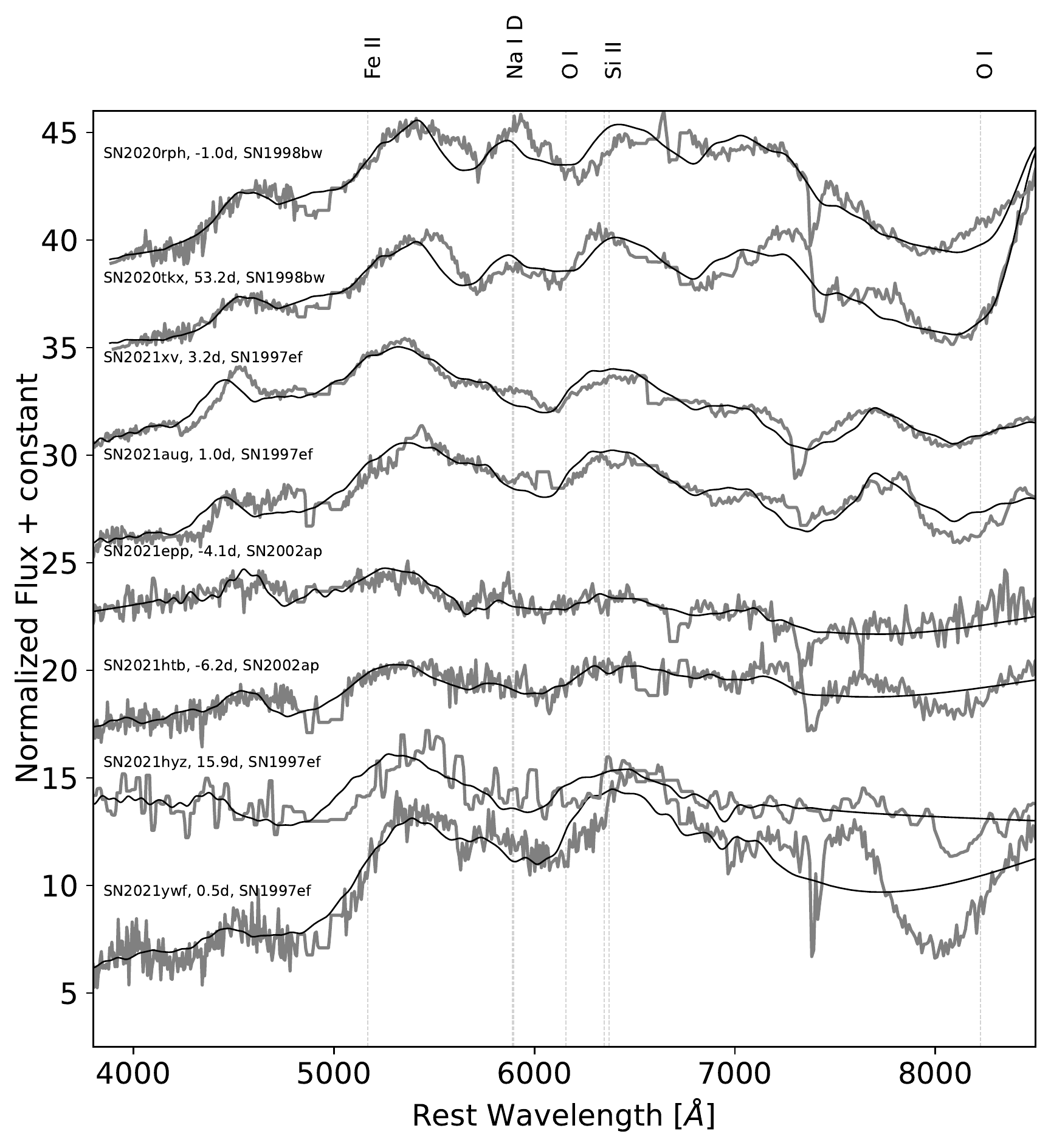}
    \caption{Same as Figure \ref{fig:spectra1} but for the second half of the SNe Ic-BL in our sample. All spectra presented in this work will be made public on WISeREP. }
    \label{fig:spectra2}
\end{figure*}

\section{Sample description}
\label{section:sample} 
\subsection{SN 2018etk}
Our first ZTF photometry of SN\,2018etk (ZTF18abklarx) was
obtained on 2018 August 1 (MJD 58331.16) with the P48. This first ZTF detection was in the $r$ band, with a host-subtracted magnitude of $19.21\pm0.12$\,mag (Figure~\ref{fig:opt-lc-mag}), at $\alpha$=15$^{\rm h}$17$^{\rm m}$02.53$^{\rm s}$,
$\delta=+03^{\circ}56'38\farcs7$ (J2000). The object was reported to the TNS by ATLAS on 2018 August 8, who discovered it on 2018 August 6  \citep{Tonry2018_ZTF18abklarx}.
The last ZTF non-detection prior to ZTF discovery was on 2018 July 16, and the last shallow ATLAS non-detection was on 2018 August 2, at $18.75\,$mag.
The transient was classified as a Type Ic SN by \citet{Fremling2018_ZTF18abklarx} based on a spectrum obtained on 2018 August 13 with the SEDM.
We re-classify this transient as a SN Type  Ic-BL most similar to SN\,2006aj based on a P200 DBSP spectrum obtained on 2018 August 21 (see Figure~\ref{fig:spectra1}).
SN\,2018etk exploded in a star-forming galaxy with a known
redshift of $z= 0.044$ derived from SDSS data.

\subsection{SN 2018hom}
Our first ZTF photometry of SN\,2018hom (ZTF18acbwxcc) was
obtained on 2018 November 1  (MJD 58423.54) with
the P48. This first ZTF detection was in the $r$ band, with a host-subtracted magnitude of $16.60\pm0.04$\,mag (Figure \ref{fig:opt-lc-mag}), at $\alpha$=22$^{\rm h}$59$^{\rm m}$22.96$^{\rm s}$,
$\delta=+08^{\circ}45'04\farcs6$ (J2000). The object was reported to the TNS by ATLAS on 2018 October 26, and discovered by ATLAS on 2018 October 24 at $o\approx 17.3\,$mag \citep{Tonry2018_ZTF18acbwxcc}.
The last ZTF non-detection prior to ZTF discovery was on 2018 October 9 at $g>20.35\,$mag, and the last ATLAS non-detection was on 2018 October 22 at $o> 18.25\,$mag.
The transient was classified as a SN Type Ic-BL by \citet{Fremling2018_ZTF18acbwxcc} based on a spectrum obtained on 2018 November 2 with the SEDM.
SN\,2018etk exploded in a galaxy with unknown redshift. We measure a redshift of $z = 0.030$ from star-forming emission lines in a Keck-I LRIS spectrum obtained on 2018 November 30. We plot this spectrum in Figure~\ref{fig:spectra1}, along with its SNID template match to the Type Ic-BL SN\,1997ef. We note that this SN was also reported in the recently released ASAS-SN bright SN catalog \citep[][]{Neumann2022}.

\subsection{SN 2018hxo}
Our first ZTF photometry of SN\,2018hxo (ZTF18acaimrb) was
obtained on 2018 October 9  (MJD 58400.14) with the P48. This first detection was in the $g$ band, with a host-subtracted magnitude of $18.89\pm0.09$\,mag (Figure~\ref{fig:opt-lc-mag}), at $\alpha$=21$^{\rm h}$09$^{\rm m}$05.80$^{\rm s}$,
$\delta=+14^{\circ}32'27\farcs8$ (J2000). The object was first reported to the TNS by ATLAS on 2018 November 6, and first detected by ATLAS on 2018 September 25 at $o=18.36\,$mag \citep{Tonry2018_ZTF18acaimrb}.
The last ZTF non-detection prior to discovery was on 2018 September 27 at $r>20.12\,$mag, and the last ATLAS non-detection was on 2018 September 24 at $o> 18.52\,$mag.
The transient was classified as a SN Type Ic-BL by \citet{Dahiwale2020_ZTF18acaimrb} based on a spectrum obtained on 2018 December 1 with Keck-I LRIS. In Figure \ref{fig:spectra1} we plot this spectrum along with its SNID match to the Type Ic-BL SN\,2002ap.
SN\,2018etk exploded in a galaxy with unknown redshift. We measure a redshift of $z=0.048$ from star-forming emission lines in the Keck spectrum. 

\subsection{SN 2018jex}
Our first ZTF photometry of SN\,2018jex (ZTF18acpeekw) was
obtained on 2018 November 16 (MJD 58438.56) with
the P48. This first detection was in the $r$ band, with a host-subtracted magnitude of $20.07\pm0.29$\,mag, at $\alpha$=11$^{\rm h}$54$^{\rm m}$13.87$^{\rm s}$,
$\delta=+20^{\circ}44'02\farcs4$ (J2000). The object was reported to the TNS by AMPEL on November 28 \citep{Nordin2018_ZTF18acpeekw}.
The last ZTF last non-detection prior to ZTF discovery was on 2018 November 16 at $r>19.85\,$mag.
The transient was classified as a SN Type Ic-BL based on a spectrum obtained on 2018 December 4 with Keck-I LRIS. In Figure~\ref{fig:spectra1} we show this spectrum plotted against the SNID template of the Type Ic-BL SN\,1997ef.
AT2018jex exploded in a galaxy with unknown redshift. We measure a redshift of $z=0.094$ from star-forming emission lines in the Keck spectrum.

\subsection{SN 2019hsx}
We refer the reader to \citet{Anand2022} for details about this SN Ic-BL.  Its P48 light curves and the spectrum used for classification are shown in Figures \ref{fig:opt-lc-mag} and \ref{fig:spectra1}. We note that this SN was also reported in the recently released ASAS-SN bright SN catalog  \citep{Neumann2022}.

\subsection{SN 2019xcc}
We refer the reader to \citet{Anand2022} for details about this SN Ic-BL.  Its P48 light curves and the spectrum used for classification are shown in Figures \ref{fig:opt-lc-mag} and \ref{fig:spectra1}.

\subsection{SN 2020jqm}
Our first ZTF photometry of SN\,2020jqm (ZTF20aazkjfv) was
obtained on 2020 May 11 (MJD 58980.27) with
the P48. This first detection was in the $r$ band, with a host-subtracted magnitude of $19.42\pm0.13$\,mag, at $\alpha$=13$^{\rm h}$49$^{\rm m}$18.57$^{\rm s}$,
$\delta=-03^{\circ}46'10\farcs4$ (J2000). The object was reported to the TNS by ALeRCE on May 11 \citep{Forster2020_ZTF20aazkjfv}.
The last ZTF  non-detection prior to ZTF discovery was on 2020 May 08 at $g>17.63\,$mag.
The transient was classified as a SN Type  Ic-BL based on a spectrum obtained on 2020 May 26 with the SEDM \citep{Dahiwale2020_ZTF20aazkjfv}.
SN\,2020jqm exploded in a galaxy with unknown redshift. We measure a redshift of $z = 0.037$ from host-galaxy emission lines in a NOT ALFOSC spectrum obtained on 2020 June 6. We plot the ALFOSC spectrum along with its SNID match to the Type Ic-BL SN\,1998bw in Figure~\ref{fig:spectra1}.

\subsection{SN 2020lao}
We refer the reader to \citet{Anand2022} for details about this SN Ic-BL.  Its P48 light curves and the spectrum used for classification are shown in Figures \ref{fig:opt-lc-mag} and \ref{fig:spectra1}. We note that this SN was also reported in the recently released ASAS-SN bright SN catalog  \citep{Neumann2022}.

\subsection{SN 2020rph}
We refer the reader to \citet{Anand2022} for details about this SN Ic-BL.  Its P48 light curves and the spectrum used for classification are shown in Figures \ref{fig:opt-lc-mag} and \ref{fig:spectra2}.

\subsection{SN 2020tkx}
We refer the reader to \citet{Anand2022} for details about this SN Ic-BL.  Its P48 light curves and the spectrum used for classification are shown in Figures \ref{fig:opt-lc-mag} and \ref{fig:spectra2}.

\subsection{SN 2021xv}
We refer the reader to \citet{Anand2022} for details about this SN Ic-BL.  Its P48 light curves and the spectrum used for classification are shown in Figures \ref{fig:opt-lc-mag} and \ref{fig:spectra2}.

\subsection{SN 2021aug}
Our first ZTF photometry of SN\,2021aug (ZTF21aafnunh) was
obtained on 2021 January 18  (MJD 59232.11) with the P48. This first detection was in the $g$ band, with a host-subtracted magnitude of $18.73\pm0.08$\,mag, at $\alpha$=01$^{\rm h}$14$^{\rm m}$04.81$^{\rm s}$,
$\delta=+19^{\circ}25'04\farcs7$ (J2000).
The last ZTF non-detection prior to ZTF discovery was on 2021 January 16 at $g>20.12\,$mag. The transient was publicly reported to the TNS by ALeRCE on 2021 January 18 \citep{MunozArancibia2021}, and classified as a SN Type Ic-BL
based on a spectrum obtained on 2021 February 09 with the SEDM \citep{Dahiwale2021_ZTF21aafnunh}.
SN\,2020jqm exploded in a galaxy with unknown redshift. We measure a redshift of $z= 0.041$  from star-forming emission lines in a P200 DBSP spectrum obtained on 2021 February 08. This spectrum is shown in Figure~\ref{fig:spectra2} along with its template match to the Type Ic-BL SN\,1997ef.

\subsection{SN 2021epp}
Our first ZTF photometry of SN\,2021epp (ZTF21aaocrlm) was
obtained on 2021 March 5 (MJD 59278.19) with the P48. This first ZTF detection was in the $r$ band, with a host-subtracted magnitude of $19.61\pm0.15$\,mag (Figure~\ref{fig:opt-lc-mag}), at $\alpha$=08$^{\rm h}$10$^{\rm m}$55.27$^{\rm s}$, $\delta=-06^{\circ}02'49\farcs3$ (J2000). 
The transient was publicly reported to the TNS by ALeRCE on 2021 March 5 \citep{MunozArancibia20210305}, and classified as a SN Type Ic-BL based on a spectrum obtained on 2021 March 13 by ePESSTO+ with the ESO Faint Object Spectrograph and Camera \citep{Kankare2021}.
The last ZTF non-detection prior to discovery was on 2021 March 2 at $r>19.72\,$mag.
In Figure~\ref{fig:spectra2} we show the classification spectrum 
plotted against the SNID template of the Type Ic-BL SN\,2002ap.
SN\,2021epp exploded in a galaxy with known redshift of $z = 0.038$.

\subsection{SN 2021htb}
Our first ZTF photometry of SN\,2021htb (ZTF21aardvol) was
obtained on 2021 March 31 (MJD 59304.164) with the P48. This first ZTF detection was in the $r$ band, with a host-subtracted magnitude of $20.13\pm0.21$\,mag (Figure~\ref{fig:opt-lc-mag}), at $\alpha$=07$^{\rm h}$45$^{\rm m}$31.19$^{\rm s}$,
$\delta=46^{\circ}40'01\farcs4$ (J2000).
The transient was publicly reported to the TNS by SGLF on 2021 April 2 \citep{Poidevin_2021}.
The last ZTF non-detection prior to ZTF discovery was on 2021 March 2, at $r>19.88\,$mag.
In Figure~\ref{fig:spectra2} we show a P200 DBSP spectrum taken on 2021 April 09   plotted against the SNID template of the Type Ic-BL SN\,2002ap. SN\, 2021htb exploded in a SDSS galaxy with redshift $z= 0.035$.

\subsection{SN 2021hyz}
Our first ZTF photometry of SN\,2021hyz (ZTF21aartgiv) was
obtained on 2021 April 03  (MJD 59307.155) with the P48. This first ZTF detection was in the $g$ band, with a host-subtracted magnitude of $20.29\pm0.17$\,mag (Figure~\ref{fig:opt-lc-mag}), at $\alpha$=09$^{\rm h}$27$^{\rm m}$36.51$^{\rm s}$,
$\delta=04^{\circ}27'11\farcs$ (J2000).
The transient was publicly reported to the TNS by ALeRCE on 2021 April 3 \citep{Forster_2021}.
The last ZTF non-detection prior to ZTF discovery was on 2021 April 1, at $g>19.15\,$mag. In Figure~\ref{fig:spectra2} we show a P60 SEDM spectrum taken on 2021 April 30  plotted against the SNID template of the Type Ic-BL SN\,1997ef.
SN\,2021hyz exploded in a galaxy with redshift $z= 0.046$.

\subsection{SN 2021ywf}
We refer the reader to \citet{Anand2022} for details about this SN Ic-BL.  Its P48 light curves and the spectrum used for classification are shown in Figures \ref{fig:opt-lc-mag} and \ref{fig:spectra2}.

\begin{footnotesize}
\begin{longtable*}{lcccllllll}
\caption{Optical properties of the 16 SNe Ic-BL in our sample. We list the SN name; the MJD of maximum light in $r$ band; the absolute magnitude at $r$-band peak; the absolute magnitude at $g$-band peak; the explosion time estimated as days since $r$-band maximum; the estimated nickel mass; the characteristic timescale of the bolometric light curve; the photospheric velocity; the ejecta mass; and the kinetic energy of the explosion.  See Sections \ref{sec:vphot} and \ref{sec:optical_properties} for discussion. \label{tab:opt_data}}\\
\toprule
\toprule
SN & T$_{r,\rm max}$ & M$^{\rm peak}_{r}$ & M$^{\rm peak}_{g}$  & ${\rm T}_{\rm exp}$-T$_{r,\rm max}$  & $M_{\rm Ni}$   & $\tau_{m}$  & v$_{\rm ph}$($^{a}$) & $M_{\rm ej}$ & $E_{\rm k}$ \\
 &  (MJD) & (AB mag)  & (AB mag) & (d) & (M$_{\odot}$) & (d) & ($10^4$\,km/s) & (M$_{\odot}$) & ($10^{51}$erg) \\
\midrule[0.5pt]
2018etk & 58337.40 & $-18.31\pm0.03$ & $-18.30\pm0.02$ & $-9\pm1$ & $0.13_{-0.02}^{+0.01}$ & $5.0_{-2}^{+2}$ &  $2.6\pm0.2$ (5)
& $0.7\pm0.5$ & $3\pm2$ \\
2018hom & 58426.31 & $-19.30\pm0.11$ & $-18.91\pm0.01$ & $-9.3_{-0.4}^{+0.7}$ & $0.4\pm0.1$ & $6.9\pm0.2$ & $1.7\pm0.2$ (27) & $> 0.7$ & $>1$ \\
2018hxo & 58403.76 & $-18.68\pm0.06$ & $-18.4\pm0.1$ & $-28.6_{-0.3}^{+0.2}$ & $0.4\pm0.2$ & $6\pm2$ & $0.6\pm0.1$ (48)
& $>0.1$ & $>0.02$ \\
2018jex & 58457.01 & $-19.06\pm0.02$ & $-18.61\pm0.04$ & $-18.49\pm0.04$ & $0.53_{-0.06}^{+0.07}$ & $13_{-3}^{+2}$ & $1.8\pm0.3$ (8)
& $3\pm1$ & $7\pm3$ \\
2019hsx & 58647.07 & $-17.08\pm0.02$ & $-16.14\pm0.04$ & $-15.6_{-0.5}^{+0.4}$ & $0.07_{-0.01}^{+0.01}$ & $12\pm1$ & $1.0\pm0.2$ (-0.2)
& $1.6\pm0.4$ & $1.0\pm0.5$ \\
2019xcc & 58844.59 & $-16.58\pm0.06$ & $-15.6\pm0.2$ & $-11\pm2$ & $0.04\pm0.01$ & $5.0_{-0.9}^{+1.4}$ & $2.4\pm0.2$ (6)
& $0.7\pm0.3$ & $2\pm1$ \\
2020jqm & 58996.21 & $-18.26\pm0.02$ & $-17.39\pm0.04$ & $-17\pm1$ & $0.29_{-0.04}^{+0.05}$ & $18\pm2$ & $1.3\pm0.3$ (-0.5) 
& $5\pm1$ & $5\pm3$ \\
2020lao & 59003.92 & $-18.66\pm0.02$ & $-18.55\pm0.02$ & $-11\pm1$ & $0.23\pm0.01$ & $7.7\pm0.2$ & $1.8\pm0.2$ (9)
& $1.2\pm0.2$ & $2.5\pm0.7$ \\
2020rph & 59092.34 & $-17.48\pm0.02$ & $-16.94\pm0.03$ & $-19.88\pm0.02$  & $0.07\pm0.01$ & $17.23_{-0.9}^{+1.2}$ & $1.2\pm0.5$ (-1)
& $4\pm2$ & $3\pm3$ \\
2020tkx & 59116.50 & $-18.49\pm0.05$ & $-18.19\pm0.03$ & $-13\pm4$  & $0.22\pm0.01$ & $10.9_{-0.8}^{+0.7}$ & $1.32\pm0.09$ (53)
& $> 1.5$ & $> 1.5$ \\
2021xv & 59235.56 & $-18.92\pm0.07$ & $-18.99\pm0.05$ & $-12.8_{-0.3}^{+0.2}$ & $0.30_{-0.02}^{+0.01}$ & $7.7_{-0.5}^{+0.7}$ & $1.3\pm0.1$ (3)
& $0.9\pm0.1$ & $1.0\pm0.2$ \\
2021aug & 59251.98 & $-19.42\pm0.01$ & $-19.32\pm0.06$ & $-24\pm2$ & $0.7\pm0.1$ & $17\pm7$ & $0.8\pm0.3$ (1) 
& $3\pm2$ & $1\pm1$ \\
2021epp & 59291.83 & $-17.49\pm0.03$ & $-17.12\pm0.09$ & $-15\pm1$ & $0.12\pm0.02$ & $17_{-3}^{+4}$ & $1.4\pm0.5$ (-4)
& $5\pm2$ & $6\pm5$ \\
2021htb & 59321.56 & $-16.55\pm0.03$ & $-15.66\pm0.07$ & $-19.38\pm0.02$  & $0.04\pm0.01$ & $13\pm2$ & $1.0\pm0.5$ (-6)
& $1.8\pm0.9$ & $1\pm1$ \\
2021hyz & 59319.10 & $-18.83\pm0.05$ & $-18.81\pm0.01$ & $-12.9\pm0.9$  & $0.29_{-0.02}^{+0.01}$ & $7.7_{-0.4}^{+0.5}$ & $2.3\pm0.3$ (16) & $> 1.3$ & $>4$ \\
2021ywf & 59478.64 & $-17.10\pm0.05$ & $-16.5\pm0.1$ & $-10.7\pm0.5$  & $0.06\pm0.01$ & $8.9\pm0.8$ & $1.2\pm0.1$ (0.5) 
& $1.1\pm0.2$ & $0.9\pm0.3$ \\
\bottomrule
\multicolumn{9}{l}{$^{a}$ Rest-frame phase days of the spectrum that was used to measure the velocity.}\\
\end{longtable*}
\end{footnotesize}

\begin{figure*}
    \centering
    \includegraphics[height=8cm]{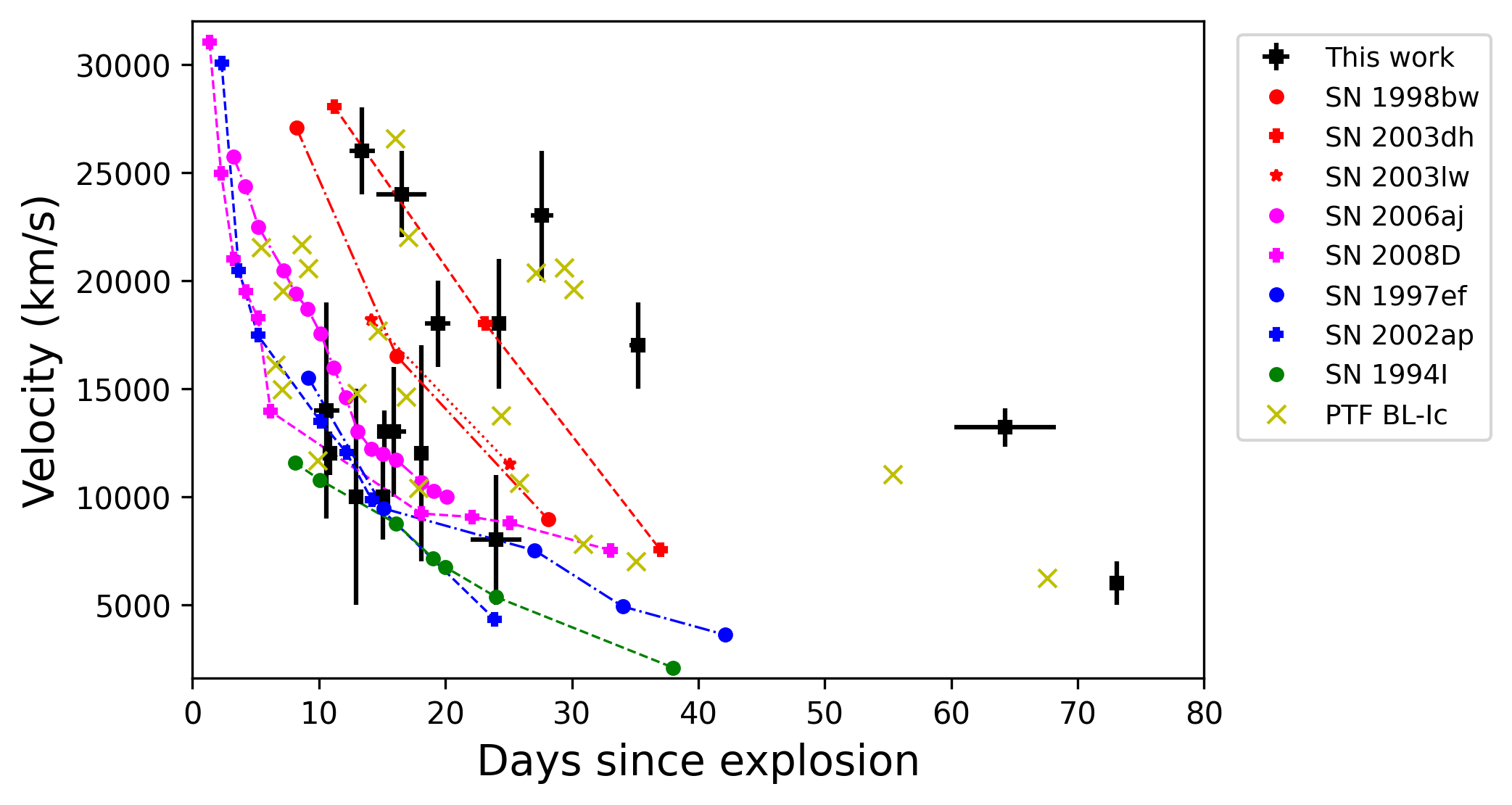}
    \caption{Photospheric velocities of the ZTF SNe in our sample (black) plotted as a function of (rest frame) time since explosion (see Table \ref{tab:opt_data}).  Velocities are measured using Fe II (5169 $\AA$); velocities quoted refer to 84\% confidence and are measured relative to the Ic template velocity using the open source software \texttt{SESNspectraLib} \citep{Liu2016,Modjaz2016}. We compare our results with photospheric velocities derived from spectroscopic modeling for a number of Ib/c SNe. Red symbols represent GRB-SNe \citep{Iwamoto1998,Mazzali2003,Mazzali2006b}; magenta is used for XRF/X-ray transients-SNe \citep{Mazzali2006a,Pian2006,Modjaz2009}; blue represents SNe Ic-BL \citep{Mazzali2000,Mazzali2002}; green is used for the ``normal'' Type Ic SN 1994I \citep{Sauer2006}. Finally, for comparison we also plot the photospheric velocities for the SNe Ic-BL in the \citet{Corsi2016} sample as measured by \citet{Taddia2019} (see their Tables 2 and A1; yellow crosses). Errors on the times since explosion account for the uncertainties on T$_{\rm exp}$ as reported in Table \ref{tab:opt_data}. \label{fig:velocities}}
\end{figure*}

\section{Multi-wavelength analysis}
\label{sec:modeling}
\subsection{Photospheric velocities}
\label{sec:vphot}
We confirm the SN Type Ic-BL classification of each object in our sample by measuring the photospheric velocities (${\rm v_{ph}}$).
SNe Ic-BL are characterized by high expansion velocities evident in the broadness of their spectral lines. A good proxy for the photospheric velocity is that derived from the maximum absorption position of the Fe II \citep[$\lambda 5169$; e.g.,][]{Modjaz2016}. We caution, however, that estimating this velocity is not easy given the strong line blending. We first pre-processed one high-quality spectrum per object using the IDL routine \texttt{WOMBAT}, then smoothed the spectrum using the IDL routine \texttt{SNspecFFTsmooth} \citep{Liu2016}, and finally ran \texttt{SESNSpectraLib} \citep{Liu2016,Modjaz2016} to obtain final velocity estimates. 

In  Figure~\ref{fig:velocities} we show a comparison of the photospheric velocities estimated for the SNe in our sample with those derived from spectroscopic modeling for a number of SNe Ib/c. The velocities measured for our sample are compatible, within the measurement errors, with what was observed for the PTF/iPTF samples.
Measured values for the photospheric velocities with the corresponding rest-frame phase in days since maximum $r$-band light of the spectra that were used to measure them are also reported in  Table~\ref{tab:opt_data}.  

We note that the spectra used to estimate the photospheric velocities of SN\,2019xcc, SN\,2020lao, and SN\,2020jqm are different from those used for the classification of those events as SNe Ic-BL (see Section \ref{section:sample} and Figure \ref{fig:spectra1}). This is because for spectral classification we prefer later-time but higher-resolution spectra, while for velocity measurements we prefer earlier-time spectra even if taken with the lower-resolution SEDM. 

\subsection{Bolometric light curve analysis}
\label{sec:optical_properties}
In our analysis we correct all ZTF photometry for Galactic extinction, using the Milky Way (MW) color excess $E(B-V)_{\mathrm{MW}}$ toward the position of the SNe.
These are all obtained from \cite{Schlafly2011}. All reddening corrections are applied using the \cite{Cardelli1989} extinction law with $R_V=3.1$. 
After correcting for Milky Way extinction, we interpolate our P48 forced-photometry light curves 
using a Gaussian Process via the {\tt GEORGE}\footnote{\href{https://george.readthedocs.io}{https://george.readthedocs.io}} package with a stationary Matern32 kernel and the analytic functions of \citet{Bazin2009} as mean for the flux form.
As shown in Figure~\ref{fig:opt-lc-mag}, the colour evolution of the SNe  in our sample are not too dissimilar with one another, which implies that the amount of additional host extinction is small. Hence, we set the host extinction to zero.
Next, we derive bolometric light curves calculating bolometric corrections from the $g$- and $r$-band data following the empirical relations by \cite{Lyman2014,Lyman2016}. For SN\,2018hxo, since there is only one $g$-band detection, we assume a constant bolometric correction to estimate its bolometric light curve. These bolometric light curves are shown in the bottom panel of Figure \ref{fig:opt-lc-flux}. 

We estimate the explosion time ${\rm T}_{\rm exp}$ of the SNe in our sample as follows.  For SN\,2021aug, we fit the early ZTF light curve data following the method presented in \citet{Miller2020}, where we fix the power-law index of the rising early-time temporal evolution to $\alpha=2$, and derive an estimate of the explosion time from the fit. For most of the other SNe in our sample, the ZTF $r$- and $g$-band light curves lack enough early-time data to determine an estimate of the explosion time following the formulation of \citet{Miller2020}. For all these SNe we instead set the explosion time to the mid-point between the last non-detection prior to discovery, and the first detection. 
Results on ${\rm T}_{\rm exp}$ are reported in Table  \ref{tab:opt_data}.

We fit the bolometric light curves around peak ($-20$ to 60 rest-frame days relative to peak) to a model using the Arnett formalism \citep{Arnett1982}, with the nickel mass ($M_{\rm Ni}$) and characteristic time scale $\tau_m$ as free parameters \citep[see e.g. Equation A1 in][]{Valenti2008}. The derived values of $M_{\rm Ni}$ (Table \ref{tab:opt_data}) have a median of $\approx 0.22$\,M$_{\odot}$, compatible with the median value found for SNe Ic-BL in the PTF sample by \citet{Taddia2019}, somewhat lower than for SN\,1998bw for which the estimated nickel mass values are in the range $(0.4-0.7)$\,M$_{\odot}$, but comparable to the $M_{\rm Ni}\approx 0.19-0.25$\,M$_{\odot}$ estimated for
SN\,2009bb \citep[see e.g.,][]{Lyman2016,Afsariardchi2021}. We note that events such as SN\,2019xcc and SN\,2021htb have relatively low values of $M_{\rm Ni}$, which are however compatible with the range of $0.02-0.05$\,M$_{\odot }$ expected for the nickel mass of magnetar-powered SNe Ic-BL \citep{Nishimura2015,Chen2017,Suwa2015}. 

Next, from the measured  characteristic timescale $\tau_m$ of the bolometric light curve, and the photospheric velocities estimated via spectral fitting  (see previous Section) we derive the ejecta mass ($M_{\rm ej}$) and the kinetic energy ($E_{\rm k}$) via the following relations \citep[see e.g. Equations 1 and 2 in][]{Lyman2016}:
\begin{eqnarray}
    \tau^2_m {\rm v_{ph, max}}=\frac{2\kappa}{13.8 c} M_{\rm ej}~~~ & 
    ~~~{\rm v}^2_{\rm ph, max}=\frac{5}{3}\frac{2 E_{\rm k}}{M_{\rm ej}},\label{eq:ejecta}
\end{eqnarray}
where we assume a constant opacity of $\kappa = 0.07$\,g\,cm$^{-2}$. 

We note that to derive $M_{\rm ej}$ and $E_{\rm k}$ as described above we assume the photospheric velocity evolution is negligible within 15 days relative to peak epoch, and use the spectral velocities measured within this time frame to estimate ${
\rm v}_{\rm ph, max}$ in Equation \ref{eq:ejecta}. However, there are four objects in our sample (SN\,2018hom, SN\,2018hxo, SN\,2020tkx, and SN\,2021hyz) for which the spectroscopic analysis constrained the photospheric velocity only after day 15 relative to peak epoch. For these events, we only provide lower limits on the ejecta mass and kinetic energy (see Table \ref{tab:opt_data}). 
 
Considering only the SNe in our sample for which we are able to measure the photospheric velocity within 15\,d since peak epoch, we derive median values for the ejecta masses and kinetic energies of  $1.7$\,M$_{\odot}$ and $2.2\times10^{51}$\,erg, respectively. These are both a factor of $\approx 2$ smaller than the median values derived for the PTF/iPTF sample of SNe Ic-BL \citep{Taddia2019}. This could be due to either an intrinsic effect, or to uncertainties on the measured photospheric velocities. In fact, we note that the photospehric velocity is expected to decrease very quickly after maximum light (see e.g. Figure \ref{fig:velocities}). Since the photospheric velocity in Equation (\ref{eq:ejecta}) of the Arnett formulation is the one at peak, our estimates of ${\rm v}_{\rm ph,max}$ could easily underestimate that velocity by a factor of $\approx 2$ for many of the SNe in our sample. This would in turn yield an underestimate of $M_{\rm ej}$ by a factor of $\approx 2$ (though the kinetic energy would be reduced by a larger factor). A more in-depth analysis of these trends and uncertainties will be presented in \citet{Gokul2022}. 
 
\subsection{Search for gamma-rays}
Based on the explosion dates derived for each object in Section \ref{sec:optical_properties} (Table \ref{tab:opt_data}), we searched for potential GRB coincidences in several online archives. No potential counterparts were identified in both spatial and temporal coincidence with either the Burst Alert Telescope (BAT; \citealt{Barthelmy2005}) on the \textit{Neil Gehrels
Swift Observatory}\footnote{See \url{https://swift.gsfc.nasa.gov/results/batgrbcat}.} or the Gamma-ray Burst Monitor  \citep[GBM;][]{Meegan2009} on \textit{Fermi}\footnote{See \url{https://heasarc.gsfc.nasa.gov/W3Browse/fermi/fermigbrst.html}.}.

Several candidate counterparts were found with \textit{temporal} coincidence in the online catalog from the KONUS instrument on the \textit{Wind} satellite (SN\,2018etk, SN\,2018hom, SN\,2019xcc, SN\,2020jqm, SN\,2020lao, SN\,2020tkx, SN\,2021aug). However, given the relatively imprecise explosion date constraints for several of the events in our sample (see Table \ref{tab:opt_data}), and the coarse localization information from the KONUS instrument, we cannot firmly associate any of these GRBs with the SNe Ic-BL. In fact, given the rate of GRB detections by KONUS ($\sim 0.42$\,d$^{-1}$) and the time window over which we searched for counterparts ($30$\,d in total; derived from the explosion date constraints), the observed number of coincidences (13) is consistent with random fluctuations. Finally, none of the possible coincidences were identified in events with explosion date constraints more precise than 1\,d. 

\begin{figure*}
    \centering
    \includegraphics[width=18cm]{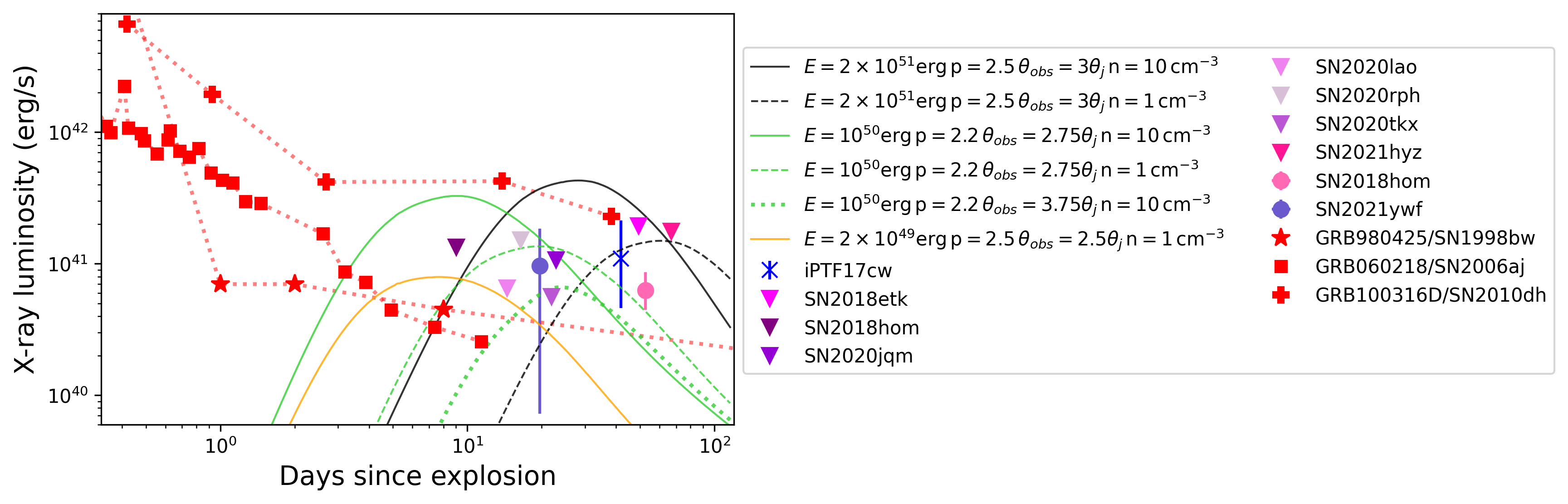}
    \caption{\textit{Swift}/XRT upper-limits and detections (downward pointing triangles and filled circles with error bars, respectively) obtained for 9 of the 16 SNe Ic-BL in our sample. We plot the observed X-ray luminosity as a function of time since explosion. We compare these observations with the X-ray light curves of the low-luminosity GRB\,980425 \citep[red stars;][]{Kouveliotou2004}, GRB\,060218 \citep[red squares;][]{Campana2006},  GRB\,100316D \citep[red crosses;][]{Margutti2013}, and with the relativistic iPTF17cw \citep[blue cross;][]{Corsi2017}. Dotted red lines connect the observed data points (some of which at early and late times are not shown in the plot) for these three low-luminosity GRBs. We also plot the observed  X-ray luminosity predicted by off-axis GRB models \citep[black, green, and orange lines;][]{vanEerten2011, VanEerten2012}. We assume  $\epsilon_B=\epsilon_e=0.1$, a constant density ISM in the range $n = 1-10\,{\rm cm}^{-3}$, a top-hat jet of opening angle $\theta_j=0.2$, and various observer's angles $\theta_{\rm obs}=(2.5-3)\theta_j$.}
    \label{fig:X-raymodel}
\end{figure*}

\subsection{X-ray constraints}
\label{sec:X-raymodel}
Seven of the 9 SNe Ic-BL observed with \textit{Swift}-XRT did not result in a significant detection. In Table \ref{tab:x-ray} we report the derived 90\% confidence flux upper limits in the 0.3--10\,keV band after correcting for Galactic absorption \citep{Willingale+2013}. 

While observations of SN\,2021ywf resulted in a  $\approx 3.2\sigma$  detection significance (Gaussian equivalent), the limited number of photons (8) precluded a meaningful spectral fit. Thus, a $\Gamma = 2$ power-law spectrum was adopted for the flux conversion for this source as well. We note that because of the relatively poor spatial resolution of the \textit{Swift}-XRT (estimated positional uncertainty of 11.7\arcsec~radius  at 90\% confidence), we cannot entirely rule out unrelated emission from the host galaxy of SN 2021ywf (e.g., AGN, X-ray binaries, diffuse host emission; see Figure \ref{fig:hosts-det} for the host galaxy).

For SN\,2019hsx we detected enough photons to perform a spectral fit for count rate to flux conversion. The spectrum is found to be relatively soft, with a best-fit power-law index of $\Gamma = 3.9^{+3.0}_{-2.1}$. 
Our \textit{Swift} observations of SN\,2019hsx do not show significant evidence for variability of the source X-ray flux over the timescales of our follow up. While the lack of temporal variability is not particularly constraining given the low signal-to-noise ratio in individual epochs, we caution that also in this case the relatively poor spatial resolution of the \textit{Swift}-XRT (7.4\arcsec~radius position uncertainty at 90\% confidence) implies that unrelated emission from the host galaxy cannot be excluded.

The constraints derived from 
the \textit{Swift}-XRT observations can be compared with the
X-ray light curves of low-luminosity GRBs, or models of GRB afterglows 
observed slightly off-axis. For the latter, we use
the numerical model by \citet{vanEerten2011, VanEerten2012}. We assume equal energies in the electrons and magnetic fields ($\epsilon_B= \epsilon_e=0.1$), and an interstellar medium (ISM) of density $n = 1-10$\,cm$^{-3}$. We note that a constant density ISM (rather than a wind profile) has been shown to fit the majority of GRB afterglow light curves, implying that most GRB progenitors might have relatively small wind termination-shock radii \citep{Schulze2011}. We generate the model light curves for a nominal redshift of $z=0.05$ and then convert the predicted flux densities into X-ray luminosities by integrating over the 0.3-10\,keV energy range and  neglecting the small redshift corrections. We plot the model light curves in Figure \ref{fig:X-raymodel}, for various energies, different power-law indices $p$ of the electron energy distribution, and various off-axis angles (relative to a jet opening angle, set to $\theta_j=0.2$). In the same Figure we also plot  the X-ray light curves of low-luminosity GRBs for comparison (neglecting redshift corrections). 
As evident from this Figure, our \textit{Swift}/XRT upper limits (downward-pointing triangles) 
exclude X-ray afterglows associated with higher-energy GRBs observed slightly off-axis.  However, X-ray emission as faint as the afterglow of
the low-luminosity GRB\,980425 cannot be excluded. As we discuss in the next Section, radio data collected with the VLA enable us to exclude GRB\,980425/SN\,1998bw-like emission for most of the SNe in our sample. 

We note that our two X-ray detections of SN\,2019hsx and SN\,2021ywf are consistent with several GRB off-axis light curve models and, in the case of SN\,2021ywf, also with GRB\,980425-like emission within the large errors. However, for this interpretation of their X-ray emission to be compatible with our radio observations (see Table \ref{tab:data}), one needs to invoke a flattening of the radio-to-X-ray spectrum, similar to what has been invoked for other stripped-envelope SNe in the context of cosmic-ray dominated shocks \citep{Ellison2000, Chevalier2006}.

\subsection{Radio constraints}
\label{sec:radioanalysis}
As evident from Table \ref{tab:data}, we have obtained at least one radio detection for 11 of the 16 SNe in our sample.  None of these 11 radio sources were found to be coincident with known radio sources in the VLA FIRST \citep{Becker1995} catalog (using a search radius of 30\arcsec~around the optical SN positions). This is not surprising since the FIRST survey had a typical RMS sensitivity of $\approx 0.15$\,mJy at 1.4\,GHz, much shallower than the deep VLA follow-up observations carried out within this follow-up program. We also checked the quick look images from the VLA Sky Survey (VLASS) which reach a typical RMS sensitivity of $\approx 0.12$\,mJy at 3\,GHz \citep{Hernandez2018,Law2018}. We could find images for all but one (SN\,2021epp) of the fields containing the 16 SNe BL-Ic in our sample. The VLASS images did not provide any radio detection at the locations of the SNe in our sample.

Five of the 11 SNe Ic-BL with radio detections are associated with extended or marginally resolved radio emission. Two other radio-detected events (SN\,2020rph and SN\,2021hyz) appear point-like in our images, but show no evidence for significant variability of the detected radio flux densities over the timescales of our observations. Thus, for a total of 7 out of 11 SNe Ic-BL with radio detections, we consider the measured flux densities as upper-limits corresponding to the brightness of their host galaxies, similarly to what was done in e.g. \citet{Soderberg2006} and \citet{Corsi2016}. The remaining 4 SNe Ic-BL with radio detections are compatible with point sources (SN\,2018hom, SN\,2020jqm, SN\,2020tkx, and SN\,2021ywf), and all but one (SN\,2018hom) had more than one observation in the radio via which we were able to establish the presence of substantial variability of the radio flux density. Hereafter we consider these 4 detections as genuine radio SN counterparts, though we stress that with only one observation of SN\,2018hom we cannot rule out a contribution from host galaxy emission, especially given that the radio follow up of this event was carried out with the VLA in its most compact (D) configuration with poorer angular resolution.

In summary, our radio follow-up campaign of 16 SNe Ic-BL resulted in 4 radio counterpart detections, and 12 deep upper-limits on associated radio counterparts.

\begin{figure*}
    \begin{center}
    \includegraphics[width=6cm]{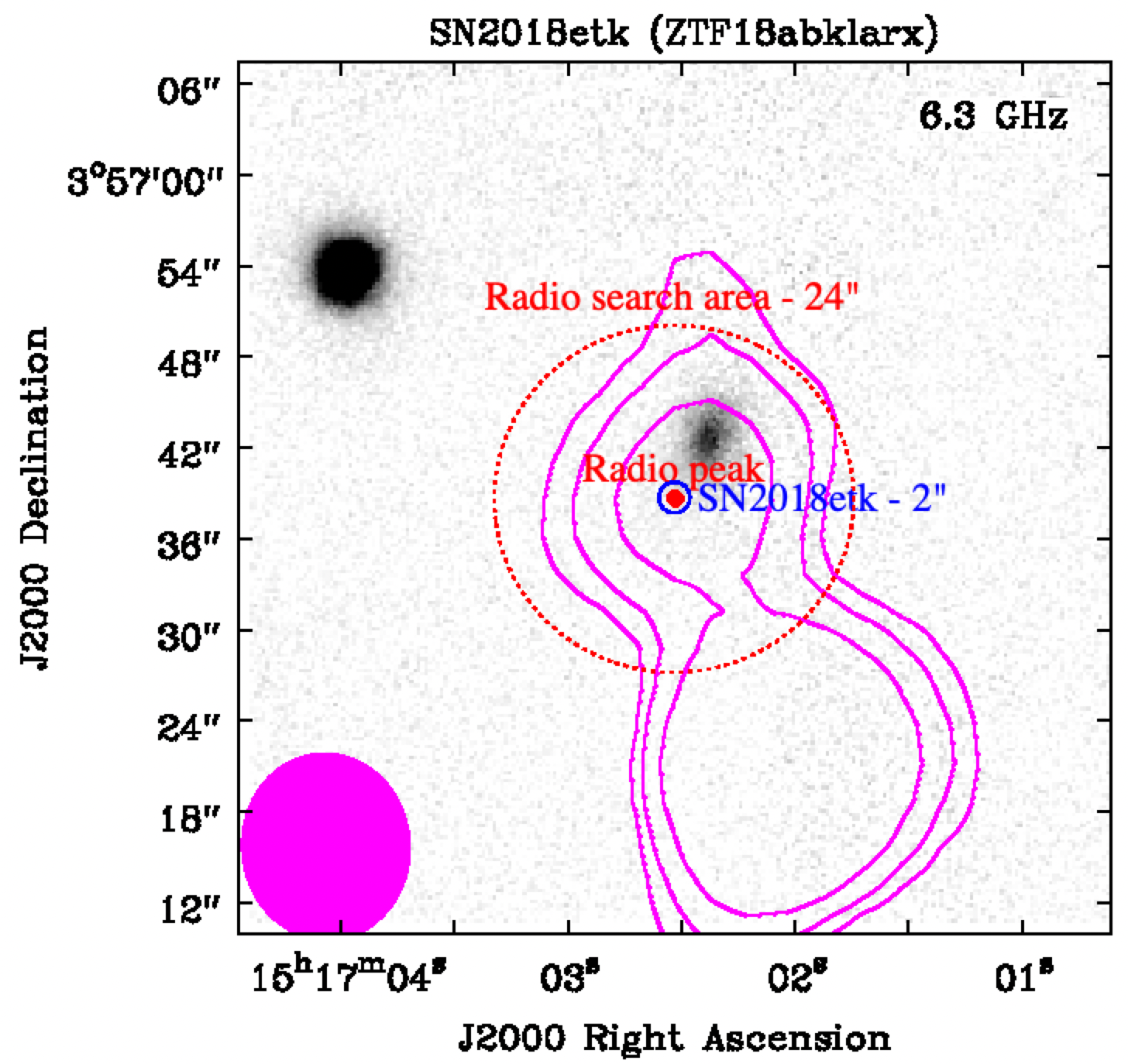}
    \includegraphics[width=6cm]{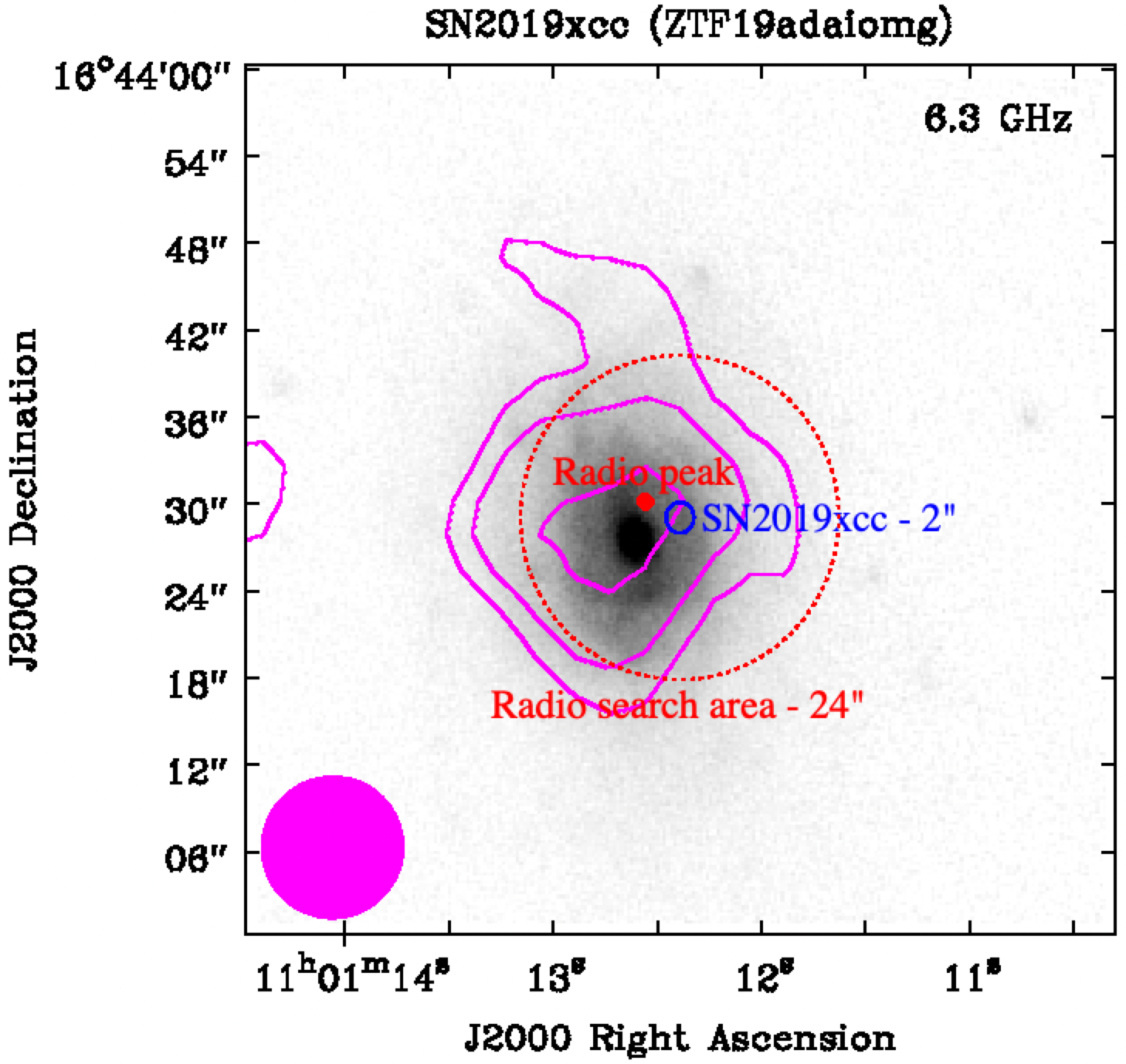}
    \includegraphics[width=6.cm,height=5.6cm]{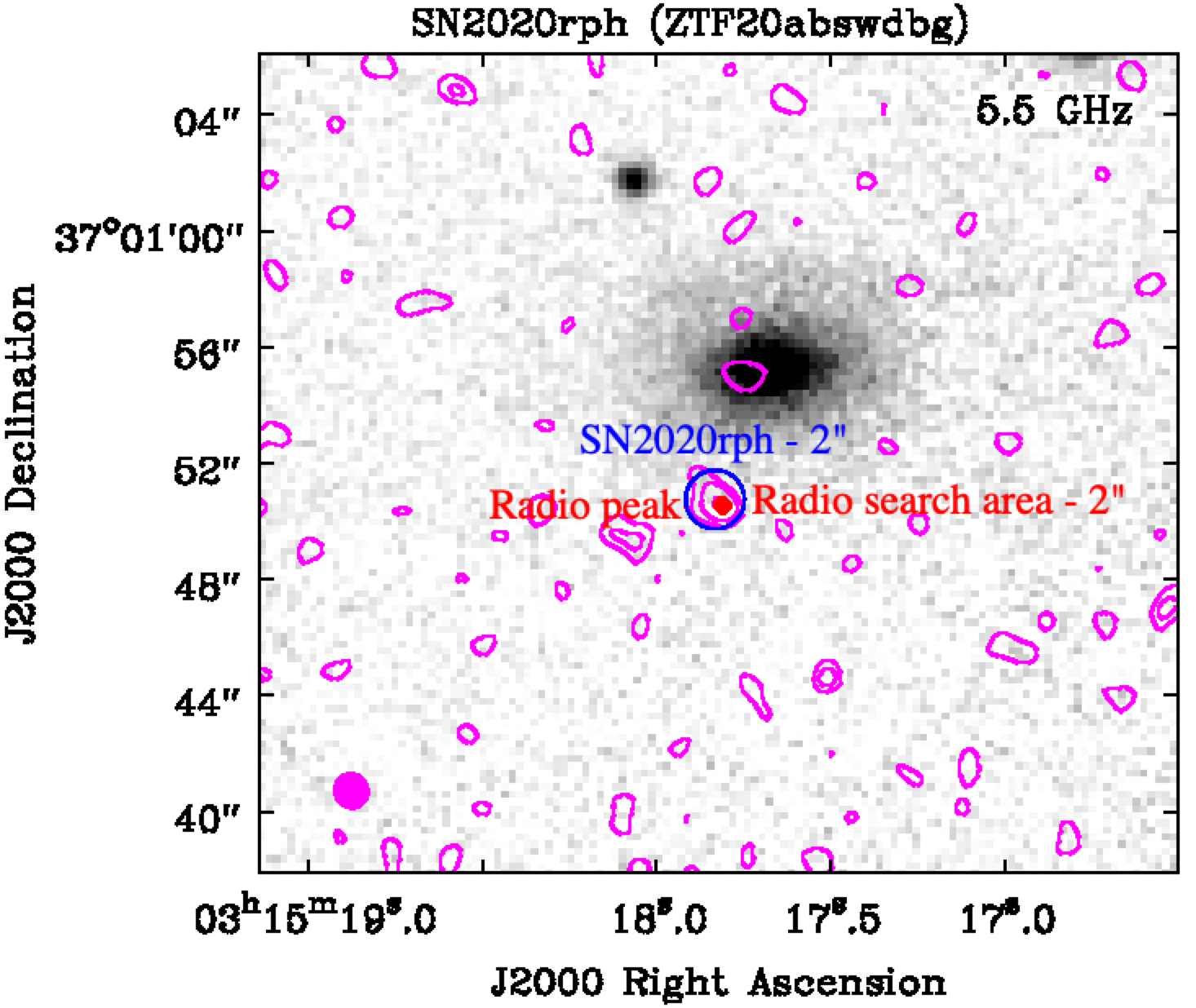}
    \includegraphics[width=6cm]{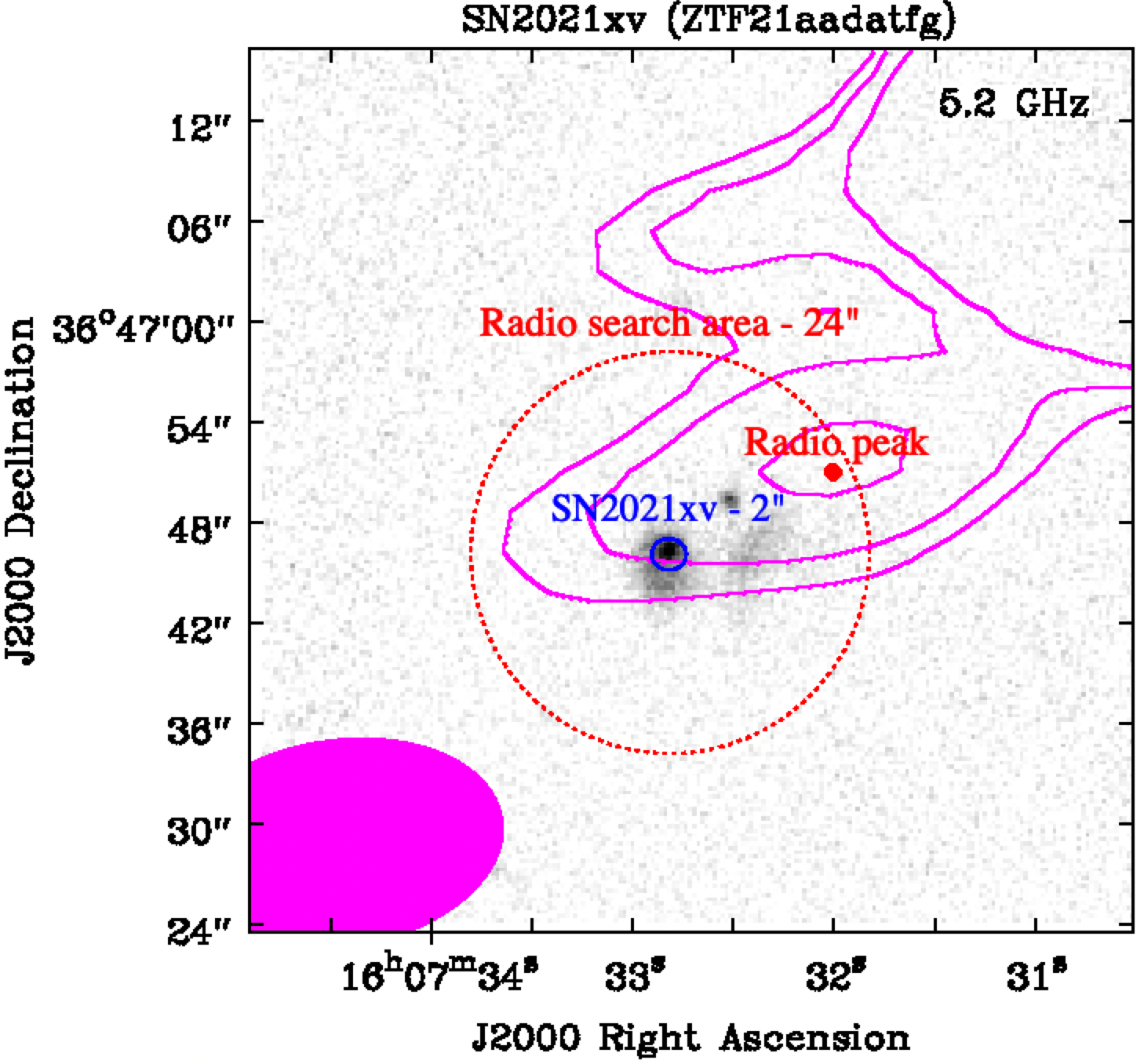}
    \includegraphics[width=6cm,height=5.6cm]{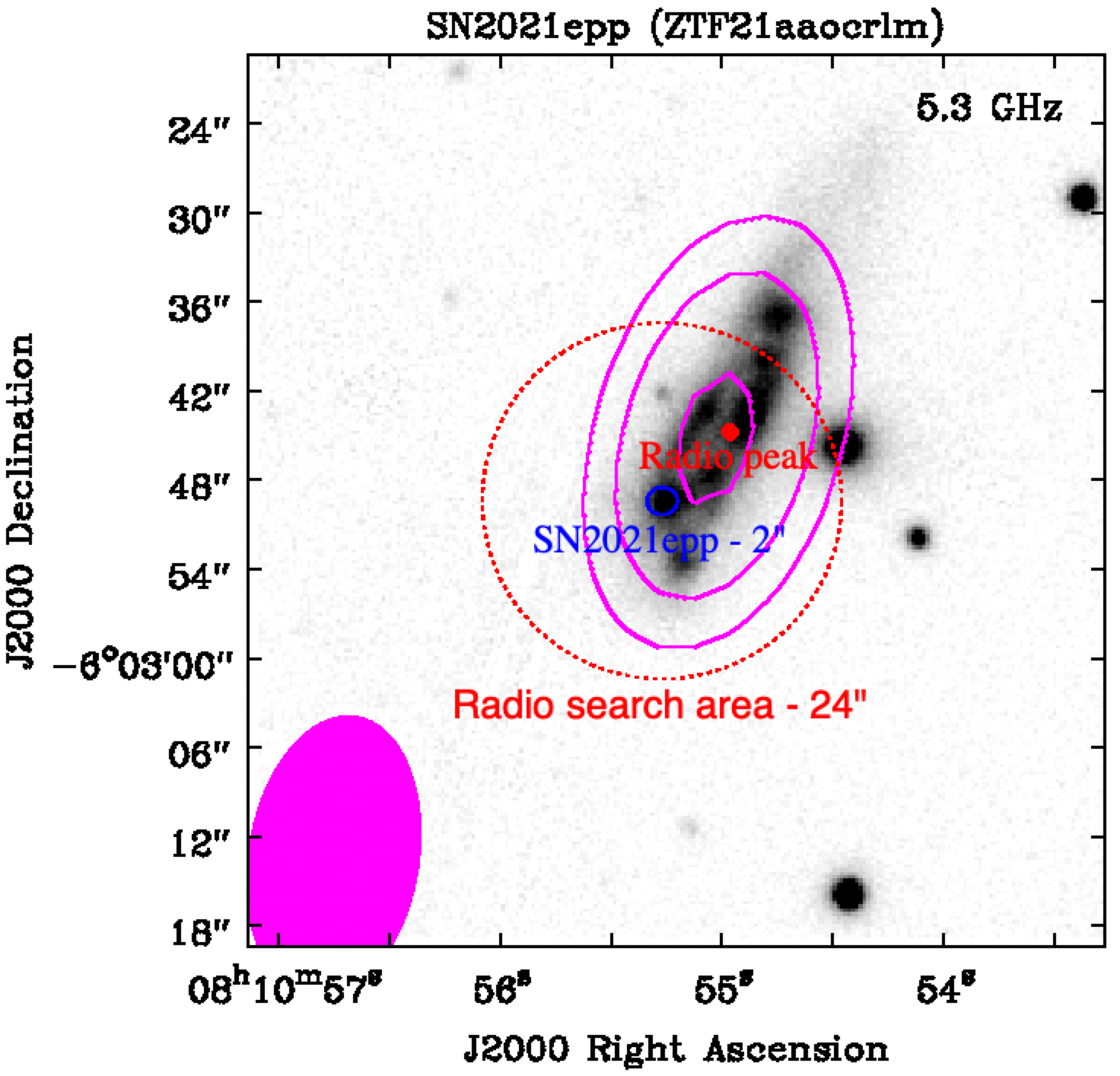}
    \includegraphics[width=6cm]{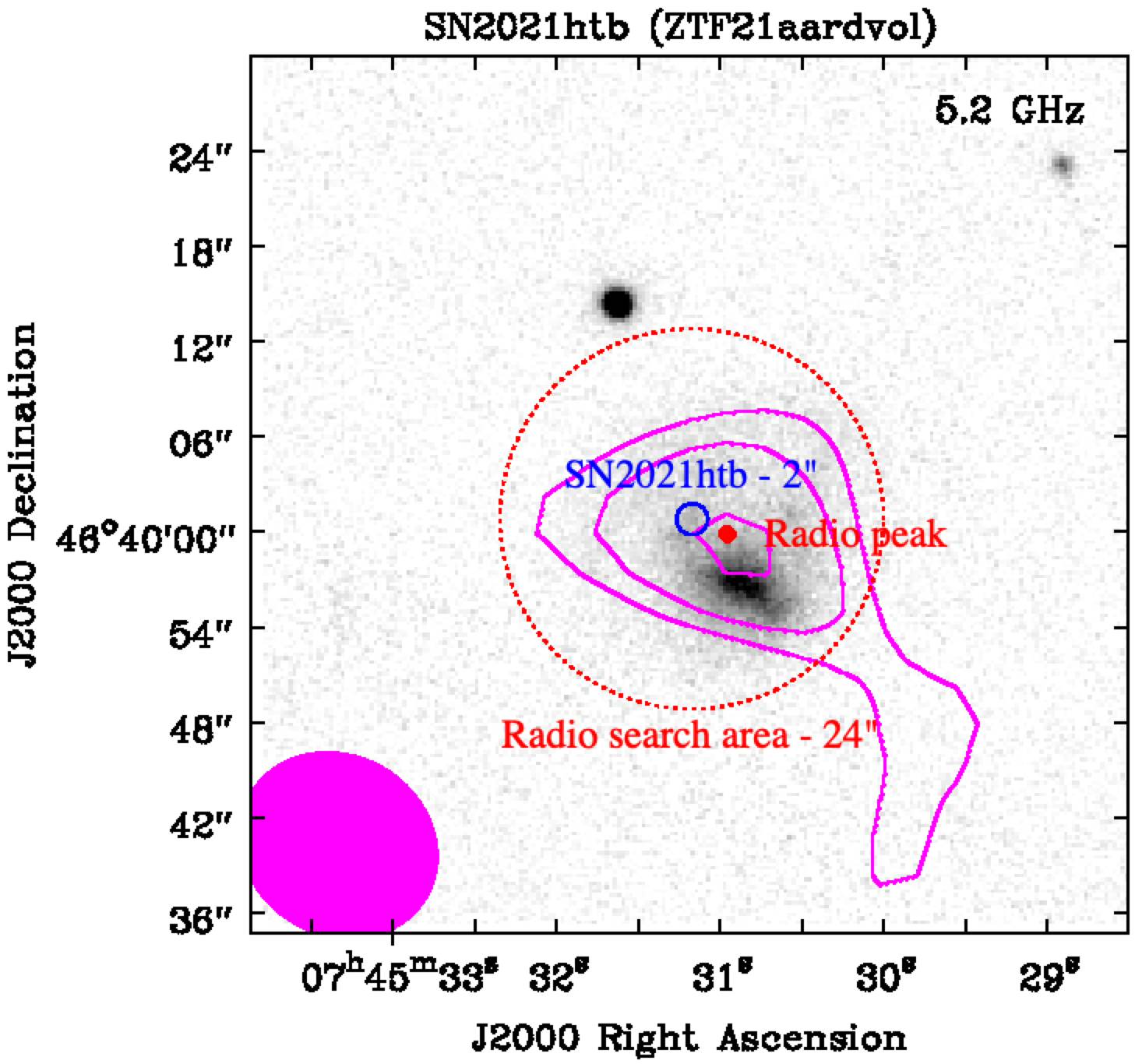}
    \includegraphics[width=6cm]{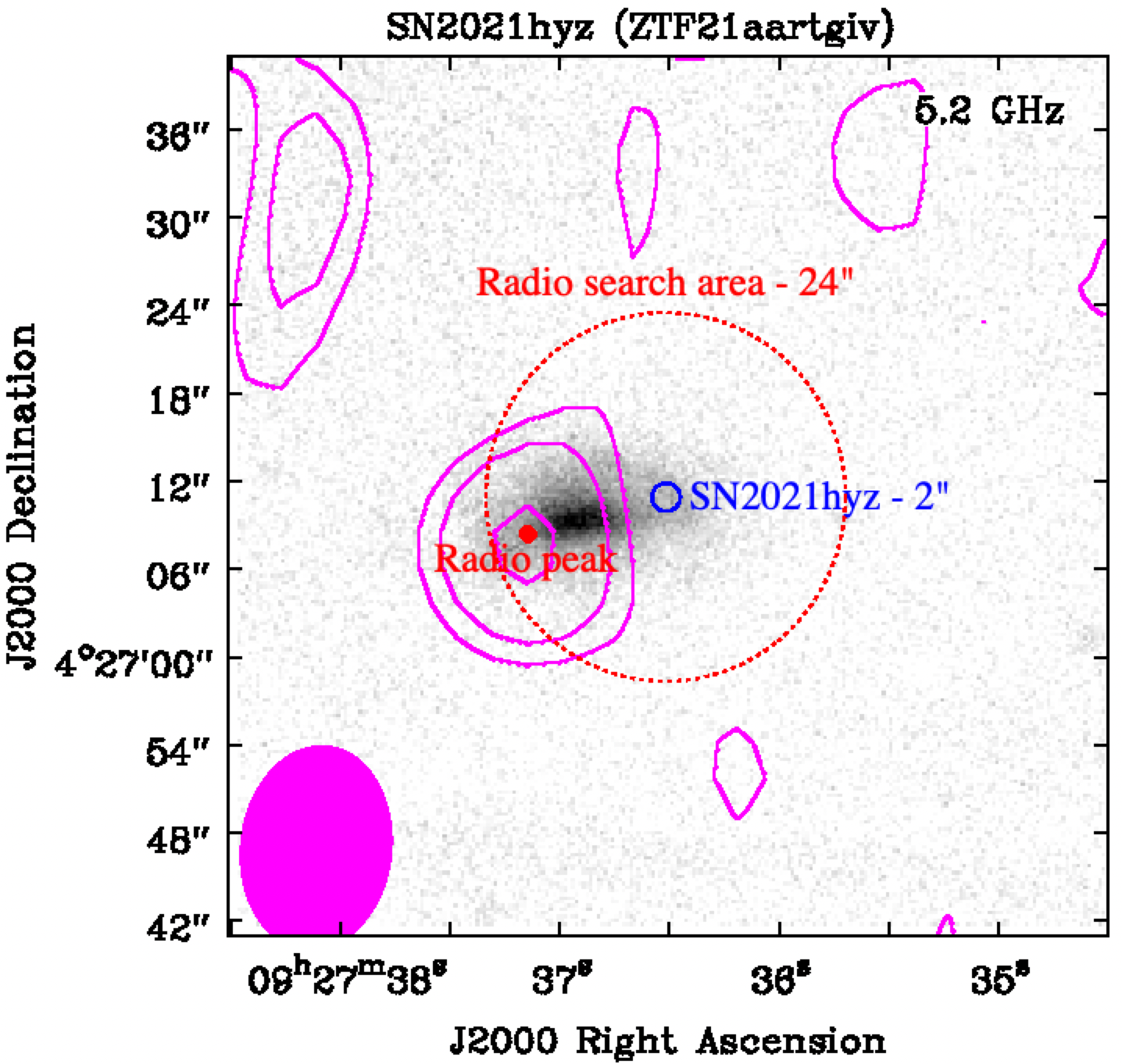}
    \caption{PanSTARRS-1 \citep{Flewelling2020} reference $r$-band images of the fields of the SNe in our sample for which host galaxy light dominates the radio emission. Contours in magenta are 30\%, 50\%, and 90\% of the radio peak flux reported in Table \ref{tab:data} for the first radio detection of each field. The blue circles centered on the optical SN positions (not shown in the images) have sizes of 2\arcsec  \citep[comparable to the ZTF PSF at average seeing;][]{Bellm2019}. The red-dotted circles enclose the region in which we search for radio counterparts (radii equal to the nominal FWHM of the VLA synthesized beams;  Table \ref{tab:data}). The sizes of the actual VLA synthesized beams  are shown as filled magenta ellipses. The red dots mark the locations of the radio peak fluxes measured in the radio search areas. \label{fig:hosts}}
    \end{center}
\end{figure*}

\begin{figure*}
    \begin{center}
    \hbox{
    \includegraphics[height=8.3cm]{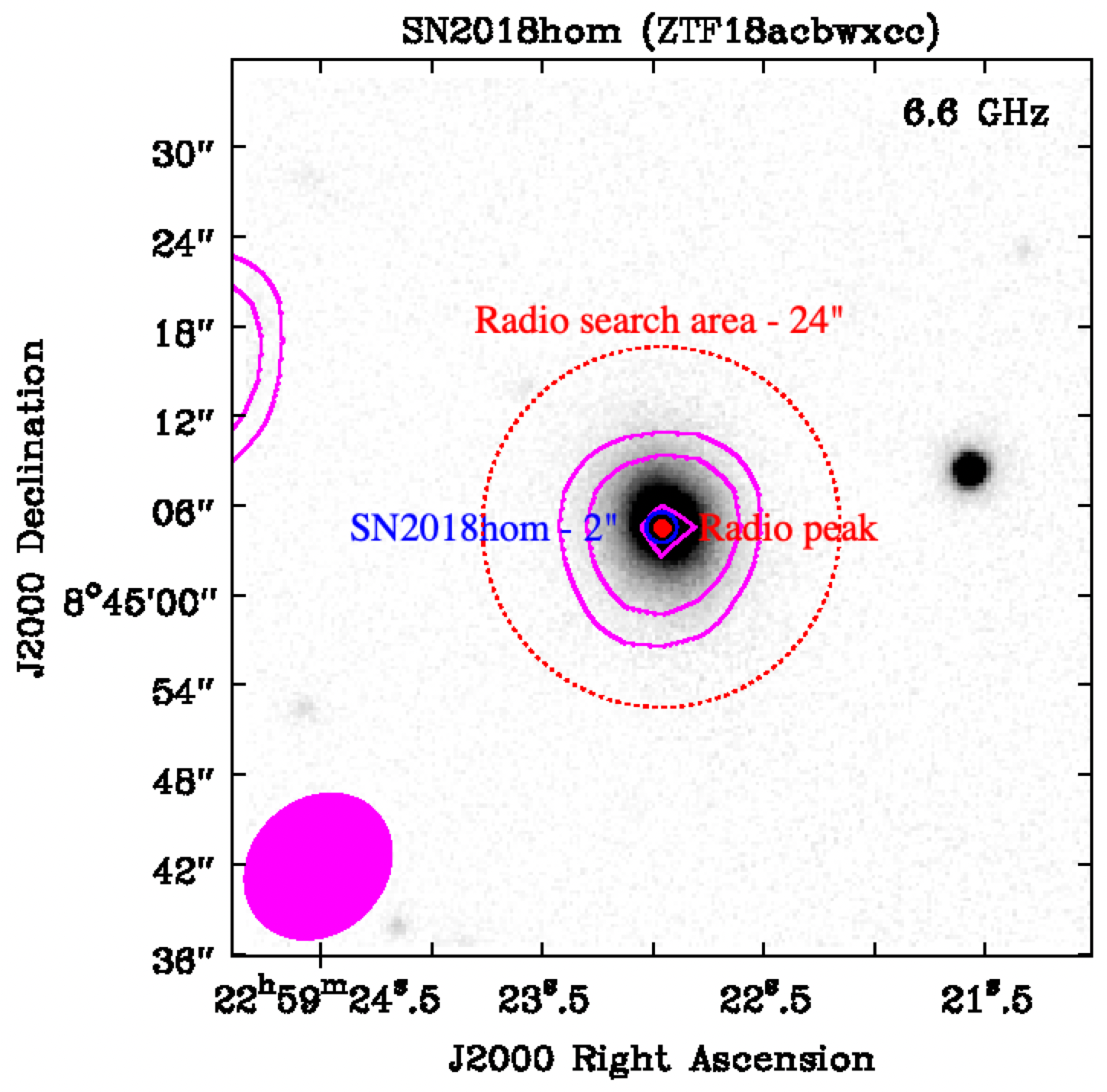}
    \hspace{0.2cm}
    \includegraphics[height=8.3cm]{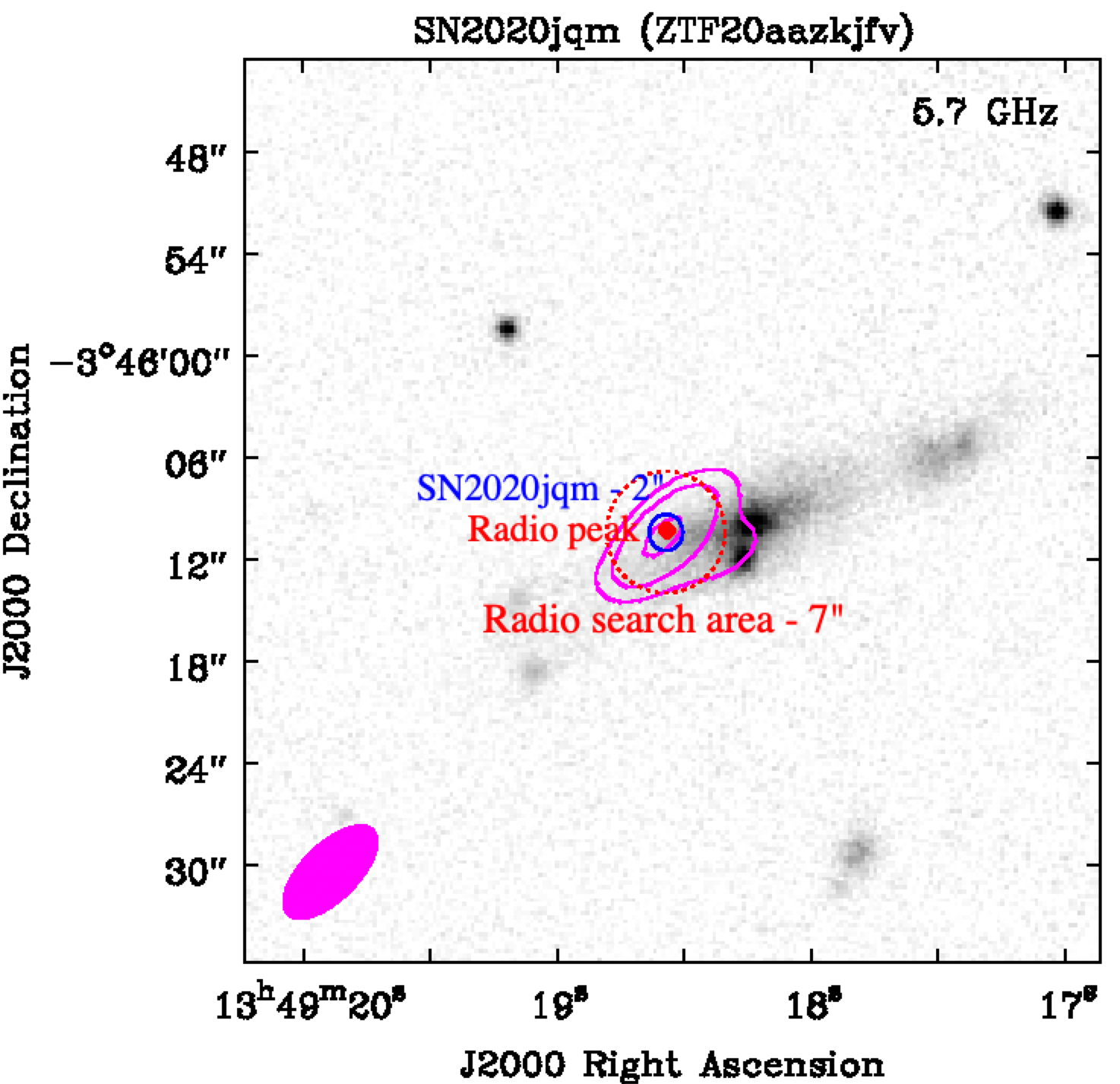}}
    \hbox{
    \includegraphics[height=7.8cm]{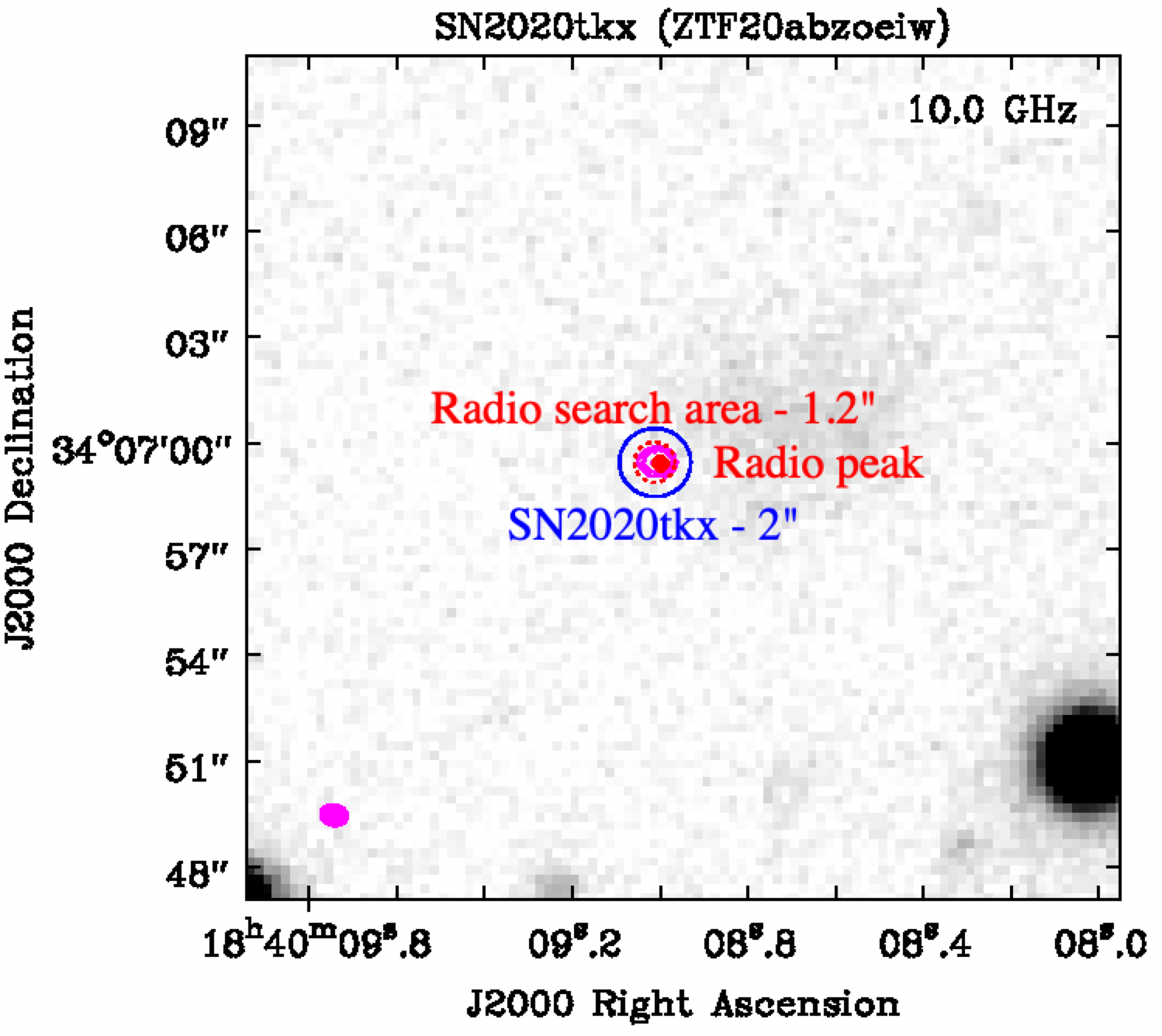}
    \includegraphics[height=7.8cm]{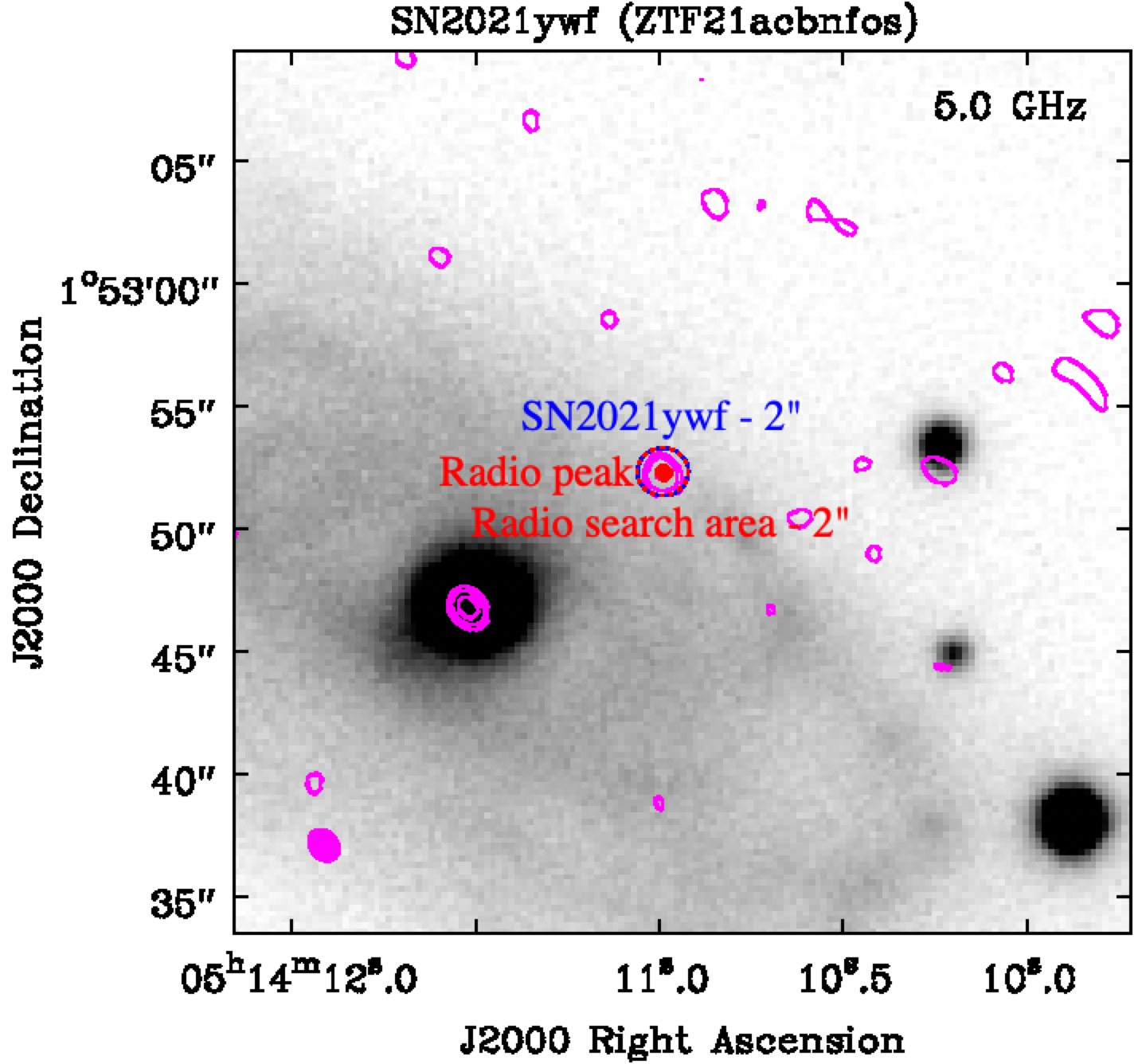}}
    \caption{Same as Figure \ref{fig:hosts} but for the fields containing the SNe in our sample for which we detected a SN radio counterpart.  We stress that with only one observation of SN\,2018hom we cannot rule out a contribution from host galaxy emission, especially given that the radio follow up of this event was carried out with the VLA in its D configuration. \label{fig:hosts-det}}
    \end{center}
\end{figure*}

\begin{figure*}
    \centering
    \includegraphics[width=18cm]{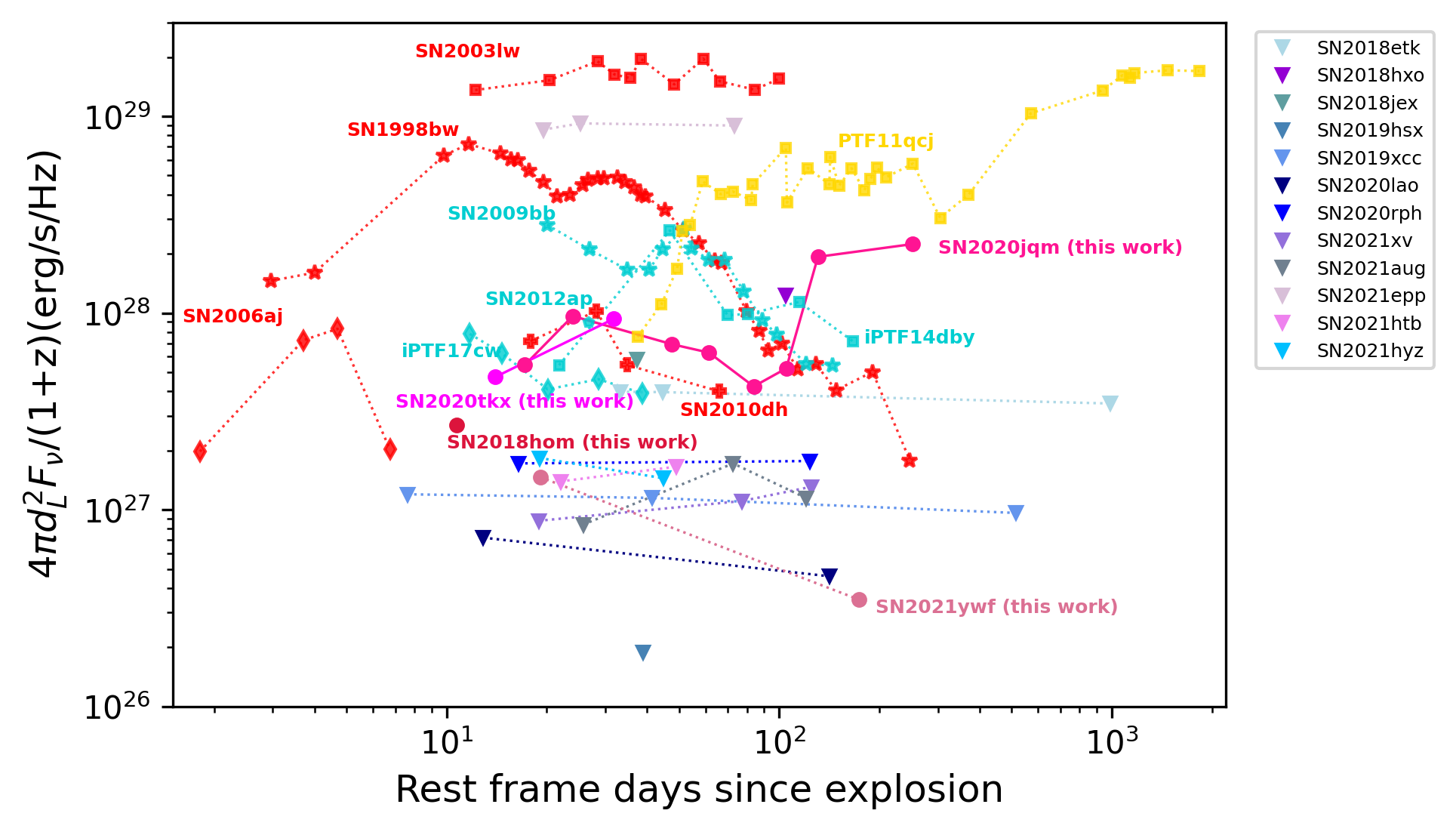}
    \caption{Radio ($\approx 6$\,GHz) observations of the 16 SNe Ic-BL in our sample (filled circles and downward pointing triangles in shades of pink, purple, and blue). Upper-limits associated with non-detections ($3\sigma$ or brightness of the host galaxy at the optical location of the SN) are plotted with downward-pointing triangles; detections are plotted with filled circles. We compare these observations with the radio light curves of GRB-SNe (red); of relativistic-to-mildly relativistic SNe Ic-BL discovered independently of a $\gamma$-ray trigger (cyan); and with PTF11qcj \citep{Corsi2014}, an example of a radio-loud non-relativistic and CSM-interacting SN Ic-BL (yellow). As evident from this Figure, our observations exclude SN\,1998bw-like radio emission for all but one (SN\,2021epp) of the events in our sample. This doubles the sample of SNe Ic-BL for which radio emission observationally similar to SN\,1998bw was previously excluded \citep{Corsi2016}, bringing the upper limit on the fraction of SNe compatible with SN\,1998bw down to $< 19\%$ \citep[compared to $< 41\%$ previously reported in][]{Corsi2016}.
    For 10 of the 16 SNe presented here we also exclude relativistic ejecta with radio luminosity densities in between $\approx 5\times10^{27}$\,erg\,s$^{-1}$\,Hz$^{-1}$ and  $\approx 10^{29}$\,erg\,s$^{-1}$\,Hz$^{-1}$ at $t\gtrsim 20$\,d, similar to SNe associated with low-luminosity GRBs such as SN\,1998bw \citep{Kulkarni1998}, SN\,2003lw \citep{Soderberg2004}, SN\,2010dh \citep{Margutti2013}, or to the relativistic SN\,2009bb \citep{Soderberg2010} and iPTF17cw \citep{Corsi2017}.  None of our observations exclude radio emission similar to that of SN\,2006aj. }
    \label{fig:radiolight}
\end{figure*}

\begin{figure}
    \centering
    \includegraphics[width=8.6cm]{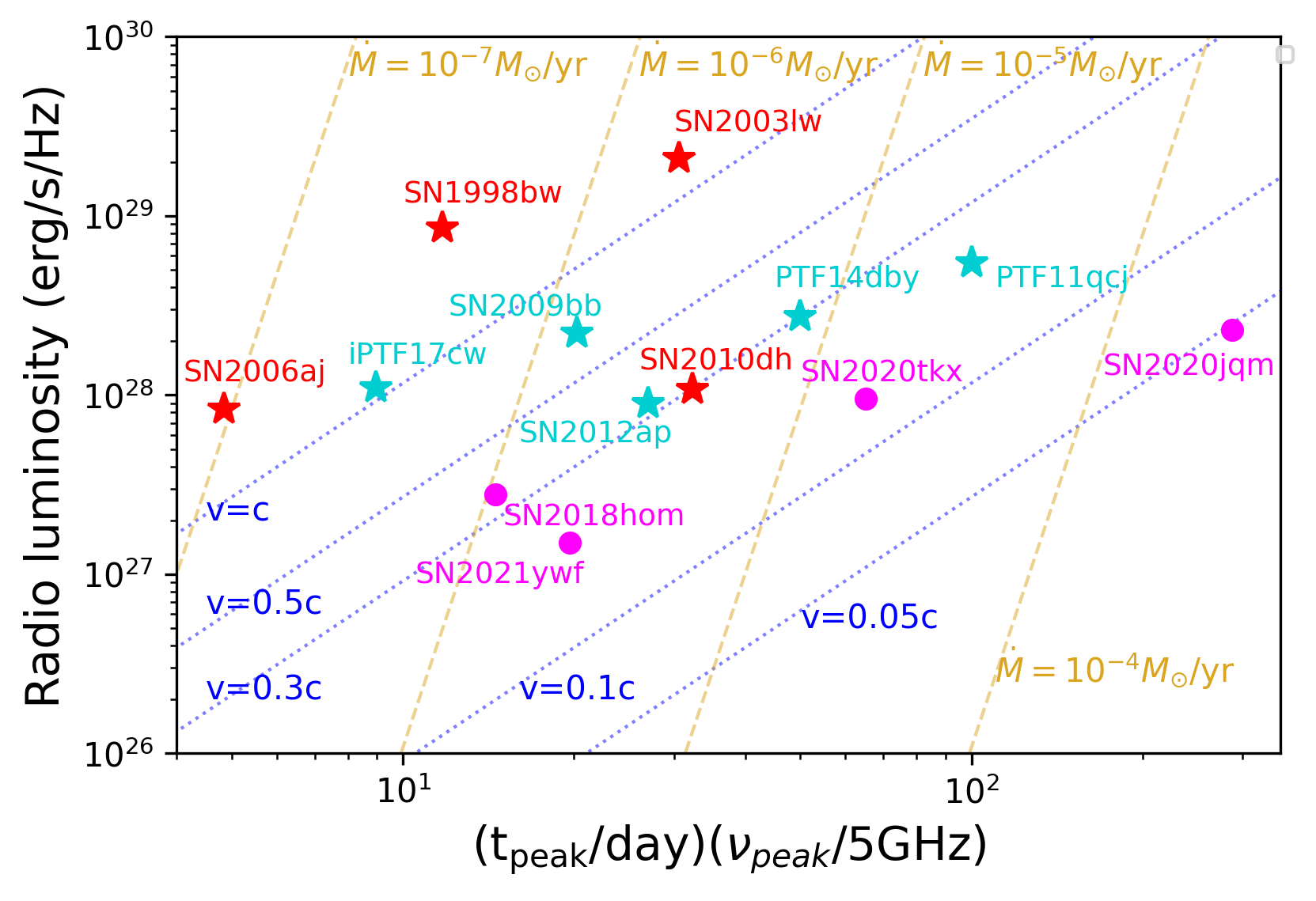}
    \caption{Properties of the radio-emitting ejecta of the SNe in our sample for which we detect a radio counterpart (magenta dots), compared with those of GRB-SNe (red stars) and of relativistic-to-mildly
relativistic SNe Ic-BL discovered independently of a $\gamma$-ray trigger (cyan stars). As evident from this Figure, only SN\,2018hom is compatible with an ejecta speed $\gtrsim 0.3c$, though with the caveat that we only have one radio observation for this SN. None of the other ZTF SNe Ic-BL in our sample shows evidence for ejecta faster than $0.3c$. We also note that SN\,2020jqm lies in the region of the parameter space occupied by radio-loud CSM-interacting SNe similar to PTF11qcj. See Section \ref{sec:radio_properties} for discussion.}
    \label{fig:radiospeed}
\end{figure}

\subsubsection{Fraction of SN\,1998bw-like SNe Ic-BL}
The local rate of SNe Ic-BL is estimated to be $\approx 5\%$ of the core-collapse SN rate \citep{Li2011,Shivvers2017,Perley2020b} or $\approx 5\times10^3$\,Gpc$^{-3}$\,yr$^{-1}$ assuming a core-collapse SN rate of $\approx 10^5$\,Gpc$^{-3}$\,yr$^{-1}$ \citep{Perley2020b}.  Observationally, we know that cosmological long GRBs are characterized by ultra-relativistic jets observed on-axis, and have an \textit{intrinsic} (corrected for beaming angle) local volumetric rate of $79^{+57}_{-33}$\,Gpc$^{-3}$\,yr$^{-1}$ \citep[e.g.,][and references therein]{Ghirlanda2022}. Hence, only ${\cal O}(1)\%$ of SNe Ic-BL can make long GRBs. For low-luminosity GRBs, the \textit{observed} local rate is affected by large errors, $230^{+490}_{-190}$\,Gpc$^{-3}$\,yr$^{-1}$ \citep[see][and references therein]{Bromberg2011}, and their typical beaming angles are largely unconstrained. Hence, the question of what fraction of SNe Ic-BL can make low-luminosity GRBs remains to be answered. 

Radio observations of SNe Ic-BL are a powerful way to constrain this fraction independently of relativistic beaming effects that preclude observations of jets in X-rays and $\gamma$-rays for off-axis observers. However, observational efforts aimed at constraining the fraction of SNe Ic-BL harboring low-luminosity GRBs independently of $\gamma$-ray observations have long been challenged by the rarity of the SN Ic-BL optical detections (compared to other core-collapse events), coupled with the small number of these rare SNe for which the community has been able to collect deep radio follow-up observations within 1\,yr since explosion \citep[see e.g.,][]{Soderberg2006catalog}. Progress in this respect has been made since the advent of the PTF, and more generally with synoptic optical surveys that have greatly boosted the rate of stripped-envelope core-collapse SN discoveries \citep[e.g.,][]{Shappee2014,Sand2018,Tonry2018}. 

In our previous work \citep{Corsi2016}, we presented one of the most extensive samples of SNe Ic-BL with deep VLA observations, largely composed of events detected by the PTF/iPTF. Combining our sample with the SN Ic-BL\,2002ap \citep{GalYam2002,Mazzali2002} and SN\,2002bl \citep{Armstrong2002,Berger2003}, and the CSM-interacting SN Ic-BL\,2007bg \citep{Salas2013}, we had overall 16 SNe Ic-BL for which radio emission observationally similar to SN\,1998bw was excluded, constraining the rate of SNe Ic-BL observationally similar to SN\,1998bw to $< 6.61/16\approx 41\%$, where we have used the fact that the Poisson 99.865\% confidence (or $3\sigma$ Gaussian equivalent for a single-sided distribution) upper-limit on zero SNe compatible with SN\,1998bw is $\approx 6.61$.

With the addition of the 16 ZTF SNe Ic-BL presented in this work, we now have doubled the sample of SNe Ic-BL with deep VLA observations presented in \citet{Corsi2016}, providing evidence for additional 15 SNe Ic-BL (all but SN\,2021epp; see Figure \ref{fig:radiolight}) that are observationally different from SN\,1998bw in the radio. Adding to our sample also SN\,2018bvw \citep{Ho2020_ZTF18aaqjovh}, AT\,2018gep \citep{Ho2019_ZTF18abukavn}, and SN\,2020bvc \citep{Ho2020_ZTF20aalxlis}, whose radio observations exclude SN\,1998bw-like emission, we are now at 34 SNe Ic-BL that are observationally different from SN\,1998bw. Hence, we can tighten our constraint on the fraction of 1998bw-like SNe Ic-BL to $< 6.61/34\approx 19\%$ (99.865\% confidence). This upper-limit implies that the \textit{intrinsic} rate of 1998bw-like GRBs is $\lesssim 950$\,Gpc$^{-3}$\,yr$^{-1}$. Combining this constraint with the rate of low-luminosity GRBs derived from their high-energy emission, we conclude that low-luminosity GRBs have inverse beaming factors $2/\theta^2\lesssim 4^{+20}_{-3}$, corresponding to  jet half-opening angles $\theta \gtrsim 40^{+40}_{-24}$\,deg. 

We note that for 10 of the SNe in the sample presented here we also exclude relativistic ejecta with radio luminosity densities in between $\approx 5\times10^{27}$\,erg\,s$^{-1}$\,Hz$^{-1}$ and  $\approx 10^{29}$\,erg\,s$^{-1}$\,Hz$^{-1}$ at $t\gtrsim 20$\,d, pointing to the fact that SNe Ic-BL similar to those associated with low-luminosity GRBs, such as SN\,1998bw \citep{Kulkarni1998}, SN\,2003lw \citep{Soderberg2004}, SN\,2010dh \citep{Margutti2013}, or to the relativistic SN\,2009bb \citep{Soderberg2010} and iPTF17cw \citep{Corsi2017}, are intrinsically rare.  However, none of our observations exclude radio emission similar to that of SN\,2006aj. This is not surprising since the afterglow of this low-luminosity GRB faded on timescales much faster than the $20-30$ days since explosion that our VLA monitoring campaign allowed us to target. To enable progress, obtaining prompt ($\lesssim 5$\,d since explosion) and accurate spectral classification paired with deep radio follow-up observations of SNe Ic-BL should be a major focus of future studies. At the same time, as discussed in \citet{Ho2020_ZTF20aalxlis}, high-cadence optical surveys can provide an alternative way to measure the rate of SNe Ic-BL that are similar to SN\,2006aj independently of $\gamma$-ray and radio observations, by catching potential optical signatures of shock-cooling emission at early times.  Based on an analysis of ZTF SNe with early high-cadence light curves, \citet{Ho2020_ZTF20aalxlis} concluded that it appears that SN\,2006aj-like events are uncommon, but more events will be needed to measure a robust rate.

\subsubsection{Properties of the radio-emitting ejecta}
\label{sec:radio_properties}
Given that none of the SNe in our sample shows evidence for relativistic ejecta, hereafter we consider their radio properties within the synchrotron self-absorption (SSA) model for radio SNe \citep{Chevalier1998}. Within this model, constraining the radio peak frequency and peak flux can provide information on the size of the radio emitting material. We start from Equations (11) and (13) of \citet{Chevalier1998}:
\begin{eqnarray}
\nonumber R_{p}\approx 8.8\times10^{15}\,{\rm cm}\left(\frac{\eta}{2\alpha}\right)^{1/(2p+13)} \left(\frac{F_p}{\rm Jy}\right)^{(p+6)/(2p+13)}\times\\\left(\frac{d_L}{\rm Mpc}\right)^{(2p+12)/(2p+13)}\left(\frac{\nu_p}{5\,\rm GHz}\right)^{-1},~~~
\label{eq:ejspeed}
\end{eqnarray}
where $\alpha \approx 1$ is the ratio of relativistic electron energy density to magnetic energy density, $F_{p}$  is the flux density at the time of SSA peak, $\nu_{p}$ is the SSA frequency, and where $R/\eta$ is the thickness of the radiating electron shell. The normalization of the above Equation has a small dependence on $p$ and in the above we assume $p\approx 3$ for the power-law index of the electron energy distribution. Setting $R_p \approx {\rm v_s} t_p$ in Equation (\ref{eq:ejspeed}), and considering that $L_p \approx 4\pi d^2_L F_p$ (neglecting redshift effects), we get:
\begin{eqnarray}
 \nonumber  \left(\frac{L_p}{\rm erg\,s^{-1}\,Hz^{-1}}\right) \approx 1.2\times10^{27} \left(\frac{\beta_s}{3.4}\right)^{(2p+13)/(p+6)} \\\times\left(\frac{\eta}{2\alpha}\right)^{-1/(p+6)} \left(\frac{\nu_p}{5\,\rm GHz}\frac{t_p}{\rm 1\,d}\right)^{(2p+13)/(p+6)}
\end{eqnarray}
where we have set $\beta_s = {\rm v_s}/c$. We plot in Figure \ref{fig:radiospeed} with blue-dotted lines the relationship above for various values of $\beta_s$ (and for $p=3$, $\eta=2$, $\alpha=1$). As evident from this Figure, relativistic events such as SN\,1998bw (for which the non-relativistic approximation used in the above Equations breaks down) are located at $\beta_s\gtrsim 1$. None of the ZTF SNe Ic-BL in our sample for which we obtained a radio counterpart detection shows evidence for ejecta faster than $0.3c$, except possibly for SN\,2018hom. However, for this event we only have one radio observation and hence contamination from the host galaxy cannot be excluded. We also note that SN\,2020jqm lies in the region of the parameter space occupied by radio-loud CSM interacting SNe similar to PTF\,11qcj. 

The magnetic field can be expressed as \citep[see Equations (12) and (14) in][]{Chevalier1998}:
\begin{eqnarray}
\nonumber B_p \approx 0.58\rm\,G \left(\frac{\eta}{2\alpha}\right)^{4/(2p+13)}\left(\frac{F_{p}}{\rm Jy}\right)^{-2/(2p+13)}\times\\\left(\frac{d_{L}}{\rm Mpc}\right)^{-4/(2p+13)}\left(\frac{\nu_p}{5\,\rm GHz}\right).
\label{eq2}
\end{eqnarray} 
Consider a SN shock expanding in a circumstellar medium (CSM) of density:
\begin{equation}
\rho\approx 5 \times10^{11} \,{\rm g\,cm}^{-1}A_{*}R^{-2}
\end{equation}
where:
\begin{equation}
  A_{*}=  \frac{\dot{M}/(10^{-5}M_{\odot}/{\rm yr})}{4\pi {\rm v}_w/(10^3{\rm km/s})}.
\label{eq:rho}
\end{equation}
Assuming that a fraction $\epsilon_B$ of the energy density $\rho{\rm v}^2_s$  goes into magnetic fields:
\begin{equation}
 \frac{B_p^2}{8\pi} = \epsilon_B \rho {\rm v_s}^2 = \epsilon_B \rho R_p^2 t_p^{-2},
 \label{eq3}
\end{equation}
one can write:
\begin{eqnarray}
\nonumber \left(\frac{L_p}{\rm erg\,s^{-1}\,Hz^{-1}}\right) \approx 1.2\times10^{27} \left(\frac{\eta}{2\alpha}\right)^{2}\left(\frac{\nu_p}{5\,\rm GHz}\frac{t_p}{1\,\rm d}\right)^{(2p+13)/2}\\\times \left(5\times10^3\epsilon_B A_*\right)^{-(2p+13)/4},~~~~~~
\end{eqnarray}
where we have used Equations  (\ref{eq2}), (\ref{eq:rho}), and (\ref{eq3}). We plot in Figure \ref{fig:radiospeed} with yellow-dashed lines the relationship above for various values of $\dot{M}$ (and for $p=3$, $\eta=2$, $\alpha=1$, $\epsilon_B =0.33$, $v_w=1000$\,km\,s$^{-1}$).  As evident from this Figure, relativistic events such as SN\,1998bw show a preference for smaller mass-loss rates. We note that while the above relationship depends strongly on the assumed values of $\eta$, $\epsilon_B$, and $v_w$, this trend for $\dot{M}$ remains true regardless of the specific values of these (uncertain) parameters. We also note that the above analysis assumes mass-loss in the form of a steady wind. While this is generally considered to be the case for relativistic SNe Ic-BL, binary interaction or eruptive mass loss in core-collapse SNe can produce denser CSM with more complex profiles \citep[e.g.][]{Montes1998,Soderberg2006,Salas2013,Corsi2014,Margutti2017,Balasubramanian2021,Maeda2021,Stroh2021}.

Finally, the total energy coupled to the fastest (radio emitting) ejecta can be expressed as \citep[e.g.,][]{Soderberg2006}:
\begin{equation}
E_r\approx \frac{4\pi R_p^3}{\eta}\frac{B_p^2}{8\pi\epsilon_B}=\frac{R_p^3}{\eta}\frac{B_p^2}{2\epsilon_B}.
\label{eq4}
\end{equation}
 In Table \ref{tab:radioproperties} we summarize the properties of the radio ejecta derived for the four SNe for which we detect a radio counterpart. These values can be compared with  $\dot{M}\approx2.5\times10^{-7}M_{\odot}{\rm yr}^{-1}$ and $E_r\approx (1-10)\times10^{49}$\,erg estimated for SN\,1998bw by \citet{Li1999}, with $\dot{M}\approx2\times10^{-6}M_{\odot}{\rm yr}^{-1}$ and $E_r\approx 1.3\times10^{49}$\,erg estimated for SN\,2009bb by \citet{Soderberg2010}, and with  with $\dot{M}\approx(0.4-1)\times10^{-5}M_{\odot}{\rm yr}^{-1}$ and $E_r\approx (0.3-4)\times10^{49}$\,erg estimated for GRB\,100316D by \citet{Margutti2013}. 

\begin{center}
\begin{table}
\caption{Properties of the radio ejecta of the SNe in our sample for which we detect a radio counterpart. We report the SN name, the estimated SN shock speed normalized to the speed of light ($\beta_s$), the mass-loss rate of the pre-SN progenitor ($\dot{M}$), the energy coupled to the fastest (radio-emitting) ejecta ($E_r$), and the ratio between the last and the kinetic energy of the explosion (estimated from the optical light curve modeling, $E_k$). See Section \ref{sec:radioanalysis} for discussion. \label{tab:radioproperties}}
\begin{tabular}{lcccc}
\hline
\hline
SN  & $\beta_s$ &  $\dot{M}$ & $E_r$ & $E_r/E_k$ \\
 & & (M$_{\odot}$\,yr$^{-1}$) & (erg) & \\
\hline
2018hom &  0.35 & $1.1\times10^{-6}$ & $3.6\times10^{47}$ & $<0.04$\%\\
2020jqm &  0.048 & $2.7\times10^{-4}$ & $5.7\times10^{48}$ & 0.1\% \\
2020tkx &  0.14 & $1.7\times10^{-5}$ & $1.1\times10^{48}$ & $<0.07$\% \\
2021ywf & 0.19 &  $2.2\times10^{-6}$ & $2.3\times10^{47}$ & 0.02\%\\
\hline
\end{tabular}
\end{table}
\end{center}

\begin{figure*}
    \centering
    \includegraphics[width=16cm]{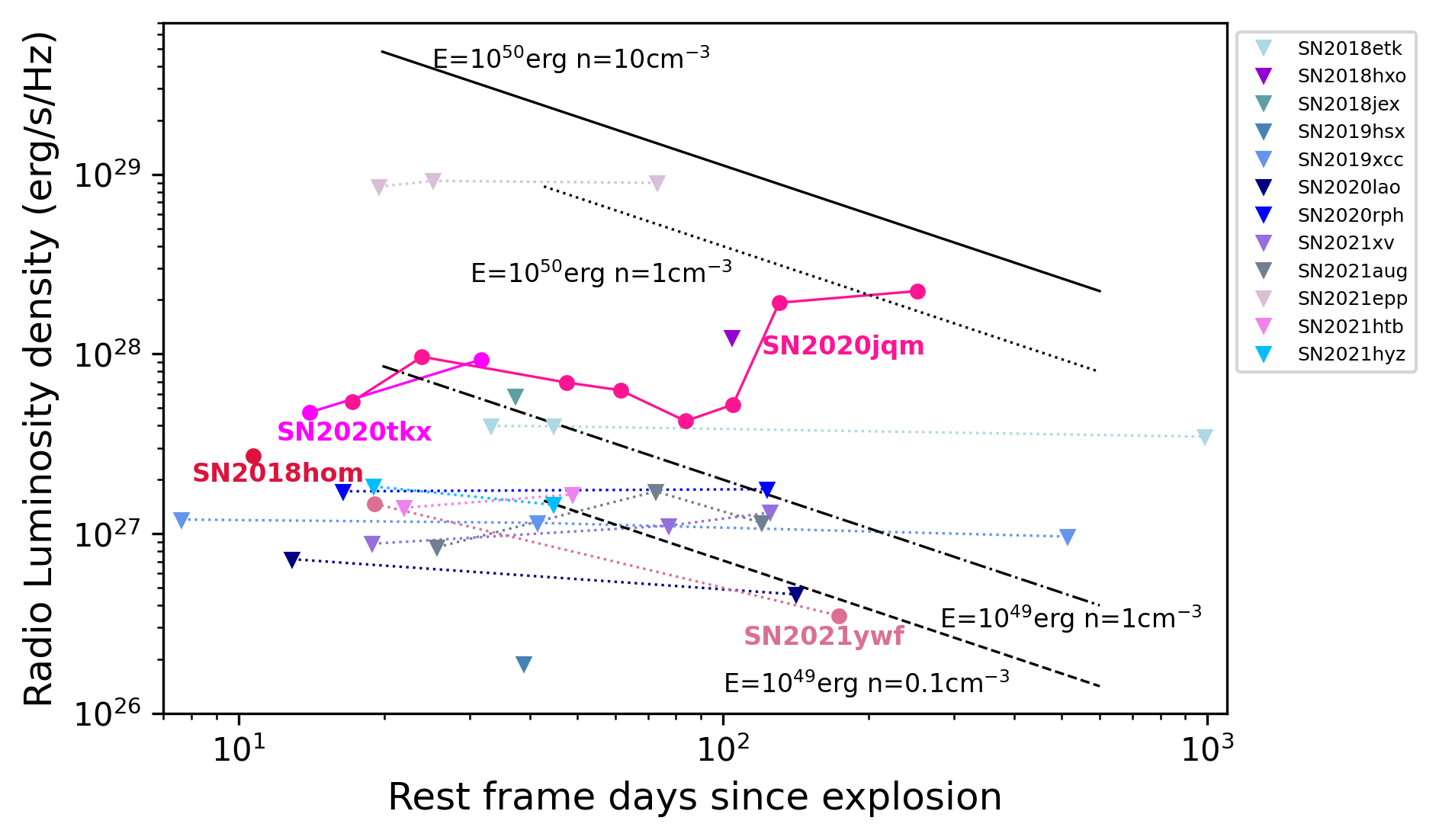}
    \caption{Approximate radio luminosity density for GRBs observed largely off-axis during the sub-relativistic phase (black solid, dotted, dashed, and dash-dotted lines) compared with the radio detections and upper-limits of the SNe Ic-BL in our sample. Most of our observations exclude fireballs with energies $E\gtrsim 10^{50}$\,erg expanding in ISM media with densities $\gtrsim 1$\,cm$^{-3}$. However, our observations become less constraining for smaller energy and ISM density values. For example, most of our radio data cannot exclude off-axis jets with energies $E\sim 10^{49}$\,erg and $n\sim 0.1$\,cm$^{-3}$. See Section \ref{sec:off-axis-radio} for discussion.}
    \label{fig:off-axis-radio}
\end{figure*}

\subsubsection{Off-axis GRB radio afterglow constraints}
\label{sec:off-axis-radio}
We finally consider what type of constraints our radio observations put on a scenario where the SNe Ic-BL in our sample could be accompanied by relativistic ejecta from 
a largely (close to 90\,deg) off-axis GRB afterglow that would become visible in the radio band when the relativistic fireball enters the sub-relativistic phase and approaches spherical symmetry.
Because our radio observations do not extend past 100-200 days since explosion, we can put only limited constraints on this scenario. Hence, hereafter we present some general order-of-magnitude considerations rather than a detailed event-by-event modeling. 

Following \citet{Corsi2016}, we can model approximately the late-time radio emission from an off-axis GRB based on the results by \citet{Livio2000}, \citet{Waxman2004b}, \citet{Zhang2009}, and \citet{VanEerten2012}. For fireballs expanding in an interstellar medium (ISM) of constant
density $n$ (in units of cm$^{-3}$), at timescales $t$ such that:
\begin{equation}
    t \gtrsim (1+z) \times t_{\rm SNT}/2
\end{equation}
where the transition time to the spherical Sedov–Neumann–Taylor (SNT) blast wave, $t_{\rm SNT}$,  reads:
\begin{equation}
    t_{\rm SNT}\approx 92\,{\rm d}\left(E_{51}/n\right)^{1/3},
\end{equation}
the luminosity density can be approximated analytically via the following formula \citep[see Equation (23) in][ where we neglect redshift corrections and assume $p=2$]{Zhang2009}:
\begin{eqnarray}
\nonumber L_{\nu}({\rm t}) \approx 4\pi d^{2}_{\rm L} F_{\nu}({\rm t}) \approx 2\times10^{30}\left(\frac{\epsilon_e}{0.1}\right)\left(\frac{\epsilon_B}{0.1}\right)^{3/4}n^{9/20}\\\times E^{13/10}_{51}\left(\frac{\nu}{\rm 1\,GHz}\right)^{-1/2}\left(\frac{t}{92\,{\rm d}}\right)^{-9/10}\,{\rm erg\,s^{-1}\,{\rm Hz}^{-1}}.~~~~~~~
\end{eqnarray}
In the above Equations, $E_{51}$ is the beaming-corrected ejecta energy in units of $10^{51}$\,erg. We note that here we assume again a constant density ISM in agreement with the majority of GRB afterglow observtions \citep[e.g.,][]{Schulze2011}. 

We plot the above luminosity in Figure \ref{fig:off-axis-radio} together with our radio observations and upper-limits, assuming $\epsilon_e=0.1$, $\epsilon_B=0.1$, and for representative values of low-luminosity GRB energies and typical values of long GRB ISM densities $n$. As evident from this Figure, our observations exclude fireballs with energies $E\gtrsim 10^{50}$\,erg expanding in ISM media with densities $\gtrsim 1$\,cm$^{-3}$. However, our observations become less constraining for smaller energy and ISM density values. 

\section{Summary and Conclusion}\label{sec:conclusion}
We have presented deep radio follow-up observations of 16 SNe Ic-BL that are part of the ZTF sample.  
Our campaign resulted in 4 radio counterpart detections and 12 deep radio upper-limits. For 9 of these 16 events we have also carried out X-ray observations with \textit{Swift}/XRT. All together, these results constrain the fraction of SN\,1998bw-like explosions to $< 19\%$ (3$\sigma$ Gaussian equivalent), tightening previous constraints by a factor of $\approx 2$. Moreover, our results exclude relativistic ejecta with radio luminosities densities in between $\approx 5\times10^{27}$\,erg\,s$^{-1}$\,Hz$^{-1}$ and  $\approx 10^{29}$\,erg\,s$^{-1}$\,Hz$^{-1}$ at $t\gtrsim 20$\,d since explosion for $\approx 60\%$ of the events in our sample, pointing to the fact that SNe Ic-BL similar to low-luminosity-GRB-SN such as SN\,1998bw, SN\,2003lw, SN\,2010dh, or to the relativistic SN\,2009bb and iPTF17cw, are intrinsically rare. This result is in line with numerical simulations that suggest that a SN Ic-BL can be triggered even if a jet
engine fails to produce a successful GRB jet.

We showed that our radio observations exclude an association of the SNe Ic-BL in our sample with largely off-axis GRB afterglows with energies $E\gtrsim 10^{50}$\,erg expanding in ISM media with densities $\gtrsim 1$\,cm$^{-3}$. On the other hand, our radio observations are less constraining for smaller energy and ISM density values, and cannot exclude off-axis jets with energies $E\sim 10^{49}$\,erg.

We noted that the main conclusion of our work is subject to the caveat that the parameter space of SN\,2006aj-like explosions (with faint radio emission peaking only a few days after explosion) is left largely unconstrained by current systematic radio follow-up efforts like the one presented here. In other words, we cannot exclude that a larger fraction of SNe Ic-BL harbors GRB\,060218/SN\,2006aj-like emission. In the future, obtaining fast and accurate spectral classification of SNe Ic-BL paired with deep radio follow-up observations executed within $5$\,d since explosion would overcome this limitation. While high-cadence optical surveys can provide an alternative way to measure the rate of SNe Ic-BL that are similar to SN\,2006aj via shock-cooling emission at early times, more optical detections are also needed to measure a robust rate. 

The Legacy Survey of Space and Time on the Vera C. Rubin Observatory \citep[LSST;][]{Rubin2019} promises to provide numerous discoveries of even the rarest type of explosive transients, such as the SNe Ic-BL discussed here. The challenge will be to recognize and classify these explosions promptly \citep[e.g.,][]{Villar2019,Villar2020}, so that they can be followed up in the radio with current and next generation radio facilities. Indeed, Rubin, paired with the increased sensitivity of the next generation VLA \citep[ngVLA;][]{ngVLA}, could provide a unique opportunity for building a large statistical sample of SNe Ic-BL with deep radio observations that may be used to guide theoretical modeling in a more systematic fashion, beyond what has been achievable over the last $\approx 25$ years (i.e., since the discovery of GRB-SN\,1998bw).  In addition, the Square Kilometer Array (SKA) will enable discoveries of radio SNe and other transients in an untargeted and optically-unbiased way \citep{Lien2011}. Hence, one can envision that the Rubin-LSST+ngVLA and SKA samples will, together, provide crucial information on
massive star evolution, as well as SNe Ic-BL physics and CSM properties. 

We conclude by noting that understanding the evolution of single and stripped binary stars up to core collapse is of special interest in the new era of time-domain multi-messenger (gravitational-wave and neutrino) astronomy \citep[see e.g., ][for recent reviews]{Murase2018,Scholber2012,Ernazar2020,Guepin2022}. Gravitational waves from nearby core-collapse SNe, in particular, represent an exciting prospect for expanding multi-messenger studies beyond the current realm of compact binary coalescences. While they may come into reach with the current LIGO \citep{Aasi2015} and Virgo \citep{Acernese2015} detectors, it is more likely that next generation gravitational-wave observatories, such as the Einstein Telescope \citep{Maggiore2020} and the Cosmic Explorer \citep{Evans2021}, will enable painting the first detailed multi-messenger picture of a core-collapse explosion. The physics behind massive stars' evolution and deaths also impacts the estimated rates and mass distribution of compact object mergers \citep[e.g.,][]{Schneider2021} which, in turn, are current primary sources for LIGO and Virgo, and will be detected in much large numbers by next generation gravitational-wave detectors. Hence, continued and coordinated efforts dedicated to understanding massive stars' deaths and the link between pre-SN progenitors and properties of SN explosions, using multiple messengers, undoubtedly represent an exciting path forward.

\begin{acknowledgements}
\small
A.C. and A.B. acknowledge support from NASA \textit{Swift} Guest investigator programs (Cycles 16 and 17 via grants \#80NSSC20K1482 and \#80NSSC22K0203). S.A. gratefully acknowledges support from the National Science Foundation GROWTH PIRE grant No. 1545949. S.Y. has been supported by the research project grant ``Understanding the Dynamic Universe'' funded by the Knut and Alice Wallenberg Foundation under Dnr KAW 2018.0067, and the G.R.E.A.T research environment, funded by {\em Vetenskapsr\aa det}, the Swedish Research Council, project number 2016-06012.
Based on observations obtained with the Samuel Oschin Telescope 48-inch and the 60-inch Telescope at the Palomar
Observatory as part of the Zwicky Transient Facility project. ZTF is supported by the National Science Foundation under Grants
No. AST-1440341 and No. AST-2034437, and a collaboration including Caltech, IPAC, the Weizmann Institute for Science, the Oskar Klein Center at
Stockholm University, the University of Maryland, Deutsches Elektronen-Synchrotron and Humboldt University, Los Alamos
National Laboratories, the TANGO
Consortium of Taiwan, the University of Wisconsin at Milwaukee, Trinity College Dublin, Lawrence Berkeley
National Laboratories, Lawrence Livermore National
Laboratories, and IN2P3, France. Operations are conducted by COO, IPAC, and UW. The SED Machine is based upon work supported by the National Science Foundation under Grant No. 1106171. The ZTF forced-photometry service was funded under the Heising-Simons Foundation grant \#12540303 (PI: Graham).
The National Radio Astronomy Observatory is a facility of the National Science Foundation operated under cooperative agreement by Associated Universities, Inc.
The Pan-STARRS1 Surveys (PS1) and the PS1 public science archive have been made possible through contributions by the Institute for Astronomy, the University of Hawaii, the Pan-STARRS Project Office, the Max-Planck Society and its participating institutes, the Max Planck Institute for Astronomy, Heidelberg and the Max Planck Institute for Extraterrestrial Physics, Garching, The Johns Hopkins University, Durham University, the University of Edinburgh, the Queen's University Belfast, the Harvard-Smithsonian Center for Astrophysics, the Las Cumbres Observatory Global Telescope Network Incorporated, the National Central University of Taiwan, the Space Telescope Science Institute, the National Aeronautics and Space Administration under Grant No. NNX08AR22G issued through the Planetary Science Division of the NASA Science Mission Directorate, the National Science Foundation Grant No. AST-1238877, the University of Maryland, Eotvos Lorand University (ELTE), the Los Alamos National Laboratory, and the Gordon and Betty Moore Foundation. Based in part on observations made with the Nordic Optical Telescope, owned in collaboration by the University of Turku and Aarhus University, and operated jointly by Aarhus University, the University of Turku and the University of Oslo, representing Denmark, Finland and Norway, the University of Iceland and Stockholm University at the Observatorio del Roque de los Muchachos, La Palma, Spain, of the Instituto de Astrofisica de Canarias.
\end{acknowledgements}

\bibliographystyle{aasjournal}
\bibliography{references}{}

\begin{thebibliography}{}
\expandafter\ifx\csname natexlab\endcsname\relax\def\natexlab#1{#1}\fi
\providecommand{\url}[1]{\href{#1}{#1}}
\providecommand{\dodoi}[1]{doi:~\href{http://doi.org/#1}{\nolinkurl{#1}}}
\providecommand{\doeprint}[1]{\href{http://ascl.net/#1}{\nolinkurl{http://ascl.net/#1}}}
\providecommand{\doarXiv}[1]{\href{https://arxiv.org/abs/#1}{\nolinkurl{https://arxiv.org/abs/#1}}}

\bibitem[{{Abdikamalov} {et~al.}(2020){Abdikamalov}, {Pagliaroli}, \&
  {Radice}}]{Ernazar2020}
{Abdikamalov}, E., {Pagliaroli}, G., \& {Radice}, D. 2020, arXiv e-prints,
  arXiv:2010.04356.
\newblock \doarXiv{2010.04356}

\bibitem[{{Acernese} {et~al.}(2015){Acernese}, {Agathos}, {Agatsuma}, {Aisa},
  {Allemandou}, {Allocca}, {Amarni}, {Astone}, {Balestri}, {Ballardin},
  {Barone}, {Baronick}, {Barsuglia}, {Basti}, {Basti}, {Bauer}, {Bavigadda},
  {Bejger}, {Beker}, {Belczynski}, {Bersanetti}, {Bertolini}, {Bitossi},
  {Bizouard}, {Bloemen}, {Blom}, {Boer}, {Bogaert}, {Bondi}, {Bondu},
  {Bonelli}, {Bonnand}, {Boschi}, {Bosi}, {Bouedo}, {Bradaschia}, {Branchesi},
  {Briant}, {Brillet}, {Brisson}, {Bulik}, {Bulten}, {Buskulic}, {Buy},
  {Cagnoli}, {Calloni}, {Campeggi}, {Canuel}, {Carbognani}, {Cavalier},
  {Cavalieri}, {Cella}, {Cesarini}, {Mottin}, {Chincarini}, {Chiummo}, {Chua},
  {Cleva}, {Coccia}, {Cohadon}, {Colla}, {Colombini}, {Conte}, {Coulon},
  {Cuoco}, {Dalmaz}, {D'Antonio}, {Dattilo}, {Davier}, {Day}, {Debreczeni},
  {Degallaix}, {Del{\'e}glise}, {Pozzo}, {Dereli}, {Rosa}, {Fiore}, {Lieto},
  {Virgilio}, {Doets}, {Dolique}, {Drago}, {Ducrot}, {Endr{\H{o}}czi},
  {Fafone}, {Farinon}, {Ferrante}, {Ferrini}, {Fidecaro}, {Fiori}, {Flaminio},
  {Fournier}, {Franco}, {Frasca}, {Frasconi}, {Gammaitoni}, {Garufi},
  {Gaspard}, {Gatto}, {Gemme}, {Gendre}, {Genin}, {Gennai}, {Ghosh},
  {Giacobone}, {Giazotto}, {Gouaty}, {Granata}, {Greco}, {Groot}, {Guidi},
  {Harms}, {Heidmann}, {Heitmann}, {Hello}, {Hemming}, {Hennes}, {Hofman},
  {Jaranowski}, {Jonker}, {Kasprzack}, {K{\'e}f{\'e}lian}, {Kowalska}, {Kraan},
  {Kr{\'o}lak}, {Kutynia}, {Lazzaro}, {Leonardi}, {Leroy}, {Letendre}, {Li},
  {Lieunard}, {Lorenzini}, {Loriette}, {Losurdo}, {Magazz{\`u}}, {Majorana},
  {Maksimovic}, {Malvezzi}, {Man}, {Mangano}, {Mantovani}, {Marchesoni},
  {Marion}, {Marque}, {Martelli}, {Martellini}, {Masserot}, {Meacher},
  {Meidam}, {Mezzani}, {Michel}, {Milano}, {Minenkov}, {Moggi}, {Mohan},
  {Montani}, {Morgado}, {Mours}, {Mul}, {Nagy}, {Nardecchia}, {Naticchioni},
  {Nelemans}, {Neri}, {Neri}, {Nocera}, {Pacaud}, {Palomba}, {Paoletti},
  {Paoli}, {Pasqualetti}, {Passaquieti}, {Passuello}, {Perciballi}, {Petit},
  {Pichot}, {Piergiovanni}, {Pillant}, {Piluso}, {Pinard}, {Poggiani},
  {Prijatelj}, {Prodi}, {Punturo}, {Puppo}, {Rabeling}, {R{\'a}cz},
  {Rapagnani}, {Razzano}, {Re}, {Regimbau}, {Ricci}, {Robinet}, {Rocchi},
  {Rolland}, {Romano}, {Rosi{\'n}ska}, {Ruggi}, {Saracco}, {Sassolas},
  {Schimmel}, {Sentenac}, {Sequino}, {Shah}, {Siellez}, {Straniero},
  {Swinkels}, {Tacca}, {Tonelli}, {Travasso}, {Turconi}, {Vajente}, {van
  Bakel}, {van Beuzekom}, {van den Brand}, {Van Den Broeck}, {van der Sluys},
  {van Heijningen}, {Vas{\'u}th}, {Vedovato}, {Veitch}, {Verkindt}, {Vetrano},
  {Vicer{\'e}}, {Vinet}, {Visser}, {Vocca}, {Ward}, {Was}, {Wei}, {Yvert},
  {{\.z}ny}, \& {Zendri}}]{Acernese2015}
{Acernese}, F., {Agathos}, M., {Agatsuma}, K., {et~al.} 2015, Classical and
  Quantum Gravity, 32, 024001, \dodoi{10.1088/0264-9381/32/2/024001}

\bibitem[{{Afsariardchi} {et~al.}(2021){Afsariardchi}, {Drout}, {Khatami},
  {Matzner}, {Moon}, \& {Ni}}]{Afsariardchi2021}
{Afsariardchi}, N., {Drout}, M.~R., {Khatami}, D.~K., {et~al.} 2021, \apj, 918,
  89, \dodoi{10.3847/1538-4357/ac0aeb}

\bibitem[{{Anand} {et~al.}(2022)}]{Anand2022}
{Anand}, S., {et~al.} 2022, in preparation

\bibitem[{{Armstrong}(2002)}]{Armstrong2002}
{Armstrong}, M. 2002, \iaucirc, 7845, 1

\bibitem[{{Arnett}(1982)}]{Arnett1982}
{Arnett}, W.~D. 1982, \apj, 253, 785, \dodoi{10.1086/159681}

\bibitem[{{Balasubramanian} {et~al.}(2021){Balasubramanian}, {Corsi},
  {Polisensky}, {Clarke}, \& {Kassim}}]{Balasubramanian2021}
{Balasubramanian}, A., {Corsi}, A., {Polisensky}, E., {Clarke}, T.~E., \&
  {Kassim}, N.~E. 2021, \apj, 923, 32, \dodoi{10.3847/1538-4357/ac2154}

\bibitem[{{Barnes} {et~al.}(2018){Barnes}, {Duffell}, {Liu}, {Modjaz},
  {Bianco}, {Kasen}, \& {MacFadyen}}]{Barnes2018}
{Barnes}, J., {Duffell}, P.~C., {Liu}, Y., {et~al.} 2018, \apj, 860, 38,
  \dodoi{10.3847/1538-4357/aabf84}

\bibitem[{{Barthelmy} {et~al.}(2005){Barthelmy}, {Barbier}, {Cummings},
  {Fenimore}, {Gehrels}, {Hullinger}, {Krimm}, {Markwardt}, {Palmer},
  {Parsons}, {Sato}, {Suzuki}, {Takahashi}, {Tashiro}, \&
  {Tueller}}]{Barthelmy2005}
{Barthelmy}, S.~D., {Barbier}, L.~M., {Cummings}, J.~R., {et~al.} 2005, \ssr,
  120, 143, \dodoi{10.1007/s11214-005-5096-3}

\bibitem[{{Bazin} {et~al.}(2009){Bazin}, {Palanque-Delabrouille}, {Rich},
  {Ruhlmann-Kleider}, {Aubourg}, {Le Guillou}, {Astier}, {Balland}, {Basa},
  {Carlberg}, {Conley}, {Fouchez}, {Guy}, {Hardin}, {Hook}, {Howell}, {Pain},
  {Perrett}, {Pritchet}, {Regnault}, {Sullivan}, {Antilogus}, {Arsenijevic},
  {Baumont}, {Fabbro}, {Le Du}, {Lidman}, {Mouchet}, {Mour{\~a}o}, \&
  {Walker}}]{Bazin2009}
{Bazin}, G., {Palanque-Delabrouille}, N., {Rich}, J., {et~al.} 2009, \aap, 499,
  653, \dodoi{10.1051/0004-6361/200911847}

\bibitem[{{Becker} {et~al.}(1995){Becker}, {White}, \& {Helfand}}]{Becker1995}
{Becker}, R.~H., {White}, R.~L., \& {Helfand}, D.~J. 1995, \apj, 450, 559,
  \dodoi{10.1086/176166}

\bibitem[{{Bellm} {et~al.}(2019){Bellm}, {Kulkarni}, {Graham}, {Dekany},
  {Smith}, {Riddle}, {Masci}, {Helou}, {Prince}, {Adams}, {Barbarino},
  {Barlow}, {Bauer}, {Beck}, {Belicki}, {Biswas}, {Blagorodnova}, {Bodewits},
  {Bolin}, {Brinnel}, {Brooke}, {Bue}, {Bulla}, {Burruss}, {Cenko}, {Chang},
  {Connolly}, {Coughlin}, {Cromer}, {Cunningham}, {De}, {Delacroix}, {Desai},
  {Duev}, {Eadie}, {Farnham}, {Feeney}, {Feindt}, {Flynn}, {Franckowiak},
  {Frederick}, {Fremling}, {Gal-Yam}, {Gezari}, {Giomi}, {Goldstein},
  {Golkhou}, {Goobar}, {Groom}, {Hacopians}, {Hale}, {Henning}, {Ho}, {Hover},
  {Howell}, {Hung}, {Huppenkothen}, {Imel}, {Ip}, {Ivezi{\'c}}, {Jackson},
  {Jones}, {Juric}, {Kasliwal}, {Kaspi}, {Kaye}, {Kelley}, {Kowalski},
  {Kramer}, {Kupfer}, {Landry}, {Laher}, {Lee}, {Lin}, {Lin}, {Lunnan},
  {Giomi}, {Mahabal}, {Mao}, {Miller}, {Monkewitz}, {Murphy}, {Ngeow},
  {Nordin}, {Nugent}, {Ofek}, {Patterson}, {Penprase}, {Porter}, {Rauch},
  {Rebbapragada}, {Reiley}, {Rigault}, {Rodriguez}, {van Roestel}, {Rusholme},
  {van Santen}, {Schulze}, {Shupe}, {Singer}, {Soumagnac}, {Stein}, {Surace},
  {Sollerman}, {Szkody}, {Taddia}, {Terek}, {Van Sistine}, {van Velzen},
  {Vestrand}, {Walters}, {Ward}, {Ye}, {Yu}, {Yan}, \& {Zolkower}}]{Bellm2019}
{Bellm}, E.~C., {Kulkarni}, S.~R., {Graham}, M.~J., {et~al.} 2019, \pasp, 131,
  018002, \dodoi{10.1088/1538-3873/aaecbe}

\bibitem[{{Bennett} {et~al.}(2014){Bennett}, {Larson}, {Weiland}, \&
  {Hinshaw}}]{Bennett2014}
{Bennett}, C.~L., {Larson}, D., {Weiland}, J.~L., \& {Hinshaw}, G. 2014, \apj,
  794, 135, \dodoi{10.1088/0004-637X/794/2/135}

\bibitem[{{Berger} {et~al.}(2003){Berger}, {Kulkarni}, {Frail}, \&
  {Soderberg}}]{Berger2003}
{Berger}, E., {Kulkarni}, S.~R., {Frail}, D.~A., \& {Soderberg}, A.~M. 2003,
  \apj, 599, 408, \dodoi{10.1086/379214}

\bibitem[{{Blagorodnova} {et~al.}(2018){Blagorodnova}, {Neill}, {Walters},
  {Kulkarni}, {Fremling}, {Ben-Ami}, {Dekany}, {Fucik}, {Konidaris}, {Nash},
  {Ngeow}, {Ofek}, {O' Sullivan}, {Quimby}, {Ritter}, \&
  {Vyhmeister}}]{Blagorodnova2018}
{Blagorodnova}, N., {Neill}, J.~D., {Walters}, R., {et~al.} 2018, \pasp, 130,
  035003, \dodoi{10.1088/1538-3873/aaa53f}

\bibitem[{{Blondin} \& {Tonry}(2007)}]{Blondin2007}
{Blondin}, S., \& {Tonry}, J.~L. 2007, \apj, 666, 1024, \dodoi{10.1086/520494}

\bibitem[{{Bromberg} {et~al.}(2011){Bromberg}, {Nakar}, \&
  {Piran}}]{Bromberg2011}
{Bromberg}, O., {Nakar}, E., \& {Piran}, T. 2011, \apjl, 739, L55,
  \dodoi{10.1088/2041-8205/739/2/L55}

\bibitem[{{Bugli} {et~al.}(2021){Bugli}, {Guilet}, \&
  {Obergaulinger}}]{Bugli2021}
{Bugli}, M., {Guilet}, J., \& {Obergaulinger}, M. 2021, \mnras, 507, 443,
  \dodoi{10.1093/mnras/stab2161}

\bibitem[{{Bugli} {et~al.}(2020){Bugli}, {Guilet}, {Obergaulinger},
  {Cerd{\'a}-Dur{\'a}n}, \& {Aloy}}]{Bugli2020}
{Bugli}, M., {Guilet}, J., {Obergaulinger}, M., {Cerd{\'a}-Dur{\'a}n}, P., \&
  {Aloy}, M.~A. 2020, \mnras, 492, 58, \dodoi{10.1093/mnras/stz3483}

\bibitem[{{Burrows} {et~al.}(2007){Burrows}, {Dessart}, {Livne}, {Ott}, \&
  {Murphy}}]{Burrows2007}
{Burrows}, A., {Dessart}, L., {Livne}, E., {Ott}, C.~D., \& {Murphy}, J. 2007,
  \apj, 664, 416, \dodoi{10.1086/519161}

\bibitem[{{Burrows} {et~al.}(2005){Burrows}, {Hill}, {Nousek}, {Kennea},
  {Wells}, {Osborne}, {Abbey}, {Beardmore}, {Mukerjee}, {Short}, {Chincarini},
  {Campana}, {Citterio}, {Moretti}, {Pagani}, {Tagliaferri}, {Giommi},
  {Capalbi}, {Tamburelli}, {Angelini}, {Cusumano}, {Br{\"a}uninger}, {Burkert},
  \& {Hartner}}]{Burrows+2005}
{Burrows}, D.~N., {Hill}, J.~E., {Nousek}, J.~A., {et~al.} 2005, \ssr, 120,
  165, \dodoi{10.1007/s11214-005-5097-2}

\bibitem[{{Campana} {et~al.}(2006){Campana}, {Mangano}, {Blustin}, {Brown},
  {Burrows}, {Chincarini}, {Cummings}, {Cusumano}, {Della Valle}, {Malesani},
  {M{\'e}sz{\'a}ros}, {Nousek}, {Page}, {Sakamoto}, {Waxman}, {Zhang}, {Dai},
  {Gehrels}, {Immler}, {Marshall}, {Mason}, {Moretti}, {O'Brien}, {Osborne},
  {Page}, {Romano}, {Roming}, {Tagliaferri}, {Cominsky}, {Giommi}, {Godet},
  {Kennea}, {Krimm}, {Angelini}, {Barthelmy}, {Boyd}, {Palmer}, {Wells}, \&
  {White}}]{Campana2006}
{Campana}, S., {Mangano}, V., {Blustin}, A.~J., {et~al.} 2006, \nat, 442, 1008,
  \dodoi{10.1038/nature04892}

\bibitem[{{Cano} {et~al.}(2017){Cano}, {Wang}, {Dai}, \& {Wu}}]{Cano2017}
{Cano}, Z., {Wang}, S.-Q., {Dai}, Z.-G., \& {Wu}, X.-F. 2017, Advances in
  Astronomy, 2017, 8929054, \dodoi{10.1155/2017/8929054}

\bibitem[{{Cardelli} {et~al.}(1989){Cardelli}, {Clayton}, \&
  {Mathis}}]{Cardelli1989}
{Cardelli}, J.~A., {Clayton}, G.~C., \& {Mathis}, J.~S. 1989, \apj, 345, 245,
  \dodoi{10.1086/167900}

\bibitem[{{Chen} {et~al.}(2017){Chen}, {Moriya}, {Woosley}, {Sukhbold},
  {Whalen}, {Suwa}, \& {Bromm}}]{Chen2017}
{Chen}, K.-J., {Moriya}, T.~J., {Woosley}, S., {et~al.} 2017, \apj, 839, 85,
  \dodoi{10.3847/1538-4357/aa68a4}

\bibitem[{{Chevalier}(1998)}]{Chevalier1998}
{Chevalier}, R.~A. 1998, \apj, 499, 810, \dodoi{10.1086/305676}

\bibitem[{{Chevalier} \& {Fransson}(2006)}]{Chevalier2006}
{Chevalier}, R.~A., \& {Fransson}, C. 2006, \apj, 651, 381,
  \dodoi{10.1086/507606}

\bibitem[{{Corsi} {et~al.}(2014){Corsi}, {Ofek}, {Gal-Yam}, {Frail},
  {Kulkarni}, {Fox}, {Kasliwal}, {Sullivan}, {Horesh}, {Carpenter}, {Maguire},
  {Arcavi}, {Cenko}, {Cao}, {Mooley}, {Pan}, {Sesar}, {Sternberg}, {Xu},
  {Bersier}, {James}, {Bloom}, \& {Nugent}}]{Corsi2014}
{Corsi}, A., {Ofek}, E.~O., {Gal-Yam}, A., {et~al.} 2014, \apj, 782, 42,
  \dodoi{10.1088/0004-637X/782/1/42}

\bibitem[{{Corsi} {et~al.}(2016){Corsi}, {Gal-Yam}, {Kulkarni}, {Frail},
  {Mazzali}, {Cenko}, {Kasliwal}, {Cao}, {Horesh}, {Palliyaguru}, {Perley},
  {Laher}, {Taddia}, {Leloudas}, {Maguire}, {Nugent}, {Sollerman}, \&
  {Sullivan}}]{Corsi2016}
{Corsi}, A., {Gal-Yam}, A., {Kulkarni}, S.~R., {et~al.} 2016, \apj, 830, 42,
  \dodoi{10.3847/0004-637X/830/1/42}

\bibitem[{{Corsi} {et~al.}(2017){Corsi}, {Cenko}, {Kasliwal}, {Quimby},
  {Kulkarni}, {Frail}, {Goldstein}, {Blagorodnova}, {Connaughton}, {Perley},
  {Singer}, {Copperwheat}, {Fremling}, {Kupfer}, {Piascik}, {Steele}, {Taddia},
  {Vedantham}, {Kutyrev}, {Palliyaguru}, {Roberts}, {Sollerman}, {Troja}, \&
  {Veilleux}}]{Corsi2017}
{Corsi}, A., {Cenko}, S.~B., {Kasliwal}, M.~M., {et~al.} 2017, \apj, 847, 54,
  \dodoi{10.3847/1538-4357/aa85e5}

\bibitem[{{Dahiwale} \&
  {Fremling}(2020{\natexlab{a}})}]{Dahiwale2020_ZTF18acaimrb}
{Dahiwale}, A., \& {Fremling}, C. 2020{\natexlab{a}}, Transient Name Server
  Classification Report, 2020-1924, 1

\bibitem[{{Dahiwale} \&
  {Fremling}(2020{\natexlab{b}})}]{Dahiwale2020_ZTF20aazkjfv}
---. 2020{\natexlab{b}}, Transient Name Server Classification Report,
  2020-1592, 1

\bibitem[{{Dahiwale} \& {Fremling}(2021)}]{Dahiwale2021_ZTF21aafnunh}
---. 2021, Transient Name Server Classification Report, 2021-405, 1

\bibitem[{{Dark Energy Survey Collaboration} {et~al.}(2016){Dark Energy Survey
  Collaboration}, {Abbott}, {Abdalla}, {Aleksi{\'c}}, {Allam}, {Amara},
  {Bacon}, {Balbinot}, {Banerji}, {Bechtol}, {Benoit-L{\'e}vy}, {Bernstein},
  {Bertin}, {Blazek}, {Bonnett}, {Bridle}, {Brooks}, {Brunner}, {Buckley-Geer},
  {Burke}, {Caminha}, {Capozzi}, {Carlsen}, {Carnero-Rosell}, {Carollo},
  {Carrasco-Kind}, {Carretero}, {Castander}, {Clerkin}, {Collett}, {Conselice},
  {Crocce}, {Cunha}, {D'Andrea}, {da Costa}, {Davis}, {Desai}, {Diehl},
  {Dietrich}, {Dodelson}, {Doel}, {Drlica-Wagner}, {Estrada}, {Etherington},
  {Evrard}, {Fabbri}, {Finley}, {Flaugher}, {Foley}, {Fosalba}, {Frieman},
  {Garc{\'\i}a-Bellido}, {Gaztanaga}, {Gerdes}, {Giannantonio}, {Goldstein},
  {Gruen}, {Gruendl}, {Guarnieri}, {Gutierrez}, {Hartley}, {Honscheid}, {Jain},
  {James}, {Jeltema}, {Jouvel}, {Kessler}, {King}, {Kirk}, {Kron}, {Kuehn},
  {Kuropatkin}, {Lahav}, {Li}, {Lima}, {Lin}, {Maia}, {Makler}, {Manera},
  {Maraston}, {Marshall}, {Martini}, {McMahon}, {Melchior}, {Merson}, {Miller},
  {Miquel}, {Mohr}, {Morice-Atkinson}, {Naidoo}, {Neilsen}, {Nichol}, {Nord},
  {Ogando}, {Ostrovski}, {Palmese}, {Papadopoulos}, {Peiris}, {Peoples},
  {Percival}, {Plazas}, {Reed}, {Refregier}, {Romer}, {Roodman}, {Ross},
  {Rozo}, {Rykoff}, {Sadeh}, {Sako}, {S{\'a}nchez}, {Sanchez}, {Santiago},
  {Scarpine}, {Schubnell}, {Sevilla-Noarbe}, {Sheldon}, {Smith}, {Smith},
  {Soares-Santos}, {Sobreira}, {Soumagnac}, {Suchyta}, {Sullivan}, {Swanson},
  {Tarle}, {Thaler}, {Thomas}, {Thomas}, {Tucker}, {Vieira}, {Vikram},
  {Walker}, {Wechsler}, {Weller}, {Wester}, {Whiteway}, {Wilcox}, {Yanny},
  {Zhang}, \& {Zuntz}}]{Abbott2016}
{Dark Energy Survey Collaboration}, {Abbott}, T., {Abdalla}, F.~B., {et~al.}
  2016, \mnras, 460, 1270, \dodoi{10.1093/mnras/stw641}

\bibitem[{{Dekany} {et~al.}(2020){Dekany}, {Smith}, {Riddle}, {Feeney},
  {Porter}, {Hale}, {Zolkower}, {Belicki}, {Kaye}, {Henning}, {Walters},
  {Cromer}, {Delacroix}, {Rodriguez}, {Reiley}, {Mao}, {Hover}, {Murphy},
  {Burruss}, {Baker}, {Kowalski}, {Reif}, {Mueller}, {Bellm}, {Graham}, \&
  {Kulkarni}}]{dekany2020}
{Dekany}, R., {Smith}, R.~M., {Riddle}, R., {et~al.} 2020, \pasp, 132, 038001,
  \dodoi{10.1088/1538-3873/ab4ca2}

\bibitem[{{Djupvik} \& {Andersen}(2010)}]{Djupvik2010}
{Djupvik}, A.~A., \& {Andersen}, J. 2010, in Astrophysics and Space Science
  Proceedings, Vol.~14, Highlights of Spanish Astrophysics V, 211,
  \dodoi{10.1007/978-3-642-11250-8\_21}

\bibitem[{{Drake} {et~al.}(2009){Drake}, {Djorgovski}, {Mahabal}, {Beshore},
  {Larson}, {Graham}, {Williams}, {Christensen}, {Catelan}, {Boattini},
  {Gibbs}, {Hill}, \& {Kowalski}}]{Drake2009}
{Drake}, A.~J., {Djorgovski}, S.~G., {Mahabal}, A., {et~al.} 2009, \apj, 696,
  870, \dodoi{10.1088/0004-637X/696/1/870}

\bibitem[{{Eisenberg} {et~al.}(2022){Eisenberg}, {Gottlieb}, \&
  {Nakar}}]{Eisenberg2022}
{Eisenberg}, M., {Gottlieb}, O., \& {Nakar}, E. 2022, \mnras,
  \dodoi{10.1093/mnras/stac2184}

\bibitem[{{Ellison} {et~al.}(2000){Ellison}, {Berezhko}, \&
  {Baring}}]{Ellison2000}
{Ellison}, D.~C., {Berezhko}, E.~G., \& {Baring}, M.~G. 2000, \apj, 540, 292,
  \dodoi{10.1086/309324}

\bibitem[{{Evans} {et~al.}(2021){Evans}, {Adhikari}, {Afle}, {Ballmer},
  {Biscoveanu}, {Borhanian}, {Brown}, {Chen}, {Eisenstein}, {Gruson}, {Gupta},
  {Hall}, {Huxford}, {Kamai}, {Kashyap}, {Kissel}, {Kuns}, {Landry}, {Lenon},
  {Lovelace}, {McCuller}, {Ng}, {Nitz}, {Read}, {Sathyaprakash}, {Shoemaker},
  {Slagmolen}, {Smith}, {Srivastava}, {Sun}, {Vitale}, \& {Weiss}}]{Evans2021}
{Evans}, M., {Adhikari}, R.~X., {Afle}, C., {et~al.} 2021, arXiv e-prints,
  arXiv:2109.09882.
\newblock \doarXiv{2109.09882}

\bibitem[{{Evans} {et~al.}(2009){Evans}, {Beardmore}, {Page}, {Osborne},
  {O'Brien}, {Willingale}, {Starling}, {Burrows}, {Godet}, {Vetere}, {Racusin},
  {Goad}, {Wiersema}, {Angelini}, {Capalbi}, {Chincarini}, {Gehrels}, {Kennea},
  {Margutti}, {Morris}, {Mountford}, {Pagani}, {Perri}, {Romano}, \&
  {Tanvir}}]{Evans+2009}
{Evans}, P.~A., {Beardmore}, A.~P., {Page}, K.~L., {et~al.} 2009, \mnras, 397,
  1177, \dodoi{10.1111/j.1365-2966.2009.14913.x}

\bibitem[{{Filippenko}(1997)}]{Filippenko1997}
{Filippenko}, A.~V. 1997, \araa, 35, 309,
  \dodoi{10.1146/annurev.astro.35.1.309}

\bibitem[{{Flewelling} {et~al.}(2020){Flewelling}, {Magnier}, {Chambers},
  {Heasley}, {Holmberg}, {Huber}, {Sweeney}, {Waters}, {Calamida}, {Casertano},
  {Chen}, {Farrow}, {Hasinger}, {Henderson}, {Long}, {Metcalfe}, {Narayan},
  {Nieto-Santisteban}, {Norberg}, {Rest}, {Saglia}, {Szalay}, {Thakar},
  {Tonry}, {Valenti}, {Werner}, {White}, {Denneau}, {Draper}, {Hodapp},
  {Jedicke}, {Kaiser}, {Kudritzki}, {Price}, {Wainscoat}, {Chastel}, {McLean},
  {Postman}, \& {Shiao}}]{Flewelling2020}
{Flewelling}, H.~A., {Magnier}, E.~A., {Chambers}, K.~C., {et~al.} 2020, \apjs,
  251, 7, \dodoi{10.3847/1538-4365/abb82d}

\bibitem[{{Foglizzo} {et~al.}(2015){Foglizzo}, {Kazeroni}, {Guilet}, {Masset},
  {Gonz{\'a}lez}, {Krueger}, {Novak}, {Oertel}, {Margueron}, {Faure}, {Martin},
  {Blottiau}, {Peres}, \& {Durand}}]{Foglizzo2015}
{Foglizzo}, T., {Kazeroni}, R., {Guilet}, J., {et~al.} 2015, \pasa, 32, e009,
  \dodoi{10.1017/pasa.2015.9}

\bibitem[{{Forster} {et~al.}(2020){Forster}, {Bauer}, {Galbany}, {Pignata},
  {Camacho}, {Silva-Farfan}, {Arredondo}, {Cabrera-Vives}, {Carrasco-Davis},
  {Estevez}, {Huijse}, {Reyes}, {Reyes}, {Sanchez-Saez}, {Valenzuela},
  {Castillo}, {Ruz-Mieres}, {Rodriguez-Mancini}, {Bauer}, {Catelan},
  {Eyheramendy}, \& {Graham}}]{Forster2020_ZTF20aazkjfv}
{Forster}, F., {Bauer}, F.~E., {Galbany}, L., {et~al.} 2020, Transient Name
  Server Discovery Report, 2020-1297, 1

\bibitem[{{Forster} {et~al.}(2021){Forster}, {Bauer}, {Pignata},
  {Munoz-Arancibia}, {Hernandez-Garcia}, {Galbany}, {Camacho}, {Silva-Farfan},
  {Mourao}, {Arredondo}, {Cabrera-Vives}, {Carrasco-Davis}, {Estevez},
  {Huijse}, {Reyes}, {Reyes}, {Sanchez-Saez}, {Valenzuela}, {Castillo},
  {Ruz-Mieres}, {Rodriguez-Mancini}, {Catelan}, {Eyheramendy}, \&
  {Graham}}]{Forster_2021}
{Forster}, F., {Bauer}, F.~E., {Pignata}, G., {et~al.} 2021, Transient Name
  Server Discovery Report, 2021-1011, 1

\bibitem[{{Fremling} {et~al.}(2018{\natexlab{a}}){Fremling}, {Dugas}, \&
  {Sharma}}]{Fremling2018_ZTF18abklarx}
{Fremling}, C., {Dugas}, A., \& {Sharma}, Y. 2018{\natexlab{a}}, Transient Name
  Server Classification Report, 2018-1169, 1

\bibitem[{{Fremling} {et~al.}(2018{\natexlab{b}}){Fremling}, {Dugas}, \&
  {Sharma}}]{Fremling2018_ZTF18acbwxcc}
---. 2018{\natexlab{b}}, Transient Name Server Classification Report,
  2018-1720, 1

\bibitem[{{Frohmaier} {et~al.}(2021){Frohmaier}, {Angus}, {Vincenzi},
  {Sullivan}, {Smith}, {Nugent}, {Cenko}, {Gal-Yam}, {Kulkarni}, {Law}, \&
  {Quimby}}]{Frohmaier2021}
{Frohmaier}, C., {Angus}, C.~R., {Vincenzi}, M., {et~al.} 2021, \mnras, 500,
  5142, \dodoi{10.1093/mnras/staa3607}

\bibitem[{{Gal-Yam}(2017)}]{GalYam2017}
{Gal-Yam}, A. 2017, in Handbook of Supernovae, ed. A.~W. {Alsabti} \&
  P.~{Murdin}, 195, \dodoi{10.1007/978-3-319-21846-5\_35}

\bibitem[{{Gal-Yam} {et~al.}(2002){Gal-Yam}, {Ofek}, \& {Shemmer}}]{GalYam2002}
{Gal-Yam}, A., {Ofek}, E.~O., \& {Shemmer}, O. 2002, \mnras, 332, L73,
  \dodoi{10.1046/j.1365-8711.2002.05535.x}

\bibitem[{{Galama} {et~al.}(1998){Galama}, {Vreeswijk}, {van Paradijs},
  {Kouveliotou}, {Augusteijn}, {B{\"o}hnhardt}, {Brewer}, {Doublier},
  {Gonzalez}, {Leibundgut}, {Lidman}, {Hainaut}, {Patat}, {Heise}, {in't Zand},
  {Hurley}, {Groot}, {Strom}, {Mazzali}, {Iwamoto}, {Nomoto}, {Umeda},
  {Nakamura}, {Young}, {Suzuki}, {Shigeyama}, {Koshut}, {Kippen}, {Robinson},
  {de Wildt}, {Wijers}, {Tanvir}, {Greiner}, {Pian}, {Palazzi}, {Frontera},
  {Masetti}, {Nicastro}, {Feroci}, {Costa}, {Piro}, {Peterson}, {Tinney},
  {Boyle}, {Cannon}, {Stathakis}, {Sadler}, {Begam}, \& {Ianna}}]{Galama1998}
{Galama}, T.~J., {Vreeswijk}, P.~M., {van Paradijs}, J., {et~al.} 1998, \nat,
  395, 670, \dodoi{10.1038/27150}

\bibitem[{{Gehrels} {et~al.}(2004){Gehrels}, {Chincarini}, {Giommi}, {Mason},
  {Nousek}, {Wells}, {White}, {Barthelmy}, {Burrows}, {Cominsky}, {Hurley},
  {Marshall}, {M{\'e}sz{\'a}ros}, {Roming}, {Angelini}, {Barbier}, {Belloni},
  {Campana}, {Caraveo}, {Chester}, {Citterio}, {Cline}, {Cropper}, {Cummings},
  {Dean}, {Feigelson}, {Fenimore}, {Frail}, {Fruchter}, {Garmire}, {Gendreau},
  {Ghisellini}, {Greiner}, {Hill}, {Hunsberger}, {Krimm}, {Kulkarni}, {Kumar},
  {Lebrun}, {Lloyd-Ronning}, {Markwardt}, {Mattson}, {Mushotzky}, {Norris},
  {Osborne}, {Paczynski}, {Palmer}, {Park}, {Parsons}, {Paul}, {Rees},
  {Reynolds}, {Rhoads}, {Sasseen}, {Schaefer}, {Short}, {Smale}, {Smith},
  {Stella}, {Tagliaferri}, {Takahashi}, {Tashiro}, {Townsley}, {Tueller},
  {Turner}, {Vietri}, {Voges}, {Ward}, {Willingale}, {Zerbi}, \&
  {Zhang}}]{Gehrels+2004}
{Gehrels}, N., {Chincarini}, G., {Giommi}, P., {et~al.} 2004, \apj, 611, 1005,
  \dodoi{10.1086/422091}

\bibitem[{{Ghirlanda} \& {Salvaterra}(2022)}]{Ghirlanda2022}
{Ghirlanda}, G., \& {Salvaterra}, R. 2022, \apj, 932, 10,
  \dodoi{10.3847/1538-4357/ac6e43}

\bibitem[{{Gilkis} \& {Soker}(2014)}]{Gilkis2014}
{Gilkis}, A., \& {Soker}, N. 2014, \mnras, 439, 4011,
  \dodoi{10.1093/mnras/stu257}

\bibitem[{{Gilkis} {et~al.}(2016){Gilkis}, {Soker}, \& {Papish}}]{Gilkis2016}
{Gilkis}, A., {Soker}, N., \& {Papish}, O. 2016, \apj, 826, 178,
  \dodoi{10.3847/0004-637X/826/2/178}

\bibitem[{{Gottlieb} {et~al.}(2022){Gottlieb}, {Lalakos}, {Bromberg}, {Liska},
  \& {Tchekhovskoy}}]{Gottlieb2022}
{Gottlieb}, O., {Lalakos}, A., {Bromberg}, O., {Liska}, M., \& {Tchekhovskoy},
  A. 2022, \mnras, 510, 4962, \dodoi{10.1093/mnras/stab3784}

\bibitem[{{Graham} {et~al.}(2019){Graham}, {Kulkarni}, {Bellm}, {Adams},
  {Barbarino}, {Blagorodnova}, {Bodewits}, {Bolin}, {Brady}, {Cenko}, {Chang},
  {Coughlin}, {De}, {Eadie}, {Farnham}, {Feindt}, {Franckowiak}, {Fremling},
  {Gezari}, {Ghosh}, {Goldstein}, {Golkhou}, {Goobar}, {Ho}, {Huppenkothen},
  {Ivezi{\'c}}, {Jones}, {Juric}, {Kaplan}, {Kasliwal}, {Kelley}, {Kupfer},
  {Lee}, {Lin}, {Lunnan}, {Mahabal}, {Miller}, {Ngeow}, {Nugent}, {Ofek},
  {Prince}, {Rauch}, {van Roestel}, {Schulze}, {Singer}, {Sollerman}, {Taddia},
  {Yan}, {Ye}, {Yu}, {Barlow}, {Bauer}, {Beck}, {Belicki}, {Biswas}, {Brinnel},
  {Brooke}, {Bue}, {Bulla}, {Burruss}, {Connolly}, {Cromer}, {Cunningham},
  {Dekany}, {Delacroix}, {Desai}, {Duev}, {Feeney}, {Flynn}, {Frederick},
  {Gal-Yam}, {Giomi}, {Groom}, {Hacopians}, {Hale}, {Helou}, {Henning},
  {Hover}, {Hillenbrand}, {Howell}, {Hung}, {Imel}, {Ip}, {Jackson}, {Kaspi},
  {Kaye}, {Kowalski}, {Kramer}, {Kuhn}, {Landry}, {Laher}, {Mao}, {Masci},
  {Monkewitz}, {Murphy}, {Nordin}, {Patterson}, {Penprase}, {Porter},
  {Rebbapragada}, {Reiley}, {Riddle}, {Rigault}, {Rodriguez}, {Rusholme}, {van
  Santen}, {Shupe}, {Smith}, {Soumagnac}, {Stein}, {Surace}, {Szkody}, {Terek},
  {Van Sistine}, {van Velzen}, {Vestrand}, {Walters}, {Ward}, {Zhang}, \&
  {Zolkower}}]{Graham2019}
{Graham}, M.~J., {Kulkarni}, S.~R., {Bellm}, E.~C., {et~al.} 2019, \pasp, 131,
  078001, \dodoi{10.1088/1538-3873/ab006c}

\bibitem[{{Gu{\'e}pin} {et~al.}(2022){Gu{\'e}pin}, {Kotera}, \&
  {Oikonomou}}]{Guepin2022}
{Gu{\'e}pin}, C., {Kotera}, K., \& {Oikonomou}, F. 2022, arXiv e-prints,
  arXiv:2207.12205.
\newblock \doarXiv{2207.12205}

\bibitem[{{Heger} {et~al.}(2003){Heger}, {Fryer}, {Woosley}, {Langer}, \&
  {Hartmann}}]{Heger2003}
{Heger}, A., {Fryer}, C.~L., {Woosley}, S.~E., {Langer}, N., \& {Hartmann},
  D.~H. 2003, \apj, 591, 288, \dodoi{10.1086/375341}

\bibitem[{{Ho} {et~al.}(2019){Ho}, {Goldstein}, {Schulze}, {Khatami}, {Perley},
  {Ergon}, {Gal-Yam}, {Corsi}, {Andreoni}, {Barbarino}, {Bellm},
  {Blagorodnova}, {Bright}, {Burns}, {Cenko}, {Cunningham}, {De}, {Dekany},
  {Dugas}, {Fender}, {Fransson}, {Fremling}, {Goldstein}, {Graham}, {Hale},
  {Horesh}, {Hung}, {Kasliwal}, {Kuin}, {Kulkarni}, {Kupfer}, {Lunnan},
  {Masci}, {Ngeow}, {Nugent}, {Ofek}, {Patterson}, {Petitpas}, {Rusholme},
  {Sai}, {Sfaradi}, {Shupe}, {Sollerman}, {Soumagnac}, {Tachibana}, {Taddia},
  {Walters}, {Wang}, {Yao}, \& {Zhang}}]{Ho2019_ZTF18abukavn}
{Ho}, A. Y.~Q., {Goldstein}, D.~A., {Schulze}, S., {et~al.} 2019, \apj, 887,
  169, \dodoi{10.3847/1538-4357/ab55ec}

\bibitem[{{Ho} {et~al.}(2020{\natexlab{a}}){Ho}, {Corsi}, {Cenko}, {Taddia},
  {Kulkarni}, {Adams}, {De}, {Dekany}, {Frederiks}, {Fremling}, {Golkhou},
  {Graham}, {Hung}, {Kupfer}, {Laher}, {Mahabal}, {Masci}, {Miller}, {Neill},
  {Reiley}, {Riddle}, {Ridnaia}, {Rusholme}, {Sharma}, {Sollerman},
  {Soumagnac}, {Svinkin}, \& {Shupe}}]{Ho2020_ZTF18aaqjovh}
{Ho}, A. Y.~Q., {Corsi}, A., {Cenko}, S.~B., {et~al.} 2020{\natexlab{a}}, \apj,
  893, 132, \dodoi{10.3847/1538-4357/ab7f3b}

\bibitem[{{Ho} {et~al.}(2020{\natexlab{b}}){Ho}, {Kulkarni}, {Perley}, {Cenko},
  {Corsi}, {Schulze}, {Lunnan}, {Sollerman}, {Gal-Yam}, {Anand}, {Barbarino},
  {Bellm}, {Bruch}, {Burns}, {De}, {Dekany}, {Delacroix}, {Duev}, {Frederiks},
  {Fremling}, {Goldstein}, {Golkhou}, {Graham}, {Hale}, {Kasliwal}, {Kupfer},
  {Laher}, {Martikainen}, {Masci}, {Neill}, {Ridnaia}, {Rusholme}, {Savchenko},
  {Shupe}, {Soumagnac}, {Strotjohann}, {Svinkin}, {Taggart}, {Tartaglia},
  {Yan}, \& {Zolkower}}]{Ho2020_ZTF20aalxlis}
{Ho}, A. Y.~Q., {Kulkarni}, S.~R., {Perley}, D.~A., {et~al.}
  2020{\natexlab{b}}, \apj, 902, 86, \dodoi{10.3847/1538-4357/aba630}

\bibitem[{{Ivezi{\'c}} {et~al.}(2019){Ivezi{\'c}}, {Kahn}, {Tyson}, {Abel},
  {Acosta}, {Allsman}, {Alonso}, {AlSayyad}, {Anderson}, {Andrew}, {Angel},
  {Angeli}, {Ansari}, {Antilogus}, {Araujo}, {Armstrong}, {Arndt}, {Astier},
  {Aubourg}, {Auza}, {Axelrod}, {Bard}, {Barr}, {Barrau}, {Bartlett}, {Bauer},
  {Bauman}, {Baumont}, {Bechtol}, {Bechtol}, {Becker}, {Becla}, {Beldica},
  {Bellavia}, {Bianco}, {Biswas}, {Blanc}, {Blazek}, {Blandford}, {Bloom},
  {Bogart}, {Bond}, {Booth}, {Borgland}, {Borne}, {Bosch}, {Boutigny},
  {Brackett}, {Bradshaw}, {Brandt}, {Brown}, {Bullock}, {Burchat}, {Burke},
  {Cagnoli}, {Calabrese}, {Callahan}, {Callen}, {Carlin}, {Carlson},
  {Chandrasekharan}, {Charles-Emerson}, {Chesley}, {Cheu}, {Chiang}, {Chiang},
  {Chirino}, {Chow}, {Ciardi}, {Claver}, {Cohen-Tanugi}, {Cockrum}, {Coles},
  {Connolly}, {Cook}, {Cooray}, {Covey}, {Cribbs}, {Cui}, {Cutri}, {Daly},
  {Daniel}, {Daruich}, {Daubard}, {Daues}, {Dawson}, {Delgado}, {Dellapenna},
  {de Peyster}, {de Val-Borro}, {Digel}, {Doherty}, {Dubois},
  {Dubois-Felsmann}, {Durech}, {Economou}, {Eifler}, {Eracleous}, {Emmons},
  {Fausti Neto}, {Ferguson}, {Figueroa}, {Fisher-Levine}, {Focke}, {Foss},
  {Frank}, {Freemon}, {Gangler}, {Gawiser}, {Geary}, {Gee}, {Geha}, {Gessner},
  {Gibson}, {Gilmore}, {Glanzman}, {Glick}, {Goldina}, {Goldstein}, {Goodenow},
  {Graham}, {Gressler}, {Gris}, {Guy}, {Guyonnet}, {Haller}, {Harris},
  {Hascall}, {Haupt}, {Hernandez}, {Herrmann}, {Hileman}, {Hoblitt}, {Hodgson},
  {Hogan}, {Howard}, {Huang}, {Huffer}, {Ingraham}, {Innes}, {Jacoby}, {Jain},
  {Jammes}, {Jee}, {Jenness}, {Jernigan}, {Jevremovi{\'c}}, {Johns}, {Johnson},
  {Johnson}, {Jones}, {Juramy-Gilles}, {Juri{\'c}}, {Kalirai}, {Kallivayalil},
  {Kalmbach}, {Kantor}, {Karst}, {Kasliwal}, {Kelly}, {Kessler}, {Kinnison},
  {Kirkby}, {Knox}, {Kotov}, {Krabbendam}, {Krughoff}, {Kub{\'a}nek},
  {Kuczewski}, {Kulkarni}, {Ku}, {Kurita}, {Lage}, {Lambert}, {Lange},
  {Langton}, {Le Guillou}, {Levine}, {Liang}, {Lim}, {Lintott}, {Long},
  {Lopez}, {Lotz}, {Lupton}, {Lust}, {MacArthur}, {Mahabal}, {Mandelbaum},
  {Markiewicz}, {Marsh}, {Marshall}, {Marshall}, {May}, {McKercher}, {McQueen},
  {Meyers}, {Migliore}, {Miller}, {Mills}, {Miraval}, {Moeyens}, {Moolekamp},
  {Monet}, {Moniez}, {Monkewitz}, {Montgomery}, {Morrison}, {Mueller},
  {Muller}, {Mu{\~n}oz Arancibia}, {Neill}, {Newbry}, {Nief}, {Nomerotski},
  {Nordby}, {O'Connor}, {Oliver}, {Olivier}, {Olsen}, {O'Mullane}, {Ortiz},
  {Osier}, {Owen}, {Pain}, {Palecek}, {Parejko}, {Parsons}, {Pease},
  {Peterson}, {Peterson}, {Petravick}, {Libby Petrick}, {Petry},
  {Pierfederici}, {Pietrowicz}, {Pike}, {Pinto}, {Plante}, {Plate}, {Plutchak},
  {Price}, {Prouza}, {Radeka}, {Rajagopal}, {Rasmussen}, {Regnault}, {Reil},
  {Reiss}, {Reuter}, {Ridgway}, {Riot}, {Ritz}, {Robinson}, {Roby}, {Roodman},
  {Rosing}, {Roucelle}, {Rumore}, {Russo}, {Saha}, {Sassolas}, {Schalk},
  {Schellart}, {Schindler}, {Schmidt}, {Schneider}, {Schneider}, {Schoening},
  {Schumacher}, {Schwamb}, {Sebag}, {Selvy}, {Sembroski}, {Seppala}, {Serio},
  {Serrano}, {Shaw}, {Shipsey}, {Sick}, {Silvestri}, {Slater}, {Smith},
  {Smith}, {Sobhani}, {Soldahl}, {Storrie-Lombardi}, {Stover}, {Strauss},
  {Street}, {Stubbs}, {Sullivan}, {Sweeney}, {Swinbank}, {Szalay}, {Takacs},
  {Tether}, {Thaler}, {Thayer}, {Thomas}, {Thornton}, {Thukral}, {Tice},
  {Trilling}, {Turri}, {Van Berg}, {Vanden Berk}, {Vetter}, {Virieux},
  {Vucina}, {Wahl}, {Walkowicz}, {Walsh}, {Walter}, {Wang}, {Wang}, {Warner},
  {Wiecha}, {Willman}, {Winters}, {Wittman}, {Wolff}, {Wood-Vasey}, {Wu},
  {Xin}, {Yoachim}, \& {Zhan}}]{Rubin2019}
{Ivezi{\'c}}, {\v{Z}}., {Kahn}, S.~M., {Tyson}, J.~A., {et~al.} 2019, \apj,
  873, 111, \dodoi{10.3847/1538-4357/ab042c}

\bibitem[{{Iwamoto} {et~al.}(1998){Iwamoto}, {Mazzali}, {Nomoto}, {Umeda},
  {Nakamura}, {Patat}, {Danziger}, {Young}, {Suzuki}, {Shigeyama},
  {Augusteijn}, {Doublier}, {Gonzalez}, {Boehnhardt}, {Brewer}, {Hainaut},
  {Lidman}, {Leibundgut}, {Cappellaro}, {Turatto}, {Galama}, {Vreeswijk},
  {Kouveliotou}, {van Paradijs}, {Pian}, {Palazzi}, \&
  {Frontera}}]{Iwamoto1998}
{Iwamoto}, K., {Mazzali}, P.~A., {Nomoto}, K., {et~al.} 1998, \nat, 395, 672,
  \dodoi{10.1038/27155}

\bibitem[{{Izzard} {et~al.}(2004){Izzard}, {Ramirez-Ruiz}, \&
  {Tout}}]{Izzard2004}
{Izzard}, R.~G., {Ramirez-Ruiz}, E., \& {Tout}, C.~A. 2004, \mnras, 348, 1215,
  \dodoi{10.1111/j.1365-2966.2004.07436.x}

\bibitem[{{Janka}(2012)}]{Janka2012}
{Janka}, H.-T. 2012, Annual Review of Nuclear and Particle Science, 62, 407,
  \dodoi{10.1146/annurev-nucl-102711-094901}

\bibitem[{{Janka} {et~al.}(2007){Janka}, {Langanke}, {Marek},
  {Mart{\'\i}nez-Pinedo}, \& {M{\"u}ller}}]{Janka2007}
{Janka}, H.~T., {Langanke}, K., {Marek}, A., {Mart{\'\i}nez-Pinedo}, G., \&
  {M{\"u}ller}, B. 2007, \physrep, 442, 38,
  \dodoi{10.1016/j.physrep.2007.02.002}

\bibitem[{{Japelj} {et~al.}(2018){Japelj}, {Vergani}, {Salvaterra}, {Renzo},
  {Zapartas}, {de Mink}, {Kaper}, \& {Zibetti}}]{Japelj2018}
{Japelj}, J., {Vergani}, S.~D., {Salvaterra}, R., {et~al.} 2018, \aap, 617,
  A105, \dodoi{10.1051/0004-6361/201833209}

\bibitem[{{Jerkstrand} {et~al.}(2015){Jerkstrand}, {Timmes}, {Magkotsios},
  {Sim}, {Fransson}, {Spyromilio}, {M{\"u}ller}, {Heger}, {Sollerman}, \&
  {Smartt}}]{Jerkstrand2015}
{Jerkstrand}, A., {Timmes}, F.~X., {Magkotsios}, G., {et~al.} 2015, \apj, 807,
  110, \dodoi{10.1088/0004-637X/807/1/110}

\bibitem[{{Kaiser} {et~al.}(2010){Kaiser}, {Burgett}, {Chambers}, {Denneau},
  {Heasley}, {Jedicke}, {Magnier}, {Morgan}, {Onaka}, \& {Tonry}}]{Kaiser2010}
{Kaiser}, N., {Burgett}, W., {Chambers}, K., {et~al.} 2010, in Society of
  Photo-Optical Instrumentation Engineers (SPIE) Conference Series, Vol. 7733,
  Ground-based and Airborne Telescopes III, ed. L.~M. {Stepp}, R.~{Gilmozzi},
  \& H.~J. {Hall}, 77330E, \dodoi{10.1117/12.859188}

\bibitem[{{Kankare} {et~al.}(2021){Kankare}, {Nagao}, {Koivisto}, {Gonzalez},
  {Silvestre}, \& {Zimmerman}}]{Kankare2021}
{Kankare}, E., {Nagao}, T., {Koivisto}, N., {et~al.} 2021, Transient Name
  Server Classification Report, 2021-762, 1

\bibitem[{{Kelly} {et~al.}(2014){Kelly}, {Filippenko}, {Modjaz}, \&
  {Kocevski}}]{Kelly2014}
{Kelly}, P.~L., {Filippenko}, A.~V., {Modjaz}, M., \& {Kocevski}, D. 2014,
  \apj, 789, 23, \dodoi{10.1088/0004-637X/789/1/23}

\bibitem[{{Kouveliotou} {et~al.}(2004){Kouveliotou}, {Woosley}, {Patel},
  {Levan}, {Blandford}, {Ramirez-Ruiz}, {Wijers}, {Weisskopf}, {Tennant},
  {Pian}, \& {Giommi}}]{Kouveliotou2004}
{Kouveliotou}, C., {Woosley}, S.~E., {Patel}, S.~K., {et~al.} 2004, \apj, 608,
  872, \dodoi{10.1086/420878}

\bibitem[{{Kulkarni} {et~al.}(1998){Kulkarni}, {Frail}, {Wieringa}, {Ekers},
  {Sadler}, {Wark}, {Higdon}, {Phinney}, \& {Bloom}}]{Kulkarni1998}
{Kulkarni}, S.~R., {Frail}, D.~A., {Wieringa}, M.~H., {et~al.} 1998, \nat, 395,
  663, \dodoi{10.1038/27139}

\bibitem[{{Langer}(2012)}]{Langer2012}
{Langer}, N. 2012, \araa, 50, 107, \dodoi{10.1146/annurev-astro-081811-125534}

\bibitem[{{Law} {et~al.}(2018){Law}, {Gaensler}, {Metzger}, {Ofek}, \&
  {Sironi}}]{Law2018}
{Law}, C.~J., {Gaensler}, B.~M., {Metzger}, B.~D., {Ofek}, E.~O., \& {Sironi},
  L. 2018, \apjl, 866, L22, \dodoi{10.3847/2041-8213/aae5f3}

\bibitem[{{Law} {et~al.}(2009){Law}, {Kulkarni}, {Dekany}, {Ofek}, {Quimby},
  {Nugent}, {Surace}, {Grillmair}, {Bloom}, {Kasliwal}, {Bildsten}, {Brown},
  {Cenko}, {Ciardi}, {Croner}, {Djorgovski}, {van Eyken}, {Filippenko}, {Fox},
  {Gal-Yam}, {Hale}, {Hamam}, {Helou}, {Henning}, {Howell}, {Jacobsen},
  {Laher}, {Mattingly}, {McKenna}, {Pickles}, {Poznanski}, {Rahmer}, {Rau},
  {Rosing}, {Shara}, {Smith}, {Starr}, {Sullivan}, {Velur}, {Walters}, \&
  {Zolkower}}]{Law2009}
{Law}, N.~M., {Kulkarni}, S.~R., {Dekany}, R.~G., {et~al.} 2009, \pasp, 121,
  1395, \dodoi{10.1086/648598}

\bibitem[{{Lazzati} {et~al.}(2012){Lazzati}, {Morsony}, {Blackwell}, \&
  {Begelman}}]{Lazzati2012}
{Lazzati}, D., {Morsony}, B.~J., {Blackwell}, C.~H., \& {Begelman}, M.~C. 2012,
  \apj, 750, 68, \dodoi{10.1088/0004-637X/750/1/68}

\bibitem[{Li {et~al.}(2011)Li, Leaman, Chornock, Filippenko, Poznanski,
  Ganeshalingam, Wang, Modjaz, Jha, Foley, \& Smith}]{Li2011}
Li, W., Leaman, J., Chornock, R., {et~al.} 2011, Monthly Notices of the Royal
  Astronomical Society, 412, 1441, \dodoi{10.1111/j.1365-2966.2011.18160.x}

\bibitem[{{Li} \& {Chevalier}(1999)}]{Li1999}
{Li}, Z.-Y., \& {Chevalier}, R.~A. 1999, \apj, 526, 716, \dodoi{10.1086/308031}

\bibitem[{{Lien} {et~al.}(2011){Lien}, {Chakraborty}, {Fields}, \&
  {Kemball}}]{Lien2011}
{Lien}, A., {Chakraborty}, N., {Fields}, B.~D., \& {Kemball}, A. 2011, \apj,
  740, 23, \dodoi{10.1088/0004-637X/740/1/23}

\bibitem[{{Liu} {et~al.}(2016){Liu}, {Modjaz}, {Bianco}, \& {Graur}}]{Liu2016}
{Liu}, Y.-Q., {Modjaz}, M., {Bianco}, F.~B., \& {Graur}, O. 2016, \apj, 827,
  90, \dodoi{10.3847/0004-637X/827/2/90}

\bibitem[{{Livio} \& {Waxman}(2000)}]{Livio2000}
{Livio}, M., \& {Waxman}, E. 2000, \apj, 538, 187, \dodoi{10.1086/309120}

\bibitem[{{Lyman} {et~al.}(2014){Lyman}, {Bersier}, \& {James}}]{Lyman2014}
{Lyman}, J.~D., {Bersier}, D., \& {James}, P.~A. 2014, \mnras, 437, 3848,
  \dodoi{10.1093/mnras/stt2187}

\bibitem[{{Lyman} {et~al.}(2016){Lyman}, {Bersier}, {James}, {Mazzali},
  {Eldridge}, {Fraser}, \& {Pian}}]{Lyman2016}
{Lyman}, J.~D., {Bersier}, D., {James}, P.~A., {et~al.} 2016, \mnras, 457, 328,
  \dodoi{10.1093/mnras/stv2983}

\bibitem[{{MacFadyen} \& {Woosley}(1999)}]{MacFadyen1999}
{MacFadyen}, A.~I., \& {Woosley}, S.~E. 1999, \apj, 524, 262,
  \dodoi{10.1086/307790}

\bibitem[{{Maeda} {et~al.}(2021){Maeda}, {Chandra}, {Matsuoka}, {Ryder},
  {Moriya}, {Kuncarayakti}, {Lee}, {Kundu}, {Patnaude}, {Saito}, \&
  {Folatelli}}]{Maeda2021}
{Maeda}, K., {Chandra}, P., {Matsuoka}, T., {et~al.} 2021, \apj, 918, 34,
  \dodoi{10.3847/1538-4357/ac0dbc}

\bibitem[{{Maggiore} {et~al.}(2020){Maggiore}, {Van Den Broeck}, {Bartolo},
  {Belgacem}, {Bertacca}, {Bizouard}, {Branchesi}, {Clesse}, {Foffa},
  {Garc{\'\i}a-Bellido}, {Grimm}, {Harms}, {Hinderer}, {Matarrese}, {Palomba},
  {Peloso}, {Ricciardone}, \& {Sakellariadou}}]{Maggiore2020}
{Maggiore}, M., {Van Den Broeck}, C., {Bartolo}, N., {et~al.} 2020, \jcap,
  2020, 050, \dodoi{10.1088/1475-7516/2020/03/050}

\bibitem[{{Margutti} {et~al.}(2013){Margutti}, {Soderberg}, {Wieringa},
  {Edwards}, {Chevalier}, {Morsony}, {Barniol Duran}, {Sironi}, {Zauderer},
  {Milisavljevic}, {Kamble}, \& {Pian}}]{Margutti2013}
{Margutti}, R., {Soderberg}, A.~M., {Wieringa}, M.~H., {et~al.} 2013, \apj,
  778, 18, \dodoi{10.1088/0004-637X/778/1/18}

\bibitem[{{Margutti} {et~al.}(2017){Margutti}, {Kamble}, {Milisavljevic},
  {Zapartas}, {de Mink}, {Drout}, {Chornock}, {Risaliti}, {Zauderer},
  {Bietenholz}, {Cantiello}, {Chakraborti}, {Chomiuk}, {Fong}, {Grefenstette},
  {Guidorzi}, {Kirshner}, {Parrent}, {Patnaude}, {Soderberg}, {Gehrels}, \&
  {Harrison}}]{Margutti2017}
{Margutti}, R., {Kamble}, A., {Milisavljevic}, D., {et~al.} 2017, \apj, 835,
  140, \dodoi{10.3847/1538-4357/835/2/140}

\bibitem[{{Masci} {et~al.}(2019){Masci}, {Laher}, {Rusholme}, {Shupe}, {Groom},
  {Surace}, {Jackson}, {Monkewitz}, {Beck}, {Flynn}, {Terek}, {Landry},
  {Hacopians}, {Desai}, {Howell}, {Brooke}, {Imel}, {Wachter}, {Ye}, {Lin},
  {Cenko}, {Cunningham}, {Rebbapragada}, {Bue}, {Miller}, {Mahabal}, {Bellm},
  {Patterson}, {Juri{\'c}}, {Golkhou}, {Ofek}, {Walters}, {Graham}, {Kasliwal},
  {Dekany}, {Kupfer}, {Burdge}, {Cannella}, {Barlow}, {Van Sistine}, {Giomi},
  {Fremling}, {Blagorodnova}, {Levitan}, {Riddle}, {Smith}, {Helou}, {Prince},
  \& {Kulkarni}}]{Masci2019}
{Masci}, F.~J., {Laher}, R.~R., {Rusholme}, B., {et~al.} 2019, \pasp, 131,
  018003, \dodoi{10.1088/1538-3873/aae8ac}

\bibitem[{{Matheson} {et~al.}(2001){Matheson}, {Filippenko}, {Li}, {Leonard},
  \& {Shields}}]{Matheson2001}
{Matheson}, T., {Filippenko}, A.~V., {Li}, W., {Leonard}, D.~C., \& {Shields},
  J.~C. 2001, \aj, 121, 1648, \dodoi{10.1086/319390}

\bibitem[{{Mazzali} {et~al.}(2000){Mazzali}, {Iwamoto}, \&
  {Nomoto}}]{Mazzali2000}
{Mazzali}, P.~A., {Iwamoto}, K., \& {Nomoto}, K. 2000, \apj, 545, 407,
  \dodoi{10.1086/317808}

\bibitem[{{Mazzali} {et~al.}(2014){Mazzali}, {McFadyen}, {Woosley}, {Pian}, \&
  {Tanaka}}]{Mazzali2014}
{Mazzali}, P.~A., {McFadyen}, A.~I., {Woosley}, S.~E., {Pian}, E., \& {Tanaka},
  M. 2014, \mnras, 443, 67, \dodoi{10.1093/mnras/stu1124}

\bibitem[{{Mazzali} {et~al.}(2002){Mazzali}, {Deng}, {Maeda}, {Nomoto},
  {Umeda}, {Hatano}, {Iwamoto}, {Yoshii}, {Kobayashi}, {Minezaki}, {Doi},
  {Enya}, {Tomita}, {Smartt}, {Kinugasa}, {Kawakita}, {Ayani}, {Kawabata},
  {Yamaoka}, {Qiu}, {Motohara}, {Gerardy}, {Fesen}, {Kawabata}, {Iye},
  {Kashikawa}, {Kosugi}, {Ohyama}, {Takada-Hidai}, {Zhao}, {Chornock},
  {Filippenko}, {Benetti}, \& {Turatto}}]{Mazzali2002}
{Mazzali}, P.~A., {Deng}, J., {Maeda}, K., {et~al.} 2002, \apjl, 572, L61,
  \dodoi{10.1086/341504}

\bibitem[{{Mazzali} {et~al.}(2003){Mazzali}, {Deng}, {Tominaga}, {Maeda},
  {Nomoto}, {Matheson}, {Kawabata}, {Stanek}, \& {Garnavich}}]{Mazzali2003}
{Mazzali}, P.~A., {Deng}, J., {Tominaga}, N., {et~al.} 2003, \apjl, 599, L95,
  \dodoi{10.1086/381259}

\bibitem[{{Mazzali} {et~al.}(2006{\natexlab{a}}){Mazzali}, {Deng}, {Pian},
  {Malesani}, {Tominaga}, {Maeda}, {Nomoto}, {Chincarini}, {Covino}, {Della
  Valle}, {Fugazza}, {Tagliaferri}, \& {Gal-Yam}}]{Mazzali2006b}
{Mazzali}, P.~A., {Deng}, J., {Pian}, E., {et~al.} 2006{\natexlab{a}}, \apj,
  645, 1323, \dodoi{10.1086/504415}

\bibitem[{{Mazzali} {et~al.}(2006{\natexlab{b}}){Mazzali}, {Deng}, {Nomoto},
  {Sauer}, {Pian}, {Tominaga}, {Tanaka}, {Maeda}, \&
  {Filippenko}}]{Mazzali2006a}
{Mazzali}, P.~A., {Deng}, J., {Nomoto}, K., {et~al.} 2006{\natexlab{b}}, \nat,
  442, 1018, \dodoi{10.1038/nature05081}

\bibitem[{{McMullin} {et~al.}(2007){McMullin}, {Waters}, {Schiebel}, {Young},
  \& {Golap}}]{McMullin2007}
{McMullin}, J.~P., {Waters}, B., {Schiebel}, D., {Young}, W., \& {Golap}, K.
  2007, in Astronomical Society of the Pacific Conference Series, Vol. 376,
  Astronomical Data Analysis Software and Systems XVI, ed. R.~A. {Shaw},
  F.~{Hill}, \& D.~J. {Bell}, 127

\bibitem[{{Meegan} {et~al.}(2009){Meegan}, {Lichti}, {Bhat}, {Bissaldi},
  {Briggs}, {Connaughton}, {Diehl}, {Fishman}, {Greiner}, {Hoover}, {van der
  Horst}, {von Kienlin}, {Kippen}, {Kouveliotou}, {McBreen}, {Paciesas},
  {Preece}, {Steinle}, {Wallace}, {Wilson}, \& {Wilson-Hodge}}]{Meegan2009}
{Meegan}, C., {Lichti}, G., {Bhat}, P.~N., {et~al.} 2009, \apj, 702, 791,
  \dodoi{10.1088/0004-637X/702/1/791}

\bibitem[{{M{\'e}sz{\'a}ros}(2006)}]{Meszaros2006}
{M{\'e}sz{\'a}ros}, P. 2006, Reports on Progress in Physics, 69, 2259,
  \dodoi{10.1088/0034-4885/69/8/R01}

\bibitem[{{Mezzacappa} {et~al.}(1998){Mezzacappa}, {Calder}, {Bruenn},
  {Blondin}, {Guidry}, {Strayer}, \& {Umar}}]{Mezzacappa1998}
{Mezzacappa}, A., {Calder}, A.~C., {Bruenn}, S.~W., {et~al.} 1998, \apj, 495,
  911, \dodoi{10.1086/305338}

\bibitem[{{Miller} {et~al.}(2020){Miller}, {Yao}, {Bulla}, {Pankow}, {Bellm},
  {Cenko}, {Dekany}, {Fremling}, {Graham}, {Kupfer}, {Laher}, {Mahabal},
  {Masci}, {Nugent}, {Riddle}, {Rusholme}, {Smith}, {Shupe}, {van Roestel}, \&
  {Kulkarni}}]{Miller2020}
{Miller}, A.~A., {Yao}, Y., {Bulla}, M., {et~al.} 2020, \apj, 902, 47,
  \dodoi{10.3847/1538-4357/abb13b}

\bibitem[{{Modjaz} {et~al.}(2016){Modjaz}, {Liu}, {Bianco}, \&
  {Graur}}]{Modjaz2016}
{Modjaz}, M., {Liu}, Y.~Q., {Bianco}, F.~B., \& {Graur}, O. 2016, \apj, 832,
  108, \dodoi{10.3847/0004-637X/832/2/108}

\bibitem[{{Modjaz} {et~al.}(2009){Modjaz}, {Li}, {Butler}, {Chornock},
  {Perley}, {Blondin}, {Bloom}, {Filippenko}, {Kirshner}, {Kocevski},
  {Poznanski}, {Hicken}, {Foley}, {Stringfellow}, {Berlind}, {Barrado y
  Navascues}, {Blake}, {Bouy}, {Brown}, {Challis}, {Chen}, {de Vries},
  {Dufour}, {Falco}, {Friedman}, {Ganeshalingam}, {Garnavich}, {Holden},
  {Illingworth}, {Lee}, {Liebert}, {Marion}, {Olivier}, {Prochaska},
  {Silverman}, {Smith}, {Starr}, {Steele}, {Stockton}, {Williams}, \&
  {Wood-Vasey}}]{Modjaz2009}
{Modjaz}, M., {Li}, W., {Butler}, N., {et~al.} 2009, \apj, 702, 226,
  \dodoi{10.1088/0004-637X/702/1/226}

\bibitem[{Modjaz {et~al.}(2014)Modjaz, Blondin, Kirshner, Matheson, Berlind,
  Bianco, Calkins, Challis, Garnavich, Hicken, Jha, Liu, \&
  Marion}]{Modjaz2014}
Modjaz, M., Blondin, S., Kirshner, R.~P., {et~al.} 2014, The Astronomical
  Journal, 147, 99, \dodoi{10.1088/0004-6256/147/5/99}

\bibitem[{{Modjaz} {et~al.}(2020){Modjaz}, {Bianco}, {Siwek}, {Huang},
  {Perley}, {Fierroz}, {Liu}, {Arcavi}, {Gal-Yam}, {Filippenko},
  {Blagorodnova}, {Cenko}, {Kasliwal}, {Kulkarni}, {Schulze}, {Taggart}, \&
  {Zheng}}]{Modjaz2020}
{Modjaz}, M., {Bianco}, F.~B., {Siwek}, M., {et~al.} 2020, \apj, 892, 153,
  \dodoi{10.3847/1538-4357/ab4185}

\bibitem[{{Montes} {et~al.}(1998){Montes}, {Van Dyk}, {Weiler}, {Sramek}, \&
  {Panagia}}]{Montes1998}
{Montes}, M.~J., {Van Dyk}, S.~D., {Weiler}, K.~W., {Sramek}, R.~A., \&
  {Panagia}, N. 1998, \apj, 506, 874, \dodoi{10.1086/306261}

\bibitem[{{M{\"u}ller}(2020)}]{Muller2020}
{M{\"u}ller}, B. 2020, Living Reviews in Computational Astrophysics, 6, 3,
  \dodoi{10.1007/s41115-020-0008-5}

\bibitem[{{Munoz-Arancibia} {et~al.}(2021{\natexlab{a}}){Munoz-Arancibia},
  {Forster}, {Bauer}, {Hernandez-Garcia}, {Pignata}, {Galbany}, {Camacho},
  {Silva-Farfan}, {Mourao}, {Arredondo}, {Cabrera-Vives}, {Carrasco-Davis},
  {Estevez}, {Huijse}, {Reyes}, {Reyes}, {Sanchez-Saez}, {Valenzuela},
  {Castillo}, {Ruz-Mieres}, {Rodriguez-Mancini}, {Catelan}, {Eyheramendy}, \&
  {Graham}}]{MunozArancibia2021}
{Munoz-Arancibia}, A., {Forster}, F., {Bauer}, F.~E., {et~al.}
  2021{\natexlab{a}}, Transient Name Server Discovery Report, 2021-180, 1

\bibitem[{{Munoz-Arancibia} {et~al.}(2021{\natexlab{b}}){Munoz-Arancibia},
  {Forster}, {Bauer}, {Pignata}, {Hernandez-Garcia}, {Galbany}, {Camacho},
  {Silva-Farfan}, {Mourao}, {Arredondo}, {Cabrera-Vives}, {Carrasco-Davis},
  {Estevez}, {Huijse}, {Reyes}, {Reyes}, {Sanchez-Saez}, {Valenzuela},
  {Castillo}, {Ruz-Mieres}, {Rodriguez-Mancini}, {Catelan}, {Eyheramendy}, \&
  {Graham}}]{MunozArancibia20210305}
---. 2021{\natexlab{b}}, Transient Name Server Discovery Report, 2021-685, 1

\bibitem[{{Murase}(2018)}]{Murase2018}
{Murase}, K. 2018, \prd, 97, 081301, \dodoi{10.1103/PhysRevD.97.081301}

\bibitem[{{Neumann} {et~al.}(2022){Neumann}, {Holoien}, {Kochanek}, {Stanek},
  {Vallely}, {Shappee}, {Prieto}, {Pessi}, {Jayasinghe}, {Brimacombe},
  {Bersier}, {Aydi}, {Basinger}, {Beacom}, {Bose}, {Brown}, {Chen},
  {Clocchiatti}, {Desai}, {Dong}, {Falco}, {Holmbo}, {Morrell}, {Shields},
  {Sokolovsky}, {Strader}, {Stritzinger}, {Swihart}, {Thompson}, {Way},
  {Aslan}, {Bishop}, {Bock}, {Bradshaw}, {Cacella}, {Castro}, {Conseil},
  {Cornect}, {Cruz}, {Farfan}, {Fernandez}, {Gabuya}, {Gonzalez-Carballo},
  {Kendurkar}, {Kiyota}, {Koff}, {Krannich}, {Marples}, {Masi}, {Monard},
  {Mu{\~n}oz}, {Nicholls}, {Post}, {Pujic}, {Stone}, {Tomasella}, {Trappett},
  \& {Wiethoff}}]{Neumann2022}
{Neumann}, K.~D., {Holoien}, T.~W.~S., {Kochanek}, C.~S., {et~al.} 2022, arXiv
  e-prints, arXiv:2210.06492.
\newblock \doarXiv{2210.06492}

\bibitem[{{Nishimura} {et~al.}(2015){Nishimura}, {Takiwaki}, \&
  {Thielemann}}]{Nishimura2015}
{Nishimura}, N., {Takiwaki}, T., \& {Thielemann}, F.-K. 2015, \apj, 810, 109,
  \dodoi{10.1088/0004-637X/810/2/109}

\bibitem[{{Nordin} {et~al.}(2018){Nordin}, {Brinnel}, {Giomi}, {Santen},
  {Gal-Yam}, {Yaron}, \& {Schulze}}]{Nordin2018_ZTF18acpeekw}
{Nordin}, J., {Brinnel}, V., {Giomi}, M., {et~al.} 2018, Transient Name Server
  Discovery Report, 2018-2043, 1

\bibitem[{{Oke} \& {Gunn}(1982)}]{Oke1982}
{Oke}, J.~B., \& {Gunn}, J.~E. 1982, \pasp, 94, 586, \dodoi{10.1086/131027}

\bibitem[{{Oke} {et~al.}(1995){Oke}, {Cohen}, {Carr}, {Cromer}, {Dingizian},
  {Harris}, {Labrecque}, {Lucinio}, {Schaal}, {Epps}, \& {Miller}}]{Oke1995}
{Oke}, J.~B., {Cohen}, J.~G., {Carr}, M., {et~al.} 1995, \pasp, 107, 375,
  \dodoi{10.1086/133562}

\bibitem[{{Pais} {et~al.}(2022){Pais}, {Piran}, \& {Nakar}}]{Pais2022}
{Pais}, M., {Piran}, T., \& {Nakar}, E. 2022, arXiv e-prints, arXiv:2208.14459.
\newblock \doarXiv{2208.14459}

\bibitem[{{Papish} \& {Soker}(2011)}]{Papish2011}
{Papish}, O., \& {Soker}, N. 2011, \mnras, 416, 1697,
  \dodoi{10.1111/j.1365-2966.2011.18671.x}

\bibitem[{{Perley} {et~al.}(2020){Perley}, {Fremling}, {Sollerman}, {Miller},
  {Dahiwale}, {Sharma}, {Bellm}, {Biswas}, {Brink}, {Bruch}, {De}, {Dekany},
  {Drake}, {Duev}, {Filippenko}, {Gal-Yam}, {Goobar}, {Graham}, {Graham}, {Ho},
  {Irani}, {Kasliwal}, {Kim}, {Kulkarni}, {Mahabal}, {Masci}, {Modak}, {Neill},
  {Nordin}, {Riddle}, {Soumagnac}, {Strotjohann}, {Schulze}, {Taggart},
  {Tzanidakis}, {Walters}, \& {Yan}}]{Perley2020b}
{Perley}, D.~A., {Fremling}, C., {Sollerman}, J., {et~al.} 2020, \apj, 904, 35,
  \dodoi{10.3847/1538-4357/abbd98}

\bibitem[{{Pian} {et~al.}(2006){Pian}, {Mazzali}, {Masetti}, {Ferrero},
  {Klose}, {Palazzi}, {Ramirez-Ruiz}, {Woosley}, {Kouveliotou}, {Deng},
  {Filippenko}, {Foley}, {Fynbo}, {Kann}, {Li}, {Hjorth}, {Nomoto}, {Patat},
  {Sauer}, {Sollerman}, {Vreeswijk}, {Guenther}, {Levan}, {O'Brien}, {Tanvir},
  {Wijers}, {Dumas}, {Hainaut}, {Wong}, {Baade}, {Wang}, {Amati}, {Cappellaro},
  {Castro-Tirado}, {Ellison}, {Frontera}, {Fruchter}, {Greiner}, {Kawabata},
  {Ledoux}, {Maeda}, {M{\o}ller}, {Nicastro}, {Rol}, \& {Starling}}]{Pian2006}
{Pian}, E., {Mazzali}, P.~A., {Masetti}, N., {et~al.} 2006, \nat, 442, 1011,
  \dodoi{10.1038/nature05082}

\bibitem[{{Piran}(2004)}]{Piran2004}
{Piran}, T. 2004, Reviews of Modern Physics, 76, 1143,
  \dodoi{10.1103/RevModPhys.76.1143}

\bibitem[{{Piran} {et~al.}(2019){Piran}, {Nakar}, {Mazzali}, \&
  {Pian}}]{Piran2019}
{Piran}, T., {Nakar}, E., {Mazzali}, P., \& {Pian}, E. 2019, \apjl, 871, L25,
  \dodoi{10.3847/2041-8213/aaffce}

\bibitem[{{Poidevin} {et~al.}(2021){Poidevin}, {Perez-Fournon}, {Angel},
  {Shirley}, {Marques-Chaves}, {Geier}, {Shu}, {Rodney}, {Roberts-Pierel},
  {Bolton}, {Chakrabarti}, {Craig}, \& {Alamiri}}]{Poidevin_2021}
{Poidevin}, F., {Perez-Fournon}, I., {Angel}, C.~J., {et~al.} 2021, Transient
  Name Server Discovery Report, 2021-1003, 1

\bibitem[{{Rigault} {et~al.}(2019){Rigault}, {Neill}, {Blagorodnova}, {Dugas},
  {Feeney}, {Walters}, {Brinnel}, {Copin}, {Fremling}, {Nordin}, \&
  {Sollerman}}]{Rigault2019}
{Rigault}, M., {Neill}, J.~D., {Blagorodnova}, N., {et~al.} 2019, \aap, 627,
  A115, \dodoi{10.1051/0004-6361/201935344}

\bibitem[{{Salas} {et~al.}(2013){Salas}, {Bauer}, {Stockdale}, \&
  {Prieto}}]{Salas2013}
{Salas}, P., {Bauer}, F.~E., {Stockdale}, C., \& {Prieto}, J.~L. 2013, \mnras,
  428, 1207, \dodoi{10.1093/mnras/sts104}

\bibitem[{{Sand} {et~al.}(2018){Sand}, {Valenti}, {Tartaglia}, {Yang}, \&
  {Wyatt}}]{Sand2018}
{Sand}, D., {Valenti}, S., {Tartaglia}, L., {Yang}, S., \& {Wyatt}, S. 2018, in
  American Astronomical Society Meeting Abstracts, Vol. 231, American
  Astronomical Society Meeting Abstracts \#231, 245.11

\bibitem[{{Sauer} {et~al.}(2006){Sauer}, {Mazzali}, {Deng}, {Valenti},
  {Nomoto}, \& {Filippenko}}]{Sauer2006}
{Sauer}, D.~N., {Mazzali}, P.~A., {Deng}, J., {et~al.} 2006, \mnras, 369, 1939,
  \dodoi{10.1111/j.1365-2966.2006.10438.x}

\bibitem[{{Schlafly} \& {Finkbeiner}(2011)}]{Schlafly2011}
{Schlafly}, E.~F., \& {Finkbeiner}, D.~P. 2011, \apj, 737, 103,
  \dodoi{10.1088/0004-637X/737/2/103}

\bibitem[{{Schneider} {et~al.}(2021){Schneider}, {Podsiadlowski}, \&
  {M{\"u}ller}}]{Schneider2021}
{Schneider}, F.~R.~N., {Podsiadlowski}, P., \& {M{\"u}ller}, B. 2021, \aap,
  645, A5, \dodoi{10.1051/0004-6361/202039219}

\bibitem[{{Scholberg}(2012)}]{Scholber2012}
{Scholberg}, K. 2012, Annual Review of Nuclear and Particle Science, 62, 81,
  \dodoi{10.1146/annurev-nucl-102711-095006}

\bibitem[{{Schulze} {et~al.}(2011){Schulze}, {Klose}, {Bj{\"o}rnsson},
  {Jakobsson}, {Kann}, {Rossi}, {Kr{\"u}hler}, {Greiner}, \&
  {Ferrero}}]{Schulze2011}
{Schulze}, S., {Klose}, S., {Bj{\"o}rnsson}, G., {et~al.} 2011, \aap, 526, A23,
  \dodoi{10.1051/0004-6361/201015581}

\bibitem[{{Selina} {et~al.}(2018){Selina}, {Murphy}, {McKinnon}, {Beasley},
  {Butler}, {Carilli}, {Clark}, {Durand}, {Erickson}, {Grammer}, {Hiriart},
  {Jackson}, {Kent}, {Mason}, {Morgan}, {Ojeda}, {Rosero}, {Shillue},
  {Sturgis}, \& {Urbain}}]{ngVLA}
{Selina}, R.~J., {Murphy}, E.~J., {McKinnon}, M., {et~al.} 2018, in
  Astronomical Society of the Pacific Conference Series, Vol. 517, Science with
  a Next Generation Very Large Array, ed. E.~{Murphy}, 15.
\newblock \doarXiv{1810.08197}

\bibitem[{{Shankar} {et~al.}(2021){Shankar}, {M{\"o}sta}, {Barnes}, {Duffell},
  \& {Kasen}}]{Shankar2021}
{Shankar}, S., {M{\"o}sta}, P., {Barnes}, J., {Duffell}, P.~C., \& {Kasen}, D.
  2021, \mnras, 508, 5390, \dodoi{10.1093/mnras/stab2964}

\bibitem[{{Shappee} {et~al.}(2014){Shappee}, {Prieto}, {Grupe}, {Kochanek},
  {Stanek}, {De Rosa}, {Mathur}, {Zu}, {Peterson}, {Pogge}, {Komossa}, {Im},
  {Jencson}, {Holoien}, {Basu}, {Beacom}, {Szczygie{\l}}, {Brimacombe},
  {Adams}, {Campillay}, {Choi}, {Contreras}, {Dietrich}, {Dubberley},
  {Elphick}, {Foale}, {Giustini}, {Gonzalez}, {Hawkins}, {Howell}, {Hsiao},
  {Koss}, {Leighly}, {Morrell}, {Mudd}, {Mullins}, {Nugent}, {Parrent},
  {Phillips}, {Pojmanski}, {Rosing}, {Ross}, {Sand}, {Terndrup}, {Valenti},
  {Walker}, \& {Yoon}}]{Shappee2014}
{Shappee}, B.~J., {Prieto}, J.~L., {Grupe}, D., {et~al.} 2014, \apj, 788, 48,
  \dodoi{10.1088/0004-637X/788/1/48}

\bibitem[{{Shivvers} {et~al.}(2017){Shivvers}, {Modjaz}, {Zheng}, {Liu},
  {Filippenko}, {Silverman}, {Matheson}, {Pastorello}, {Graur}, {Foley},
  {Chornock}, {Smith}, {Leaman}, \& {Benetti}}]{Shivvers2017}
{Shivvers}, I., {Modjaz}, M., {Zheng}, W., {et~al.} 2017, \pasp, 129, 054201,
  \dodoi{10.1088/1538-3873/aa54a6}

\bibitem[{{Smith}(2014)}]{Smith2014}
{Smith}, N. 2014, \araa, 52, 487, \dodoi{10.1146/annurev-astro-081913-040025}

\bibitem[{{Smith} {et~al.}(2011){Smith}, {Li}, {Filippenko}, \&
  {Chornock}}]{Smith2011}
{Smith}, N., {Li}, W., {Filippenko}, A.~V., \& {Chornock}, R. 2011, \mnras,
  412, 1522, \dodoi{10.1111/j.1365-2966.2011.17229.x}

\bibitem[{{Soderberg} {et~al.}(2006{\natexlab{a}}){Soderberg}, {Chevalier},
  {Kulkarni}, \& {Frail}}]{Soderberg2006}
{Soderberg}, A.~M., {Chevalier}, R.~A., {Kulkarni}, S.~R., \& {Frail}, D.~A.
  2006{\natexlab{a}}, \apj, 651, 1005, \dodoi{10.1086/507571}

\bibitem[{{Soderberg} {et~al.}(2006{\natexlab{b}}){Soderberg}, {Nakar},
  {Berger}, \& {Kulkarni}}]{Soderberg2006catalog}
{Soderberg}, A.~M., {Nakar}, E., {Berger}, E., \& {Kulkarni}, S.~R.
  2006{\natexlab{b}}, \apj, 638, 930, \dodoi{10.1086/499121}

\bibitem[{{Soderberg} {et~al.}(2004){Soderberg}, {Kulkarni}, {Berger}, {Fox},
  {Sako}, {Frail}, {Gal-Yam}, {Moon}, {Cenko}, {Yost}, {Phillips}, {Persson},
  {Freedman}, {Wyatt}, {Jayawardhana}, \& {Paulson}}]{Soderberg2004}
{Soderberg}, A.~M., {Kulkarni}, S.~R., {Berger}, E., {et~al.} 2004, \nat, 430,
  648, \dodoi{10.1038/nature02757}

\bibitem[{{Soderberg} {et~al.}(2010){Soderberg}, {Chakraborti}, {Pignata},
  {Chevalier}, {Chandra}, {Ray}, {Wieringa}, {Copete}, {Chaplin},
  {Connaughton}, {Barthelmy}, {Bietenholz}, {Chugai}, {Stritzinger}, {Hamuy},
  {Fransson}, {Fox}, {Levesque}, {Grindlay}, {Challis}, {Foley}, {Kirshner},
  {Milne}, \& {Torres}}]{Soderberg2010}
{Soderberg}, A.~M., {Chakraborti}, S., {Pignata}, G., {et~al.} 2010, \nat, 463,
  513, \dodoi{10.1038/nature08714}

\bibitem[{{Soker} \& {Gilkis}(2017)}]{Soker2017}
{Soker}, N., \& {Gilkis}, A. 2017, \apj, 851, 95,
  \dodoi{10.3847/1538-4357/aa9c83}

\bibitem[{{Srinivasaragavan} {et~al.}(2022)}]{Gokul2022}
{Srinivasaragavan}, G.~P., {et~al.} 2022, in preparation

\bibitem[{{Stroh} {et~al.}(2021){Stroh}, {Terreran}, {Coppejans}, {Bright},
  {Margutti}, {Bietenholz}, {De Colle}, {DeMarchi}, {Duran}, {Milisavljevic},
  {Murase}, {Paterson}, \& {Williams}}]{Stroh2021}
{Stroh}, M.~C., {Terreran}, G., {Coppejans}, D.~L., {et~al.} 2021, \apjl, 923,
  L24, \dodoi{10.3847/2041-8213/ac375e}

\bibitem[{{Suwa} \& {Tominaga}(2015)}]{Suwa2015}
{Suwa}, Y., \& {Tominaga}, N. 2015, \mnras, 451, 282,
  \dodoi{10.1093/mnras/stv901}

\bibitem[{{Taddia} {et~al.}(2019){Taddia}, {Sollerman}, {Fremling},
  {Barbarino}, {Karamehmetoglu}, {Arcavi}, {Cenko}, {Filippenko}, {Gal-Yam},
  {Hiramatsu}, {Hosseinzadeh}, {Howell}, {Kulkarni}, {Laher}, {Lunnan},
  {Masci}, {Nugent}, {Nyholm}, {Perley}, {Quimby}, \& {Silverman}}]{Taddia2019}
{Taddia}, F., {Sollerman}, J., {Fremling}, C., {et~al.} 2019, \aap, 621, A71,
  \dodoi{10.1051/0004-6361/201834429}

\bibitem[{{The~LIGO~Scientific~Collaboration}(2015)}]{Aasi2015}
{The~LIGO~Scientific~Collaboration}. 2015, Classical and Quantum Gravity, 32,
  074001, \dodoi{10.1088/0264-9381/32/7/074001}

\bibitem[{{Tonry} {et~al.}(2018{\natexlab{a}}){Tonry}, {Stalder}, {Denneau},
  {Heinze}, {Weiland}, {Rest}, {Smith}, {Smartt}, {Young}, {Fulton}, {McBrien},
  {O'Neill}, \& {Clark}}]{Tonry2018_ZTF18abklarx}
{Tonry}, J., {Stalder}, B., {Denneau}, L., {et~al.} 2018{\natexlab{a}},
  Transient Name Server Discovery Report, 2018-1123, 1

\bibitem[{{Tonry} {et~al.}(2018{\natexlab{b}}){Tonry}, {Denneau}, {Heinze},
  {Weiland}, {Flewelling}, {Stalder}, {Rest}, {Stubbs}, {Smith}, {Smartt},
  {Young}, {Maguire}, {Prentice}, {McBrien}, {O'Neill}, {Clark}, {Magee},
  {Fulton}, {Mccormack}, \& {Wright}}]{Tonry2018_ZTF18acbwxcc}
{Tonry}, J., {Denneau}, L., {Heinze}, A., {et~al.} 2018{\natexlab{b}},
  Transient Name Server Discovery Report, 2018-1634, 1

\bibitem[{{Tonry} {et~al.}(2018{\natexlab{c}}){Tonry}, {Denneau}, {Heinze},
  {Weiland}, {Flewelling}, {Stalder}, {Rest}, {Stubbs}, {Smith}, {Smartt},
  {Young}, {Maguire}, {Prentice}, {McBrien}, {O'Neill}, {Clark}, {Magee},
  {Fulton}, {Mccormack}, \& {Wright}}]{Tonry2018_ZTF18acaimrb}
---. 2018{\natexlab{c}}, Transient Name Server Discovery Report, 2018-1713, 1

\bibitem[{{Tonry} {et~al.}(2018{\natexlab{d}}){Tonry}, {Denneau}, {Heinze},
  {Stalder}, {Smith}, {Smartt}, {Stubbs}, {Weiland}, \& {Rest}}]{Tonry2018}
{Tonry}, J.~L., {Denneau}, L., {Heinze}, A.~N., {et~al.} 2018{\natexlab{d}},
  \pasp, 130, 064505, \dodoi{10.1088/1538-3873/aabadf}

\bibitem[{{Valenti} {et~al.}(2008){Valenti}, {Benetti}, {Cappellaro}, {Patat},
  {Mazzali}, {Turatto}, {Hurley}, {Maeda}, {Gal-Yam}, {Foley}, {Filippenko},
  {Pastorello}, {Challis}, {Frontera}, {Harutyunyan}, {Iye}, {Kawabata},
  {Kirshner}, {Li}, {Lipkin}, {Matheson}, {Nomoto}, {Ofek}, {Ohyama}, {Pian},
  {Poznanski}, {Salvo}, {Sauer}, {Schmidt}, {Soderberg}, \&
  {Zampieri}}]{Valenti2008}
{Valenti}, S., {Benetti}, S., {Cappellaro}, E., {et~al.} 2008, \mnras, 383,
  1485, \dodoi{10.1111/j.1365-2966.2007.12647.x}

\bibitem[{{van Eerten} {et~al.}(2012){van Eerten}, {van der Horst}, \&
  {MacFadyen}}]{VanEerten2012}
{van Eerten}, H., {van der Horst}, A., \& {MacFadyen}, A. 2012, \apj, 749, 44,
  \dodoi{10.1088/0004-637X/749/1/44}

\bibitem[{{van Eerten} \& {MacFadyen}(2011)}]{vanEerten2011}
{van Eerten}, H.~J., \& {MacFadyen}, A.~I. 2011, \apjl, 733, L37,
  \dodoi{10.1088/2041-8205/733/2/L37}

\bibitem[{{Villar} {et~al.}(2019){Villar}, {Berger}, {Miller}, {Chornock},
  {Rest}, {Jones}, {Drout}, {Foley}, {Kirshner}, {Lunnan}, {Magnier},
  {Milisavljevic}, {Sanders}, \& {Scolnic}}]{Villar2019}
{Villar}, V.~A., {Berger}, E., {Miller}, G., {et~al.} 2019, \apj, 884, 83,
  \dodoi{10.3847/1538-4357/ab418c}

\bibitem[{{Villar} {et~al.}(2020){Villar}, {Hosseinzadeh}, {Berger},
  {Ntampaka}, {Jones}, {Challis}, {Chornock}, {Drout}, {Foley}, {Kirshner},
  {Lunnan}, {Margutti}, {Milisavljevic}, {Sanders}, {Pan}, {Rest}, {Scolnic},
  {Magnier}, {Metcalfe}, {Wainscoat}, \& {Waters}}]{Villar2020}
{Villar}, V.~A., {Hosseinzadeh}, G., {Berger}, E., {et~al.} 2020, \apj, 905,
  94, \dodoi{10.3847/1538-4357/abc6fd}

\bibitem[{{Villarreal Hern{\'a}ndez} \& {Andernach}(2018)}]{Hernandez2018}
{Villarreal Hern{\'a}ndez}, A.~C., \& {Andernach}, H. 2018, arXiv e-prints,
  arXiv:1808.07178.
\newblock \doarXiv{1808.07178}

\bibitem[{{Waxman}(2004)}]{Waxman2004b}
{Waxman}, E. 2004, \apj, 602, 886, \dodoi{10.1086/381230}

\bibitem[{{Willingale} {et~al.}(2013){Willingale}, {Starling}, {Beardmore},
  {Tanvir}, \& {O'Brien}}]{Willingale+2013}
{Willingale}, R., {Starling}, R.~L.~C., {Beardmore}, A.~P., {Tanvir}, N.~R., \&
  {O'Brien}, P.~T. 2013, \mnras, 431, 394, \dodoi{10.1093/mnras/stt175}

\bibitem[{{Woosley} \& {Bloom}(2006)}]{Woosley2006}
{Woosley}, S.~E., \& {Bloom}, J.~S. 2006, \araa, 44, 507,
  \dodoi{10.1146/annurev.astro.43.072103.150558}

\bibitem[{{Woosley} \& {Heger}(2006)}]{WoosleyHeger2006}
{Woosley}, S.~E., \& {Heger}, A. 2006, \apj, 637, 914, \dodoi{10.1086/498500}

\bibitem[{Woosley {et~al.}(2002)Woosley, Heger, \& Weaver}]{Woosley2002}
Woosley, S.~E., Heger, A., \& Weaver, T.~A. 2002, Rev. Mod. Phys., 74, 1015,
  \dodoi{10.1103/RevModPhys.74.1015}

\bibitem[{{Yao} {et~al.}(2019){Yao}, {Miller}, {Kulkarni}, {Bulla}, {Masci},
  {Goldstein}, {Goobar}, {Nugent}, {Dugas}, {Blagorodnova}, {Neill}, {Rigault},
  {Sollerman}, {Nordin}, {Bellm}, {Cenko}, {De}, {Dhawan}, {Feindt},
  {Fremling}, {Gatkine}, {Graham}, {Graham}, {Ho}, {Hung}, {Kasliwal},
  {Kupfer}, {Laher}, {Perley}, {Rusholme}, {Shupe}, {Soumagnac}, {Taggart},
  {Walters}, \& {Yan}}]{Yao2019}
{Yao}, Y., {Miller}, A.~A., {Kulkarni}, S.~R., {et~al.} 2019, \apj, 886, 152,
  \dodoi{10.3847/1538-4357/ab4cf5}

\bibitem[{{York} {et~al.}(2000){York}, {Adelman}, {Anderson}, {Anderson},
  {Annis}, {Bahcall}, {Bakken}, {Barkhouser}, {Bastian}, {Berman}, {Boroski},
  {Bracker}, {Briegel}, {Briggs}, {Brinkmann}, {Brunner}, {Burles}, {Carey},
  {Carr}, {Castander}, {Chen}, {Colestock}, {Connolly}, {Crocker}, {Csabai},
  {Czarapata}, {Davis}, {Doi}, {Dombeck}, {Eisenstein}, {Ellman}, {Elms},
  {Evans}, {Fan}, {Federwitz}, {Fiscelli}, {Friedman}, {Frieman}, {Fukugita},
  {Gillespie}, {Gunn}, {Gurbani}, {de Haas}, {Haldeman}, {Harris}, {Hayes},
  {Heckman}, {Hennessy}, {Hindsley}, {Holm}, {Holmgren}, {Huang}, {Hull},
  {Husby}, {Ichikawa}, {Ichikawa}, {Ivezi{\'c}}, {Kent}, {Kim}, {Kinney},
  {Klaene}, {Kleinman}, {Kleinman}, {Knapp}, {Korienek}, {Kron}, {Kunszt},
  {Lamb}, {Lee}, {Leger}, {Limmongkol}, {Lindenmeyer}, {Long}, {Loomis},
  {Loveday}, {Lucinio}, {Lupton}, {MacKinnon}, {Mannery}, {Mantsch}, {Margon},
  {McGehee}, {McKay}, {Meiksin}, {Merelli}, {Monet}, {Munn}, {Narayanan},
  {Nash}, {Neilsen}, {Neswold}, {Newberg}, {Nichol}, {Nicinski}, {Nonino},
  {Okada}, {Okamura}, {Ostriker}, {Owen}, {Pauls}, {Peoples}, {Peterson},
  {Petravick}, {Pier}, {Pope}, {Pordes}, {Prosapio}, {Rechenmacher}, {Quinn},
  {Richards}, {Richmond}, {Rivetta}, {Rockosi}, {Ruthmansdorfer}, {Sandford},
  {Schlegel}, {Schneider}, {Sekiguchi}, {Sergey}, {Shimasaku}, {Siegmund},
  {Smee}, {Smith}, {Snedden}, {Stone}, {Stoughton}, {Strauss}, {Stubbs},
  {SubbaRao}, {Szalay}, {Szapudi}, {Szokoly}, {Thakar}, {Tremonti}, {Tucker},
  {Uomoto}, {Vanden Berk}, {Vogeley}, {Waddell}, {Wang}, {Watanabe},
  {Weinberg}, {Yanny}, {Yasuda}, \& {SDSS Collaboration}}]{York2000}
{York}, D.~G., {Adelman}, J., {Anderson}, John~E., J., {et~al.} 2000, \aj, 120,
  1579, \dodoi{10.1086/301513}

\bibitem[{{Young}(2004)}]{Young2004}
{Young}, T.~R. 2004, \apj, 617, 1233, \dodoi{10.1086/425675}

\bibitem[{{Zhang} \& {MacFadyen}(2009)}]{Zhang2009}
{Zhang}, W., \& {MacFadyen}, A. 2009, \apj, 698, 1261,
  \dodoi{10.1088/0004-637X/698/2/1261}

\end{thebibliography}

\end{document}